\documentclass{article}

\usepackage{graphicx}
\usepackage{psfrag}

\usepackage{amssymb,amsmath}

\usepackage{amsfonts}

\def\ggamma{\overline{\gamma}}
\def\hh{\overline{A}}
\def\zz{z}
\def\U{{\cal U}}

\def\tr{\mbox{tr}}

\def\M{{\cal M}}
\def\N{{\cal N}}
\def\H{{\cal H}}
\def\D{{\cal D}}
\def\T{{\cal T}}
\def\intSigma{\Sigma^{\circ}}

\def\yy{{\cal Y}}
\def\tt{\Theta}

\def\C{C}

\DeclareFontFamily{OT1}{rsfs}{}
\DeclareFontShape{OT1}{rsfs}{m}{n}{ <-7> rsfs5 <7-10> rsfs7 <10-> rsfs10}{}
\DeclareMathAlphabet{\mycal}{OT1}{rsfs}{m}{n}

\def\scri{{\mycal I}}

\def\g{a}

\def\etagamma{\bm{\eta_{\gamma}}}

\newcommand\bm[1]{\mbox{\boldmath$#1$}}

\newcommand{\nablanor}{\nabla^{\bot}}
\newcommand{\nablafour}{{\nabla^{g}}}

\def\book#1#2#3#4#5#6#7#8{#1, ``#2'' in {\it #3}, #5 #6 (#7), #4 (#8).}

\def\JournalPrep#1#2#3{#1, ``#2'', #3.}

\def\Journal#1#2#3#4#5#6{#1, ``#2'', {\em #3} {\bf #4}, #5 (#6).}
\def\JGP{J. Geom. Phys.}
\def\JDG{J. Diff. Geom.}
\def\CQG{Class. Quantum Grav.}
\def\JPA{J. Phys. A: Math. Gen.}
\def\PRD{{Phys. Rev.} \bm{D}}
\def\PR{{Phys. Rev.} }
\def\GRG{Gen. Rel. Grav.}

\def\PR{Phys. Rev.}

\def\JMP{J. Math. Phys.}

\def\CMP{Commun. Math. Phys.}
\def\PRL{Phys. Rev. Lett.}

\def\AIHPC{Ann. Inst. H. Poincar\'e, Sect C}
\def\ANYAS{Ann. N. Y. Acad. Sci.}

\def\NC{Nuovo Cimento}
\def\AM{Ann. of Math.}
\def\APP{Acta Phys. Pol.}

\def\CPAM{Commun. Pure Appl. Math.}
\def\PJM{Pacific J. Math.}
\def\AGAG{Ann. Global Anal. Geom.}
\def\AHP{Annales Henri Poincar\'e}
\def\PRSL{Proc. Roy. Soc. London}
\def\ATMP{Adv. Theor. Math. Phys.}
\def\CRASP{C.R Acad. Sci. Paris S\'er. A-B}

\def\CAG{Commun. Anal. Geom.}
\def\IMRN{Int. Math. Res. Not.}
\def\AMM{Amer. Math. Month.}
\def\PAMS{Proc. Amer. Math.  Soc.}
\def\PLB{Physics Letters B}
\def\AnP{Annals Phys.}
\def\ASL{Adv. Sci. Lett.}
\def\NAR{New Astron. Rev}
\def\CJP{Can. J. Phys.}
\def\NAMS{Not. Amer. Math. Soc.}

\newtheorem{theorem}{Theorem}
\newtheorem{conjecture}{Conjecture}

\begin{document}

\title{Present status of the Penrose inequality}

\author{Marc Mars, \\
Facultad de Ciencias, Universidad de Salamanca, \\
Plaza de la Merced s/n,
37008 Salamanca, Spain}

\maketitle

\begin{abstract}
The Penrose inequality gives a lower bound for the total mass of a
spacetime in terms of the area of suitable surfaces that represent
black holes. Its validity is supported by the cosmic censorship 
conjecture and therefore its proof (or disproof) is an important problem
in relation with gravitational collapse. The Penrose inequality is a
very challenging problem in mathematical relativity and it has
received continuous attention since its formulation by Penrose in the
early seventies. Important breakthroughs have been made in the last
decade or so, with the complete resolution of the so-called Riemannian
Penrose inequality and a very interesting proposal to address the
general case by Bray and Khuri. In this paper, the most important
results on this field will be discussed and the main ideas
behind their proofs will be summarized, with the aim of presenting what is
the status of our present knowledge in this topic.
\end{abstract}

\section{Introduction}
\label{introduction}

In the early seventies, Penrose \cite{Penrose1972, Penrose1973}
showed that by combining several
ingredients of the ``establishment viewpoint'' of gravitational
collapse, an inequality of the form 
\begin{eqnarray}
M \geq \sqrt{\frac{A}{16 \pi}} \label{PI1}
\end{eqnarray}
follows, where $M$ is the total mass and $A$ the area
of a black hole. Cosmic censorship is one of the fundamental
ingredients of the argument, and by far the weakest one.
Thus, finding a counterexample
of (\ref{PI1}) would very likely involve a spacetime for which cosmic
censorship  fails to hold.
In fact, this was Penrose's original motivation to study the inequality. 
On the other hand, a proof of a suitable version of (\ref{PI1}) 
would give indirect support to cosmic censorship. Inequalities of this
type are collectively termed ``Penrose inequalities'' 
(sometimes also ``isoperimetric inequality for black holes'' \cite{Gibbons1972})
and finding suitable versions thereof and trying to 
prove them has become a major task in mathematical relativity. After a first
period of heuristic proofs and partial results, important
breakthroughs have been made in the last ten years. 
The aim of this
review is to try to explain the problem and describe the main 
approaches that have been followed.

The first observation to be made is the necessity of replacing 
the area of the black hole in (\ref{PI1}) by the area of a suitable alternative surface.
This is because in order to determine whether a spacetime is a black hole, detailed knowledge of its global
future behaviour is required. On the other hand, cosmic
censorship is precisely a statement on the global future evolution of a spacetime. 
In order to have an inequality logically independent of cosmic censorship (although
motivated by it) the area
on the black hole must be replaced by the area of a surface which can be located
independently of the global future behaviour of the spacetime (for instance, directly in terms
of the initial data) and which is guaranteed (or at least expected)
to have less or equal area than the event
horizon that may eventually form during the evolution.

The global setup which supports the validity of (\ref{PI1}) is well-known and, in rough terms, 
goes as follows. Assume a spacetime $(\M,g)$ which is asymptotically flat in the sense of being
strongly asymptotically predictable, admitting
a complete future null infinity $\scri^{+}$ and satisfying $J^{-}(\scri^{+}) \neq \M$
(see \cite{Wald1984} for definitions). The event
horizon $\H$ is the boundary of $J^{-} (\scri^{+})$ and it is, therefore, a null hypersurface at least 
Lipschitz continuous.
Assume, moreover, that the spacetime admits an asymptotically flat partial Cauchy surface
with total ADM energy $E_{ADM}$ and which intersects $\H$ on a cut $S$.
If $\H$ is a smooth hypersurface then this cut
is a smooth embedded surface which has a well-defined area
$|S|$. For general event horizons, the area $|S|$ still makes sense
provided it is interpreted as its 2-dimensional Hausdorff measure (the Hausdorff measurability
of $S$ is demonstrated in \cite{Chrusciel2001}). Consider now any cut $S_1$ to the causal future
of $S$ along the event horizon.
The black hole area law \cite{Hawking1971,Hawking1972}
states $|S_1| \geq |S|$ provided the null energy condition holds.
This area theorem was proven
for general event horizons in 
\cite{Chrusciel2001} under much milder asymptotic
conditions. From physical principles, the spacetime is expected to settle down to some
equilibrium configuration. Assuming also that all the matter
fields are swallowed by the black hole in the process (an external electromagnetic field would 
not alter the conclusions, see Sect. \ref{stronger}), the uniqueness theorems for stationary black holes
(see e.g. \cite{Heusler1996}) imply that the spacetime
must approach the Kerr metric (modulo several technical conditions that 
still remain open, see \cite{Chrusciel2008} for a recent account). For the
Kerr metric, the area of the event horizon  
$A_{\mbox{\tiny Kerr}}$
is independent of the cut  (as for any Killing horizon) and takes the value (in units
$G=c=1$) $A_{\mbox{\tiny Kerr}} = 8 \pi M \left (M + \sqrt{M^2 - L^2/M^2}
\right ) \leq 16 \pi M^2$ where $M$ and $L$
are respectively the total mass and total angular momentum of the spacetime
(we do not use the more common term ``$J$'' for the angular momentum
to avoid confusion with the energy flux used later). In particular,
$M$ should be the asymptotic value of the Bondi mass along $\scri^{+}$. Since
gravitational waves carry positive energy, the Bondi mass cannot increase
to the future \cite{Bondi1962, Sachs1962}. Provided the Bondi mass approaches the ADM mass
$M_{ADM}$ of the initial slice (which is only known under additional assumptions, see
\cite{Ashtekar1979, Hayward2003, Kroon2003, Zhang2006}), the inequality $M_{ADM} \geq \sqrt{|S|/16 \pi}$ follows.
This inequality is still global in the sense that the cut $S$ of the event horizon cannot be 
determined directly in terms of the initial data. Penrose's idea was to consider situations in which
one could estimate the area of the cut from below in terms of the area of some surface that could 
be located without having to solve the whole future evolution.

One such situation occurs when the initial data set is asymptotically euclidean
and contains a future trapped surface (see below for definitions).
Then, the singularity theorems of Penrose \cite{Penrose1965}, Hawking \cite{Hawking1967}
and others (see \cite{Senovilla1998} for a review) state that, provided 
the strong energy condition holds (in some cases the null energy condition suffices),
the maximal globally hyperbolic development of this data
must contain  a singularity, i.e. an incomplete inextendible causal geodesic.  
Very little is known in general about the nature of the singularity that forms.
More specifically it is not known whether
the future development admits a complete $\scri^+$ and therefore defines a black hole
spacetime. The weak cosmic censorship conjecture, first 
proposed by Penrose \cite{Penrose1969}, asserts that all singularities 
lie behind an event horizon  and therefore are invisible to an observer at infinity.
Not much  is known about the general validity of this conjecture, which remains
a fundamental open problem in gravitational collapse physics, see 
\cite{Wald1997}. Rigorous results are available only in the case of spherical symmetry, where
the conjecture has been proven for several matter models \cite{Christodoulou1999, Dafermos2005}.

Under cosmic censorship, the initial data containing a future trapped
surface $S$ develops a black hole spacetime.
A general result on black hole spacetimes \cite{HawkingEllis, Claudel2000} 
is that  no future trapped surface can enter into the causal past of 
$\scri^{+}$. 
Therefore, the intersection of the event horizon and the initial data set $\Sigma$ defines
a spacelike surface ${\H}_{\Sigma}$
that separates $S$ from the asymptotic region. In general, the
location of $\H_{\Sigma}$ cannot be determined directly 
from the initial data. Moreover, $\H_{\Sigma}$ can have smaller area than  $S$, even though it lies in its exterior.
Nevertheless, under suitable
restrictions on $S$ (details will be given below) it makes sense to consider all surfaces
in $\Sigma$ enclosing $S$. The infimum of the areas of all such surfaces, denoted by $A_{\min}(S)$,
has the obvious property that
$|\H_{\Sigma} | \geq A_{\min}(S)$. Consequently, 
the inequality
\begin{eqnarray}
M_{ADM} \geq \sqrt{\frac{A_{\min}(S)}{16 \pi}} \label{LocalPI}
\end{eqnarray}
follows from Penrose's heuristic argument. 
The need of using the minimum area enclosure of $S$ in (\ref{LocalPI})
was first noticed by Jang
and Wald  \cite{JangWald1977} and the first example showing that $S$ may have greater area
than surfaces enclosing it is due to Horowitz \cite{Horowitz1984}.

Inequality (\ref{LocalPI}) involves objects
defined solely in terms of the local geometry of the initial data set and its validity 
can therefore be addressed independently of whether weak cosmic censorship (or any other of the
ingredients entering into the argument) holds or not.
It is clear that, since $A_{\min}(S)$ depends on the hypersurface $\Sigma$
containing $S$, we can still take the supremum of the right-hand side
with respect to all asymptotically flat hypersurfaces containing $S$, and the resulting inequality still follows 
from Penrose's heuristic argument. This gives a clearly stronger inequality. However,
it depends on the piece of spacetime available and therefore looses the desirable
property of depending solely on objects defined on the initial data set. 

If the initial data set is asymptotically hyperbolic instead of asymptotically flat 
(i.e. such that it intersects future null infinity in its Cauchy development), the same
heuristic argument gives (\ref{LocalPI}) with the Bondi mass replacing the  ADM mass in the
left hand side.

Another setup where a local geometric inequality is implied by the global heuristic argument
of gravitational collapse appears in the seminal 
paper by Penrose \cite{Penrose1973} and consists of a null shell $\N$
(with compact cross sections)  of collapsing dust in 
Minkowski spacetime. As described in more detail below,
for any given shape of the shell (restricted to be convex at each instant of time in order
to avoid shell crossings in the past) and any chosen cross section $S$ on the shell, the energy density of
the collapsing dust
can be adjusted so that $S$ is 
a marginally trapped surface with respect to the geometry of the spacetime left after the shell has passed.
Since a singularity will definitely form in this setup (the shell has self-intersections in its future),
cosmic censorship predicts the formation of a black hole. Similarly as before,
the marginally trapped surface $S$ cannot enter the causal past of $\scri^{+}$
\cite{ChruscielGalloway2008}. The intersection of the event horizon with the shell $\N$ must therefore
lie in the causal past of $S$. However, since the shell is convex and collapsing, the 
event horizon cut automatically has at least the same area as $S$. Consequently, the heuristic collapse
argument implies $M \geq \sqrt{|S|/16\pi}$, where $M$ is the mass of the shell. By energy conservation,
this can be computed directly on 
$S$ in terms of its geometry as a surface in Minkowski. Thus, a geometric inequality is obtained for
a class of spacelike surfaces in Minkowski spacetime. The status of this version of the
Penrose inequality will be discussed in detail later on.

All versions of the Penrose inequality have the structure of a lower bound of 
the total mass of the spacetime in terms of the area of suitably chosen surfaces. 
Therefore, they can be regarded as strengthenings of the positive mass theorem, which says
that the total mass of an asymptotically flat spacetime cannot be negative, under suitable energy and completeness
conditions.
The positive mass theorem
also has a rigidity part, namely that the total mass vanishes 
only for the Minkowski spacetime. The Penrose inequality conjecture also has a rigidity part, which in rough terms
states that equality will only happen for the Schwarzschild spacetime. Heuristically this can
be understood because, in the case of equality,
the final mass of the spacetime must coincide with the starting one. 
Therefore, no gravitational 
waves are emitted in the process. This suggests that the whole configuration should be stationary.
However, the only
stationary, vacuum black hole is the Kerr spacetime, and equality happens for this metric only 
if the
angular momentum is zero, i.e. if the metric is, in fact, the Schwarzschild spacetime.

Since the original proposal by Penrose this topic has become an active area of research.
However, the problem has proven
to be a difficult one and relatively little progress was made during the first decades.
The most important
contribution in this period is due to Geroch \cite{Geroch1973}, who observed that a suitable functional defined 
on surfaces embedded in a spacelike hypersurface $(\Sigma,\gamma)$ is monotonically increasing
if the curvature scalar of $\gamma$ is non-negative and the surfaces are moved outwards
at a speed which is inversely proportional to the mean curvature of the surface at each point.
This is the so-called {\it inverse mean curvature flow}. This functional (now called Geroch mass) has the 
property of approaching the ADM energy of the hypersurface (provided this is asymptotically euclidean)
if the surfaces become sufficiently spherical at infinity. 
Geroch's original idea was to prove the positive mass theorem by starting the inverse mean curvature
flow at a point (where the Geroch mass vanishes). This idea was then adapted by Jang and Wald \cite{JangWald1977},
who noticed that
the Geroch mass coincides exactly with the right-hand side of (\ref{PI1}) if the starting surface is connected,
of spherical topology and minimal (i.e. with vanishing mean curvature). If the flow existed globally and the leaves
could be seen to approach round spheres at infinity, then the monotonicity of the Geroch mass would imply
the Penrose inequality in the particular case of time-symmetric initial data sets. The flow
however, generically develops singularities and therefore the argument could not be made
rigorous at the time. The only case where the Penrose inequality could be proven to hold was in
spherical symmetry (irrespectively of whether the initial data set was time-symmetric or not) using the so-called
Hawking mass, which is a generalization of the Geroch mass when the second fundamental form is not zero.

In the late nineties, however, two important breakthroughs were made. First of all, Huisken and Ilmanen 
\cite{HuiskenIlmanen2001} were able to
prove that Geroch's heuristic derivation could be turned into a rigorous proof. This required deep results
in geometric analysis and geometric measure theory. These authors established therefore the validity of the
Penrose inequality for asymptotically euclidean Riemannian manifolds of non-negative Ricci curvature and having
a boundary consisting of an outermost minimal surface (this is now called the {\it Riemannian
Penrose inequality}). Although this boundary was allowed to be disconnected,
the Penrose inequality could only be established for the area of any of its connected components. Shortly afterwards,
Bray \cite{Bray2001} was able to prove the Riemannian Penrose inequality in full generality (i.e. in terms
of the total area of the outermost minimal surface, independently of whether this is connected or not). 
Bray's argument is completely different to the  previous one and uses a deformation of the given metric
in such a way that all the non-trivial geometry gets swallowed inside the minimal surface while
the total mass of the space does not increase and the area of the horizon does not decrease. This process settles
down into an equilibrium state given by the Schwarzschild metric. Since the Penrose inequality is fulfilled
in the final state (in fact, with equality), the Penrose inequality holds also for the original space. 
These two fundamental results have boosted tremendously the interest in the Penrose inequality, which has become
a very important topic in mathematical relativity. Although the general case is still open, several ideas have been
proposed to approach it. In particular, an important step forward has been made very recently by
Bray and Khuri \cite{BrayKhuri2009}. Although several issues remain still open regarding this proposal (in particular,
the existence of solutions of certain PDE problems) the idea is indeed very promising.

The aim of this review is to present the most important developments in this field. I have chosen not to follow
a historical order, and rather I have tried to organize the presentation in a way which I consider logically
convenient. Since several approaches share some of the techniques, I have collected many preliminary results
in one section (Sect. \ref{definitions}). While this has the potential disadvantage that, in a first reading,
it is not clear why and where such results are needed, it has the advantage of simplifying the location of the
required tools. Readers wishing to enter straight into the topic of the Penrose inequality
may skip this section and refer back to it whenever necessary.

An important warning is in order. The topic ``Penrose inequalities'' is vast and has many
ramifications. Although I will try to cover the most important results in the field, I make
no claim of exhaustivity. There are several other
review papers on this topic in the literature. For the early results the reader is advised
to look at \cite{Malec1991}. The breakthroughs of Huisken \& Ilmanen and Bray triggered the publication 
of a number of interesting reviews, where the proofs of the Riemannian Penrose inequality
were treated in detail, see 
\cite{HuiskenIlmanen1997}, \cite{HuiskenIlmanen1998},
\cite{Herzlich2000}, \cite{Bray2002}, \cite{Bray2002-2}, \cite{BrayChrusciel2004}, 
\cite{Schoen2005}, \cite{Mars2007}.

The structure of the review is as follows. In section \ref{definitions} 
the  basic facts
that will be needed later are introduced. This section is divided into five subsections. In subsection
\ref{codimension-two} the geometry of $(n-2)$-dimensional surfaces as submanifolds of 
an $n$-dimensional spacetime is described, and several types of surfaces are defined. In
subsection \ref{variation_null_expansion} some relevant variational formulas for the null expansions are
summarized. In subsection \ref{embedded}, the geometry of codimension-two surfaces as submanifolds
of spacelike hypersurfaces is discussed and two important existence theorems for outermost surfaces
with suitable properties are  reviewed. In subsection \ref{UEF}, the Hawking and Geroch quasi-local masses,
for surfaces in four dimensional spacetimes and their general variation formulas are discussed. Subsection
\ref{asymptoticflatness} recalls the notion of asymptotically euclidean. The following Section (Sect. \ref{Formulations})
is devoted to the discussion of the various formulations of the 
Penrose inequality that have been proposed. Section \ref{Spher} deals with the spherically symmetric
case. In Section \ref{Riemannian} the so-called Riemannian Penrose inequality is treated. 
%This refers
%to the case when the second fundamental form of the initial data set is identically zero. 
The main ideas behind the remarkable proofs of Huisken and Ilmanen and of Bray
are discussed in subsections 
\ref{HuiskenIlmanen} and \ref{Braysproof} respectively. However, 
previous approaches based on spinors and isoperimetric methods which admit 
interesting generalizations  are also covered in subsections \ref{spinors} and \ref{isoperimetricprofile}. 
Section \ref{hyperbolicSect} discusses the Penrose inequality when the 
initial data set is asymptotically hyperbolic instead of asymptotically euclidean. Section 
\ref{GeneralPI} is devoted to describing different attempts that have been
proposed to address the 
Penrose inequality in the non time-symmetric case. Particular attention is paid to
the  very promising recent ideas put forward by Bray and Khuri \cite{BrayKhuri2009}.
Section \ref{stronger} treats several strengthenings of the Penrose
inequality when particular matter fields or symmetries are present. Section \ref{applications} is devoted
to describing some applications
of the Penrose inequality. The paper finishes with a some concluding remarks in Section \ref{concluding}.

\section{Basic facts and definitions}
\label{definitions}

A spacetime $(\M,g)$ is a connected Hausdorff 
manifold endowed with a smooth metric
of Lorentzian signature (with sign convention $\{-,+,+ \cdots\}$). 
Smoothness is assumed for simplicity, many of the results below 
hold under  weaker differentiability assumptions. We further
take $\M$ orientable and $(\M,g)$ time-orientable and that an orientation
for both has been chosen. In this review we  will be mainly concerned
with four-dimensional spacetimes, although higher dimensional results will
be mentioned at some places. $\M$ will always be four-dimensional unless 
explicitly stated.

The Penrose inequality involves the area of codimension-two surfaces.
Let us therefore start with some basic properties concerning their geometry.

\subsection{Geometry of codimension two surfaces}
\label{codimension-two}

$S$ will denote a compact, embedded, oriented, codimension-two surface
in an $n$-dimensional spacetime $(\M,g)$ defined via
an  embedding $\Phi: S \rightarrow \M$ ($S$ will usually be identified 
with its image). The induced first fundamental
form on $S$, denoted by $h$, is assumed to be positive definite, i.e.
$S$ is spacelike. Such an object will be called simply ``surface''. The
area of $S$ is denoted by $|S|$.

At $p \in S$ we denote by $T_p S$ and $N_p S$ the tangent and normal
spaces of $S$. This implies $T_p \M = T_p S \oplus N_p S$. For any 
vector $\vec{V}$ at $p$ this decomposition defines a parallel and a normal
vector
according to $ \vec{V} =
\vec{V}^{\parallel} + \vec{V}^{\bot}$.
The second fundamental form vector $\vec{K}$ 
of $S$ is defined, as usual, as $\vec{K} (\vec{X},\vec{Y}) 
= - ( \nablafour_{\vec{X}} \vec{Y} )^{\bot}$ where $\vec{X}$ and $\vec{Y}$
are tangent to $S$ and $\nablafour$ is the covariant derivative on $(\M,g)$.
This tensor is symmetric and its
trace $\vec{H} = \tr_{h} \vec{K}$ defines the mean curvature vector. 
The trace-free part of $\vec{K}$ will be called $\vec{\Pi}$ in the following.

$S$ being oriented and the ambient spacetime being oriented and time oriented,
it follows easily that there exist (globally on $S$) 
two linearly independent null vector fields $\vec{l}^{+}$ and
$\vec{l}^{-}$ orthogonal to $S$. These vectors, which 
span the normal bundle $NS = \cup_p N_pS$, will always be chosen to be
future directed and satisfying
$(\vec{l}^{+} \cdot \vec{l}^{-}) = -2$ 
(dot stands for scalar product with the spacetime metric $g$).
They are uniquely defined up to
rescalings 
$\vec{l}^{+}  \rightarrow  F \vec{l}^{+}$, $\vec{l}^{-} \rightarrow  
F^{-1} \vec{l}^{-}$ ($F>0$),
plus interchange $\vec{l}^{+} \leftrightarrow \vec{l}^{-}$. 
The null expansions of $S$
are defined as $\theta_{\pm} = (\vec{H} \cdot \vec{l}^{\pm} )$
and they
contain the same information as $\vec{H}$ since
$\vec{H} = - \frac{1}{2} \left ( \theta_- \vec{l}^{+} +
\theta_{+}\vec{l}^{-} \right )$.  

A surface $S$ can be classified according to the
causal character of $\vec{H}$. Unfortunately, there is no unique and generally accepted
agreement on how to call the various types of surfaces which arise. In this paper, 
the following notation will be used  (in our convention, the zero vector is both future null and
past null):

%I shall make no attempt
%to discuss the various definitions and their interrelationships. The paper by
%Hayward \cite{Hayward94}
%where several of these definitions appeared for the first time (with
%different names) should, however, be mentioned.
%I also refer to 
%for a very 
%
%according to $\vec{H}$.

\begin{itemize}
\item[] {\bf Future trapped:} $\vec{H}$ is timelike and future directed everywhere (equivalently
$\theta_{+}<0$ and $\theta_{-}<0$)
\item[] {\bf Weakly future trapped:} $\vec{H}$ is causal and future directed everywhere
($\theta_{+} \leq 0$, $\theta_{-} \leq 0$).
\item[] {\bf Marginally future trapped:} $\vec{H}$ is proportional to one of the null normals
$\vec{l}^{\pm}$ with a non-negative proportionality factor ($\theta_{+} =0$ and 
$\theta_{-} \leq 0$, or viceversa).
\end{itemize}
{\bf Past trapped}, {\bf weakly past trapped} and {\bf marginally past trapped} are defined
by reversing all inequalities. 

All these types of surfaces have a 
mean curvature vector of definite causal character and time orientation. For brevity, when no indication to future
or past is given, then future should be understood (i.e. a ``trapped surface'' is meant to be a 
future trapped surface, and similarly for the other cases).

It is often useful to consider surfaces for which 
one of the null normals can be geometrically selected. This preferred normal
will be called ``outer null normal'' and will always be denoted as $l^{+}$.
Notice that ``outer'' here does not necessarily refer to
any notion of exterior to $S$, so this definition must be used with
special care. Nevertheless, when the null normals can be
geometrically distinguished, the corresponding null expansions are  also
geometrically distinct and the following definitions (which place no
restriction on $\theta_{-}$) become of interest.
\begin{itemize}
\item[] {\bf Weakly outer trapped:} $\theta_{+} \leq 0$.
\item[] {\bf Marginally outer trapped or MOTS:} $\theta_{+} =0$ (equivalently,
$\vec{H}$ points along the outer null normal).
\item[] {\bf Weakly outer untrapped:} $\theta_{+} \geq 0$.
\item[] {\bf Outer untrapped:} $\theta_{+} > 0$.
\end{itemize} 

These definitions depend not only on the choice of outer direction, but also on the time
orientation of the spacetime. If the time orientation is reversed (without 
modifying the outer direction) then $-\vec{l}^{-}$ becomes the future outer null
direction. Therefore by adding the word ``past'' to any of the four definitions above, the surface
is meant to satisfy the same inequalities with $\theta_{+}$ replaced by $-\theta_{-}$. For instance,
a {\bf past weakly outer trapped} surface satisfies $\theta_{-} \geq 0$.

%Similar definitions can be given 
%By adding the word ``past'' to any of these surfaces
%will mean replacing $\theta_{+}$ by $-\theta_{-}$ in the defining
%expressions (this amounts to keeping the outer direction and changing
%the time orientation of the spacetime).

Very recently Bray and Khuri \cite{BrayKhuri2009} have proposed a 
version of the Penrose inequality which involves a type of surfaces
whose definition is insensitive to
reversals of time direction. However, they still require a preferred
outer direction.
\begin{itemize}
\item[] {\bf Generalized trapped surface:} 
At each point, either $\theta_{+} \leq 0$ or $\theta_{-} \geq 0$ (or both).
\item[] {\bf Generalized apparent horizon:} At each point, either
$\theta_+=0$ and $\theta_{-} \leq 0$, or
$\theta_-=0$ and $\theta_{+} \geq 0$. Notice that generalized
apparent horizons are in particular generalized 
trapped surfaces.
\end{itemize}
In the detailed classification of surfaces 
according to its mean curvature vector presented in \cite{Senovilla2007},
generalized apparent horizons are termed {\it null untrapped}, which,
in my opinion, has the advantage of being  more descriptive. In fact, my preferred term for these surfaces 
would be {\it null outer untrapped} because this emphasizes the fact that they
have a  privileged ``outer'' direction. However,
notation is always a very personal matter and, in this review, I will stick to the name {\it generalized
apparent horizon} put forward by Bray and Khuri in their new proposal of the Penrose inequality.

In terms of the mean curvature vector, a generalized trapped surface is one where
$\vec{H}$ is allowed to point everywhere except along a
spacelike outer direction (defined to be a spacelike direction with positive scalar
product with $\vec{l}^{+}$). A generalized apparent horizon has 
a mean curvature vector which is null everywhere and moreover, has non-negative
scalar product with any outer spacelike direction. Notice that the second condition
is ensured provided the scalar product with {\it one} outer spacelike direction is non-negative.

Generalized trapped surfaces are indeed generalizations of the previous concepts. 
In particular, weakly future trapped and weakly outer trapped surfaces are
automatically generalized trapped surfaces, and the same is true
for marginally outer trapped surfaces.

\begin{table}[h!]
\begin{center}
\begin{tabular}{|l|c|}
\hline
Name of surface & Null expansions \\
\hline
\hline
{\bf Future trapped}  &  $\theta_{+}<0$ and $\theta_{-}<0$ \\
\hline
{\bf Weakly future trapped} & $\theta_{+} \leq 0$ and $\theta_{-} \leq 0$ \\
\hline
{\bf Marginally future trapped} & ($\theta_{+} =0$ and $\theta_{-} \leq 0$) or
($\theta_{+} \leq 0$ and $\theta_{+} = 0$) \\
\hline
{\bf Weakly outer trapped} & $\theta_{+} \leq 0$  \\
\hline
{\bf Marginally outer trapped (MOTS)} & $\theta_{+} =0$  \\
\hline
{\bf Weakly outer untrapped} &  $\theta_{+} \geq 0$  \\
\hline
{\bf Outer untrapped} & $\theta_{+} > 0$  \\
\hline
{\bf Generalized trapped surface} & At no point, $\theta_{+} \geq 0$ and $\theta_{-} \leq 0$ \\
\hline
{\bf Generalized apparent horizon} &  At each point, $\theta_{+} \theta_{-}=0$ and
$ \theta_{+} - \theta_{-} \geq 0$ \\
\hline
\end{tabular}
\caption{Types of surfaces according to their null expansion(s).}
\end{center}
\end{table}

%//
%
%Notice that this definition of  outer direction is purely local and may not coincide with any
%other notion of ``outer'' which may be present (e.g. when $S$ lies in a hypersurface
%and separates a compact region from an unbounded one,
%where outer would be naturally defined to be the direction pointing outside
%the compact region). 
%
%//

\subsection{First order variations of the null expansions}
\label{variation_null_expansion}

An important technical tool that will be used often below is the first order
change of the null expansions
$\theta_{\pm}$ under variations of the surface.
In this section we write down the variations of $\theta_{\pm}$ along the null
normals $\vec{l}^{\pm}$,  which we write 
as $\pounds_{\vec{l}^{\pm}} \theta_{\pm}$. For the sake of completeness, let us first recall
briefly how  variations of a geometric object $F$ are defined. 
Any vector field $\vec{\xi}$ in a neighbourhood
of $S$ defines a {\it variation}
of the surface as follows. For small enough $\lambda \in \mathbb{R}$, let 
$\Phi_{\lambda}: S \rightarrow \M$ be the embedding defined by moving $p \in  S$ 
a parametric amount $\lambda$ along the integral curve of $\vec{\xi}$
starting at $p$. We write $S_{\lambda} = \Phi_{\lambda} (S)$. 
Assume, for definiteness,
that $F$ is a covariant tensor geometrically defined on any surface 
and take a variation $\Phi_{\lambda}$ of $S$.
Define $F_{\lambda} (p) = \Phi_{\lambda}^{\star} (F |_{\Phi_{\lambda}(p)})$,
i.e. the pull-back of the tensor $F$ 
attached to $S_{\lambda}$ at the point $\Phi_{\lambda} (p)$.  This defines
a curve of tensors at each point $p \in S$. The geometric variation is simply
$\pounds_{\vec{\xi}} F = \partial_{\lambda} F_{\lambda} (p) |_{\lambda=0}$. This derivative only depends
on the values of $\vec{\xi}$ on $S$ for truly geometric objects. 
However, when
$F$ requires additional structure for its definition, then derivatives of $\vec{\xi}$ on $S$
may also arise. This behaviour occurs for instance in $\pounds_{\vec{\xi}\,} \theta_{+}$ which 
requires making a specific choice of $\vec{l}_{+}$ on each surface $S_{\lambda}$.

Applying this procedure to the area of $S$ gives (see e.g \cite{Jost2001})
the {\it first variation of area} 
$\left . \frac{d |S_{\lambda}|}{d\lambda} \right |_{\lambda=0} = \int_S (\vec{H} \cdot \vec{\xi} ) \bm{\eta_S}$
where $\bm{\eta_S}$ is the metric volume form of $S$ 
In particular, the change of area 
of $S$ for variations along the null directions $\vec{l}^{\pm}$ are determined by the
integral of the null expansion $\theta_{\pm}$. 

Let us next write down $\pounds_{\vec{l}_{+}} \theta_{\pm}$. 
Since $\vec{l}^+$ and $\vec{l}^{-}$ are
interchangeable, the corresponding expression with also 
hold for variations along
$\vec{l}^-$ after making the substitution $+
\longleftrightarrow -$ everywhere.
We start with the variation of the  induced
metric, which is well-known to be
\begin{eqnarray*} 
\vec{l}_{+} (h_{AB}) = 2 \,  \vec{l}_{+} \cdot
\vec{K}_{AB},
\end{eqnarray*} 
where $A,B$ etc. are tensorial indices on $S$. This implies
\begin{eqnarray}
\vec{l}_{+} (\bm{\eta_S}) = \theta_{+} \bm{\eta_{S}}
\label{FirstVarArea}
\end{eqnarray}
for the variation of the volume form on $S$. For the variation of $\theta_{+}$ along $\vec{l}^{+}$, it is necessary
to relate the null normal $\vec{l}_{+}$ on each one of the varied
surfaces $S_{\lambda}$  to the corresponding null normal on the original
surface. As already stressed, this introduces a dependence on the
first derivative of $\vec{l}^{+}$ via $Q^{+} \equiv -1/2 (\vec{l}^{-} \cdot \nablafour_{\vec{l}^{+}} \vec{l}^{+} )
|_{\lambda=0}$ in the  resulting expression, which is the  well-known
Raychaudhuri equation
\begin{eqnarray} 
\pounds_{\vec{l}_{+}} \theta_{+} = Q^{+} \theta_{+} -
K^{+}_{AB} K^{+ \, AB}  - \mbox{Ric} \left ( \vec{l}_{+}, \vec{l}_{+}
\right), \label{lthetal}
\end{eqnarray} 
where $K^{\pm}_{AB} = ( \vec{l}^{\pm} \cdot
\vec{K}_{AB} )$ and   $\mbox{Ric}$ is the Ricci tensor of $(\M,g)$. A
more involved calculation gives the derivative of $\theta_{-}$ along
$\vec{l}_{+}$  (see e.g. \cite{AMS08})
\begin{eqnarray} 
\pounds_{\vec{l}_{+}} \theta_{-} =  - Q^{+}
\theta_{-} - K^{+}_{AB} K^{- \, AB}  - \mbox{Ric} (\vec{l}_{+},
\vec{l}_{-} ) + \frac{1}{2} \mbox{Riem}  ( \vec{l}_{+}, \vec{l}_{-},
\vec{l}_{+}, \vec{l}_{-} ) + 2 \left ( D_A S^A + S_A S^A \right),
\label{lthetak}
\end{eqnarray} 
where $\mbox{Riem}$ is the Riemann tensor of $(\M,g)$,
$D$ denotes the Levi-Civita covariant derivative of $(S, h)$ and the
one-form  $S_A$ is defined by
\begin{eqnarray} 
S (\vec{X}) = - \frac{1}{2} \left ( \vec{l}_{-} \cdot
\nablafour_{\vec{X}} \vec{l}_{+} \right ),
\label{Svector}
\end{eqnarray} 
where $\vec{X}$ is any tangent vector to $S$.
% = \frac{\partial
%\Phi^{\alpha}}{\partial y^A}$ with $y^A$ being local coordinates on
%$S$. 
The Gauss identity for $S$ as a submanifold of $(\M,g)$ implies
\begin{eqnarray*} 
R(g) = R(h) - ( \vec{H} \cdot \vec{H}) + \vec{K}_{AB} \cdot
\vec{K}^{AB} - 2 \mbox{Ric} (\vec{l}_{+}, \vec{l}_{-} ) + \frac{1}{2}
\mbox{Riem} (\vec{l}_{+}, \vec{l}_{-}, \vec{l}_{+}, \vec{l}_{-} ),
\end{eqnarray*}
 where $R(g)$ is the scalar curvature of
$(\M,g)$. Thus, (\ref{lthetak}) can be rewritten as
\begin{eqnarray} 
\pounds_{\vec{l}_{+}} \theta_{-} =  - Q^{+}
\theta_{-} + \mbox{Ein} (\vec{l}_{+},\vec{l}_{-} ) - \left ( R (h) -
(\vec{H}\cdot \vec{H}) \right )  +2  \left ( D_A S^A + S_A S^A \right ),
\label{lthetak2}
\end{eqnarray} 
where $\mbox{Ein}$ is the Einstein tensor of $g$. These
expressions are valid in any dimension. The expression for
$\pounds_{\vec{l}_{-}} \theta_{+}$ follows from (\ref{lthetak2}) by
interchanging $ + \leftrightarrow -$ and substituting 
$S_A \rightarrow -S_A$ (see formula (\ref{Svector}) above).

\subsection{Surfaces embedded in a spacelike hypersurface}
\label{embedded}

Codimension-two surfaces usually arise as codimension-one
surfaces embedded  in a
spacelike hypersurface $\Sigma$ of the spacetime $\M$. 
The induced metric on $\Sigma$ will be denoted by $\gamma_{ij}$ (Latin, lower case
indices run from 1 to 3) and the second fundamental form
with respect to the unit future normal $\vec{n}$ will be denoted by $A_{ij}$.
The  constraint equations relate the geometry of $\Sigma$ with
some  components of the Einstein tensor
\begin{eqnarray}
R(\gamma) - A_{ij} A^{ij} + 
(\tr_{\gamma} A)^2 = 16 \pi \rho, \label{HamiltonianConst} \\
\nabla_{j} A^{j}_{l} - \nabla_{l} \tr_{\gamma} A = - 8 \pi J_l, \label{DiffConstraint}
\end{eqnarray}
where $R(\gamma)$ is the curvature scalar of $\gamma$, $\nabla_i$ is the  
covariant derivative in $(\Sigma,\gamma)$, the total energy density is defined
as $8 \pi \rho \equiv \mbox{Ein} (\vec{n}, \vec{n} )$  and the
energy flux one-form $J_i$ is defined as
$8 \pi J_i X^i \equiv - \mbox{Ein} (\vec{X}, \vec{n} )$  on any 
vector $\vec{X}$ tangent to $\Sigma$. The initial data set satisfies the
{\it dominant energy condition} provided $\rho \geq |\vec{J}|$, where $|\vec{J}|$ is the norm
of $J^i$ with respect to the metric $\gamma_{ij}$. Notice that, despite their names,
$\rho$ and $\vec{J}$ are defined directly in terms of the Einstein tensor. No field equations
for are therefore assumed  either here or elsewhere in this paper.
The initial data set
is said to be  {\it time-symmetric} whenever $A_{ij}=0$.

%A spacetime $(\M,g)$ is said to satisfy the dominant energy condition (DEC),
%if the Einstein tensor $G^{\alpha}_{\beta}$ of $g$ maps future causal
%vectors into past causal vectors. DEC implies for the initial data the inequality
%\begin{eqnarray}
%\rho \geq \sqrt{J^i J_i} \hspace{4cm} \mbox {(DEC).} \label{DEC}
%\end{eqnarray}
%With a slight, but generally used, abuse of notation we will say that the initial
%data set satisfies DEC provided (\ref{DEC}) holds. 

%If the matter model
%is such that the Einstein-matter equations define a well-posed system, the initial data 
%set will generate a spacetime in the usual sense that there
%exists a unique globally hyperbolic
%spacetime $(\M,g)$ satisfying the Einstein-matter equations admitting a Cauchy surface
%isometric to $(\Sigma,\gamma)$ and with second fundamental form $A_{ij}$. 

Assume, as
before, that $S$ is orientable and that a preferred unit normal $\vec{m}$ tangent to  $\Sigma$ 
can be selected. In
this situation, the null normals will always be uniquely  chosen as
$\vec{l}^{\pm} = \vec{n} \pm \vec{m}$.  As a submanifold of $\Sigma$,
$S$ has second fundamental form $\kappa_{AB}$ with respect to
$\vec{m}$.  Its trace $p$ is the mean curvature of $S$ in $\Sigma$. The 
trace-free part of $\kappa_{AB}$ will be written as $\Pi^{\vec{m}}_{AB}$.

The decomposition of the second fundamental form $A_{ij}$ into
tangential and normal components to $S$ will also play a role. We will
denote by $A^S_{AB}$ the projection of $A_{ij}$ on $S$,
$\Pi^{\vec{n}}_{AB}$ the  trace-free part of this tensor and $q$ its
trace. Then, (see e.g.  \cite{Jost2001})  $\vec{K}_{AB} = - A^S_{AB}
\vec{n} + \kappa_{AB} \vec{m}$,  $\vec{\Pi}_{AB} = -
\Pi^{\vec{n}}_{AB} \vec{n} + \Pi^{\vec{m}}_{AB} \vec{m}$ and  $\vec{H}
= - q \vec{n} + p \vec{m}$, which implies $\theta_{\pm} = \pm p + q$.

The three remaining independent components of $A_{ij}$ can be encoded
in the trace $\tr_{\gamma} (A)$ and in the  normal-tangential components
\begin{eqnarray}
S_A \equiv A_{ij} m^i e^{j}_A. \label{DefS_A}
\end{eqnarray}
This definition of $S_A$ is
consistent with (\ref{Svector}) once  the choice of $\vec{l}^{\pm}$
described above is made. The general variation formulas 
(\ref{lthetal}) and (\ref{lthetak2}) can be rewritten in this context,
after a straightforward calculation which uses the constraint equation
(\ref{HamiltonianConst}), as
\begin{eqnarray}
\pounds_{\vec{m}} p & = & - \frac{1}{2} \Pi^{\vec{m}}_{AB} \Pi^{\vec{m} \, AB} - \frac{3}{4}  p^2
- \frac{1}{2} R(\gamma) + \frac{1}{2} R (h), \label{mp}\\
\pounds_{\vec{m}} q & = & p A_{ij} m^{i} m^j - A^{S}_{AB} \kappa^{AB} + D_A S^A + 8 \pi (\vec{J} \cdot \vec{m} ).
\label{mq}
\end{eqnarray}

In this $2+1+1$ context, MOTS satisfy $p
+ q =0$, generalized trapped surfaces satisfy $p \leq |q|$ and
generalized apparent horizons $p = |q|$. Notice that the latter always
have  non-negative mean curvature, which means that outward variations
do not increase the area. This property in fact generalizes to a
global statement for so-called {\it outermost}  generalized horizons,
as follows.

\begin{figure}[h!]
\begin{center}
\psfrag{S}{$S$}
\psfrag{Sigma}{$(\Sigma,\gamma_{ij},A_{ij})$}
\psfrag{Outer}{$\partial^{+} \Sigma$}
\psfrag{Inner}{$\partial^{-} \Sigma$}
\psfrag{Omega}{$\Omega$}
\psfrag{mm}{$\vec{m}^{-}$}
\psfrag{mp}{$\vec{m}^{+}$}
\includegraphics[width=8cm]{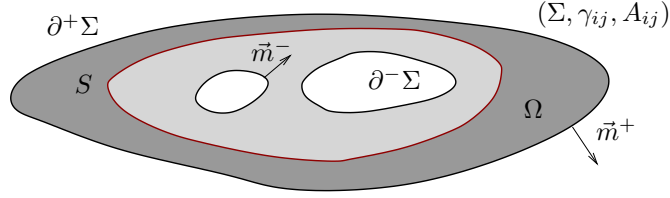}
\caption{Initial data set with an inner ($\partial^{-} \Sigma$) an outer boundary
($\partial^{+} \Sigma$) and the corresponding choice of unit normals.
A bounding surface $S$, and its exterior domain $\Omega$, are shown.}
\label{Bounding}
\end{center}
\end{figure}

Consider an initial data set $(\Sigma,\gamma_{ij},A_{ij})$ with
$\Sigma$ a compact manifold with boundary and assume that the boundary
can be split into two disjoint components $\partial^- \Sigma$ and
$\partial^{+} \Sigma$ (neither of which is necessarily connected). We
denote by  $\intSigma$ the interior of $\Sigma$, so that $\Sigma = \intSigma
\cup \partial^- \Sigma \cup
\partial^+ \Sigma$.  We want to
think of  $\partial^{-} \Sigma$ as the ``inner'' boundary, which means
that we will endow it with a unit normal $\vec{m}^{-}$ pointing
towards $\Sigma$. Similarly $\partial^{+} \Sigma$ is the ``outer''
boundary, which we endow with the unit normal $\vec{m}^{+}$ pointing
outside $\Sigma$ (see Figure \ref{Bounding}).

In many cases,  $\Sigma$ will contain plenty of
weakly outer trapped surfaces. An important question arises then: is there
an outermost weakly outer trapped surface $S$? Intuitively, this means
a surface which encloses all others, or equivalently, such that no weakly outer
trapped surface can penetrate outside $S$. In order to make this precise, it
is necessary to have a well-defined notion of ``outside'' of S. This can be
achieved by restricting the class of surfaces
to those which are homologous to the outer boundary.  More explicitly,
a surface $S^{\prime}$ is called {\it bounding} if it is contained in $\intSigma \cup
\partial^- \Sigma$ (in particular, it is disjoint with the outer
boundary) and, together with $\partial^{+} \Sigma$ bounds an open
domain $\Omega^{\prime}$.  For such surfaces, the mean curvature $p$
and the outer expansion $\theta_{+}$ is calculated with respect to the
normal  pointing towards $\Omega^{\prime}$. A surface $S$ is outermost in some class
if no other surface in the class enters the domain $\Omega$ bounded by $S$ and 
$\partial^{+} \Sigma$. The existence of outermost
MOTS in this context has been proven by Andersson and Metzger 
\cite{AnderssonMetzger2007} assuming
the inner boundary to be weakly outer trapped ($\theta_{+} \leq 0$) and
the outer boundary to be outer untrapped ($\theta_{+} >0$), so that they play
the role of barriers. The proof uses the Gauss-Bonnet theorem so the
result is valid in the 3+1 dimensional setting.  The precise statement
is as follows
\begin{theorem}[Andersson \& Metzger \cite{AnderssonMetzger2007}]
\label{AnderssonMetzger} 
Let
$(\Sigma,\gamma_{ij},A_{ij})$ by three-dimensional 
and have an inner and
outer boundary as described in the previous paragraph.
Assume that the inner boundary  has $\theta_{+}
(\partial^- \Sigma ) \leq 0$ and the outer boundary $\theta_{+}
(\partial^+ \Sigma) >0$. Then there exists a unique smooth embedded
bounding MOTS $S$, i.e. $S \cup \partial^{+} \Sigma = \partial \Omega$
and $\theta_{+} (S) = 0$,  which is outermost: any other weakly outer
trapped surface $S'$ which together with $\partial^{+} \Sigma$ bounds
a domain $\Omega^{\prime}$ satisfies $\Omega \subset \Omega^{\prime}$.
\end{theorem} 
An important issue refers to the topology of the outermost MOTS $S$.
A classic result by Hawking \cite{HawkingEllis} states that 
the topology of each connected component of the outermost MOTS
has to be either toroidal or spherical, provided the initial data set satisfies the 
dominant energy condition.
Recently, this result has been extended to 
higher dimensions by Galloway and Schoen \cite{SchoenGalloway2005}, where it is proven that,
except for exceptional cases, the outermost MOTS in any  
spacelike hypersurface of a spacetime satisfying the dominant energy condition
must by of positive Yamabe type. The exceptional cases have been ruled out by
Galloway in \cite{Galloway2007} thus leaving only the positive Yamabe case. In four
spacetime dimensions, this implies that each connected component of the outermost MOTS must be
a sphere.

An existence result involving generalized trapped surfaces
and generalized apparent horizons instead of weakly trapped
surfaces and MOTS has been proven recently by Eichmair
\cite{Eichmair2008}.
This result does not use  the Gauss-Bonnet theorem, so it it not
restricted to 3+1 dimensions. However, it relies on regularity of
minimal surfaces,
which restricts the dimension of $\Sigma$ to be at
most seven. Outermost generalized apparent horizons have one
fundamental advantage over MOTS; they are {\it area outer minimizing},
i.e. they have less or equal area than any other bounding surface fully contained in the
closure of the exterior region. This area minimizing property makes these surfaces
potentially very interesting for the Penrose inequality, as first
discussed by  Bray and Khuri \cite{BrayKhuri2009}. This theorem of
Eichmair  answers in the  affirmative a conjecture due to H.L. Bray and
T. Ilmanen \cite{BrayKhuri2009} on the existence of outermost
generalized apparent horizons and its area outer minimizing property.
\begin{theorem}[Eichmair \cite{Eichmair2008}] 
\label{Eichmair}
Let $(\Sigma,\gamma_{ij},A_{ij})$ be $m$-dimensional,
with $3 \leq m \leq 7$ and have an inner and
outer boundary as described above. Assume that
the inner boundary  is a generalized trapped surface $p \leq |q|$ and
the outer boundary satisfies $p > |q|$. Then there exists a unique
$C^2$ embedded  generalized apparent horizon $S$ ($p = |q|$) which is
bounding, $S \cup \partial^{+} \Sigma = \partial \Omega$, and outermost:
any other generalized apparent horizon $S'$ which together with
$\partial^{+} \Sigma$ bounds a domain $\Omega^{\prime}$ satisfies
$\Omega \subset \Omega^{\prime}$.  Moreover $S$ is area outer
minimizing with respect to variations in $\overline{\Omega}$.
\end{theorem}

In the purely Riemannian case ($A_{ij}=0$), MOTS and
generalized apparent horizons are simply minimal surfaces (i.e. $p=0$) and 
the two theorems above
become statements on the existence of an outermost minimal surface on
a compact domain with barrier boundaries. This particular case
was already known before (see Sect. 4 of \cite{HuiskenIlmanen2001} and references therein).
In fact, the existence of an
outermost  minimal surface played an important role in the
proof of the Riemannian Penrose inequality by G. Huisken and
T. Ilmanen \cite{HuiskenIlmanen2001}.

\subsection{Hawking mass, Geroch mass and their variation formulas}
\label{UEF}

Huisken and Ilmanen's proof of the Riemannian Penrose inequality
is based on a very interesting result by Geroch \cite{Geroch1973} who found 
a certain functional (now called Geroch mass) which is monotonic under so-called
inverse mean curvature flows (IMCF). 
This functional is defined for surfaces embedded
in a Riemannian manifold $(\Sigma,\gamma)$. For codimension-two surfaces embedded in a spacetime,
the analogous to the Geroch mass is the so-called Hawking mass 
\cite{Hawking1968} (both coincide when
the surface is embedded in a time-symmetric initial data set). Since these two masses have similar properties,
it is conceivable  that the Hawking mass may also be useful to tackle
the general Penrose inequality (i.e. for non time-symmetric initial data sets). A natural first step is to
analyze whether the
Hawking mass is also monotonic under suitable flows. Monotonicity along null directions was first 
studied by Hayward \cite{Hayward1994}.
More recently, the rate of change of the Hawking mass along an IMCF
in an arbitrary initial data set $(\Sigma,\gamma_{ij},A_{ij})$ was studied
in \cite{MalecMarsSimon2002}. The null and initial data set approaches can be unified and extended
by using a spacetime flow formulation, where the two-dimensional surface
is varied in spacetime. This is discussed in detail in \cite{BrayHayward2007}. Let us briefly
describe the main points, and how previous results fit into this framework.

For any closed spacelike surface  $S$ embedded in a four dimensional spacetime $(\M,g)$,
the Hawking mass \cite{Hawking1968}
is defined by
\begin{eqnarray}
M_{H}(S) = \sqrt{ \frac{\left | S \right |}{16 \pi}} \left (
\frac{\chi(S)}{2} - \frac{1}{16 \pi} \int_S \left ( (\vec{H}\cdot \vec{H}) + \frac{4}{3} \C \right )\bm{\eta_S} 
\right ),
\label{HawMass}
\end{eqnarray}
where $\chi(S)$ is the Euler 
characteristic of $S$. Hence $\chi(S)=2$ for a connected surface with spherical
topology. For an arbitrary compact surface $\chi(S) = \sum 2 (1- g)$ where the sum is over the
connected components and $g$ is the genus. At this point, the parameter $\C$ in (\ref{HawMass})
is a completely
arbitrary constant. Its inclusion
is relevant for settings where the spacetime has a cosmological constant, or when dealing
with hyperbolic initial data sets in an asymptotically flat spacetime (see Sect. \ref{hyperbolicSect} below).
This constant was first introduced
in \cite{BoucherGibbonsHorowitz1984} in the time-symmetric context.

Let $\vec{V}^{\star}$ denote the Hodge dual operation on the normal
space $N_p S$ (which, under our assumptions, has a Lorentzian induced metric and it is orientable, so it
admits a canonical volume element once an orientation is chosen).
This operation is idempotent and transforms any vector into an orthogonal vector with opposite norm. 
Whenever $S$ admits a notion of ``outer'', the orientation of the normal bundle will be chosen
so that  $\vec{l}^{+ \star} = \vec{l}^{+}$ holds (this implies $\vec{l}^{- \star} = - \vec{l}^{-}$).

Let us also denote by
$\nabla^{\bot}$ be the connection on the normal bundle (i.e.
if $\vec{V}$ is normal to $S$ and $\vec{X}$ is tangent to $S$,
$\nabla^{\bot}_{\vec{X}} \, \vec{V} \equiv (   \nablafour_{\vec{X}} \, \vec{V} )^{\bot})$. 
For any
orthogonal variation vector $\vec{\xi}$, the derivative of the Hawking mass along $\vec{\xi}$ reads
\cite{BrayHayward2007}
\begin{eqnarray}
\left . 
\frac{d M_{H} (S_{\lambda})}{d \lambda} \right |_{\lambda=0} & = & 
\frac{1}{8 \pi} \sqrt{ \frac{ \left | S \right |}{16 \pi}}  \int_{S}
\left[ \left ( \mbox{Ein} (\vec{H}^{\star}, \vec{\xi}^{\star} ) + \C  (\vec{H}^{\star} \cdot \vec{\xi}^{\star} \,  )
\right )
 +  
8 \pi \Theta^T (\vec{H}^{\star},\vec{\xi}^{\star} \, ) + \right. {}  \nonumber \\
 & + & \left.
\mbox{tr}_{S} \left ( \vec{H} \cdot \nablanor \nablanor \vec{\xi} \right )
-  \left ( \frac{1}{2} R(h) -  \frac{1}{4} (\vec{H}\cdot \vec{H}) \right )
\left[ \left (\vec{\xi} \cdot \vec{H} \right ) - \g  \right ] \right ] 
\bm{\eta_{S}},
\label{dMH1}
\end{eqnarray}
where $\g$ is the constant obtained by averaging $(\vec{\xi} \cdot \vec{H})$ on $S$, i.e.
\begin{eqnarray*}
\g  \equiv  \frac{\int_{S}
\left ( \vec{\xi} \cdot \vec{H} \right ) \bm{\eta_{S}}}{\left | S\right |},
\end{eqnarray*}
and
$8 \pi \Theta^T(\vec{H}^{\star},\vec{\xi}^{\star}) \equiv 
(\vec{\Pi}_{AB}\cdot\vec{H}^\star ) (\vec{\xi}^\star 
\cdot \vec{\Pi}^{AB} \, ) - 
\frac{1}{2} (\vec{\Pi}_{AB}\cdot\vec{\Pi}^{AB} ) (\vec{\xi}^\star  \cdot \vec{H}^\star \,)$
is the ``transverse part of the gravitational energy''. This object  has good positivity
properties \cite{BrayHayward2007}. In particular is non-negative if 
$\vec{H}$ and $\vec{\xi}$ are both achronal and have positive inner product.

From expression (\ref{dMH1}) is it straightforward to show that 
$d M_{H} (S_{\lambda})/d \lambda \geq 0$ whenever 
$(\mbox{Ein} + \C g)$ satisfies the
dominant energy condition
(recall that a covariant tensor satisfies this property when
it is non-negative when acting on any combination of causal future directed vectors), 
$\vec{H}$ is spacelike and
the variation vector takes the form
\begin{eqnarray}
\vec{\xi} = \frac{1}{(\vec{H} \cdot \vec{H})} \left ( \g \vec{H} + c  \vec{H}^{\star} \right )
\quad \Longleftrightarrow \quad \left (\vec{\xi} \cdot \vec{H} \right ) = \g, 
 \left (\vec{\xi} \cdot \vec{H}^{\star} \right ) = - c ,
\label{Flowvector}
\end{eqnarray}
where $|c| \leq \g$ is an arbitrary constant. Following Hayward 
\cite{Hayward1994} in the null case, these flows have been termed
{\it uniformly expanding}  in \cite{BrayHayward2007}
since the mean curvature
along $\vec{\xi}$ and its dual $\vec{\xi}^{\star}$ are both constant.

The case with $c = \pm \g$
corresponds to a null variation vector and corresponds exactly 
to the flow studied by Hayward  \cite{Hayward1994}. 
The case 
$c=0$ can be naturally called
{\it inverse mean curvature flow vector} (because $\vec{\xi} = \frac{1}{(\vec{H} \cdot \vec{H})} \vec{H}$)
and monotonicity of the Hawking mass in this
case was first mentioned
in \cite{HuiskenIlmanen2001} and studied in detail by Frauendiener \cite{Frauendiener2001}.
The case $|c | < \g$
corresponds to a spacelike
flow vector and therefore can be rephrased as variations within an initial
data set, as follows.

For a surface $S$ embedded in the initial data
set $(\Sigma,\gamma_{ij},A_{ij})$, the mean curvature
vector of $S$ reads $\vec{H} = - q \vec{n} + p \vec{m}$
and the Hawking mass takes the form 
\begin{eqnarray*}
\label{HawkingMassID}
M_H(S) = \sqrt{\frac{|S|}{16\pi}} \left ( \frac{\chi(S)}{2} - \frac{1}{16 \pi}
\int_S \left ( p^2-q^2 + \frac{4}{3} \C \right ) \bm{\eta_S}  \right ).
\end{eqnarray*}
For an outward pointing variation vector $\vec{\xi}$ tangent to
$\Sigma$, i.e. $\vec{\xi} = e^{\psi} \vec{m}$, the general variation formula (\ref{dMH1})
can be rewritten in terms of the initial data geometry as
\begin{eqnarray}
\label{dMH2}
\left . 
\frac{d M_{H} (S_{\lambda})}{d \lambda}  \right |_{\lambda=0}    =   
\frac{1}{8 \pi} \sqrt{ \frac{ \left | S \right |}{16 \pi}}  \int_{S}
\left [ e^{\psi} \left [p \left ( 8 \pi \rho - \C \right )
+ 8 \pi q J_i m^i) \right ]
+ \frac{1}{2} e^{\psi} \left ( \Pi^{\vec{n}}_{AB} \bullet \Pi^{\vec{m}\, AB} \right )
\right . \nonumber \\ 
 +  \left . 
e^{\psi} \left ( S_A \bullet D^A \psi \right )  
+ e^{\psi} q \left (D_A S^A \right) +
\left [ \Delta_S \psi - \frac{1}{2} R(h)  +
\frac{1}{4} \left ( p^2 - q^2 \right ) \right ] 
\left (p e^{\psi} - \g \right ) 
\right ] \bm{\eta_{S}},
\end{eqnarray}
where  the $\bullet$ operation acts on two tensors of the same class
$X_{A_1 \cdots A_r}$, $Y_{A_1 \cdots A_r}$ and gives
\begin{eqnarray*}
\left ( X_{A_1 \cdots A_r} \bullet Y^{A_1 \cdots A_r} \right ) = 
p X_{A_1 \cdots A_r} X^{A_1 \cdots A_r}  - q 
X_{A_1 \cdots A_r} Y^{A_1 \cdots A_r}
+ p Y_{A_1 \cdots A_r} Y^{A_1 \cdots A_r}. 
\end{eqnarray*}
This quadratic expression is non-negative provided 
$|q| \leq p$, or equivalently when $\theta_{+} \geq 0, \theta_{-} \leq 0$
at each point, which is an ``untrappedness'' condition (note that this condition
is basically the complementary at each point of the 
generalized trapped surface condition).
In the particular case of $\C=0$ and IMCF ($p e^{\psi} = \g$),
the variation formula (\ref{dMH2}) was obtained in \cite{MalecMarsSimon2002}. 

The first two terms
in (\ref{dMH2}) are non-negative provided $ 8 \pi | \vec{J}| \leq 8 \pi \rho - \C$ holds (this is automatically true
if $\mbox{Ein} + \C g$ satisfies the dominant energy condition)
and $|q| \leq p$. The last two terms have no
sign in general. However, the factor in round brackets on each of them integrates
to zero. Thus, monotonicity can be ensured \cite{BrayHayward2007} provided
the following conditions hold simultaneously: 
(i) $q e^{\psi}  = c$ \underline{or} $S^A$ is divergence-free and (ii)
$\Delta_S \psi - \frac{1}{2} R(h)  +
\frac{1}{4} \left ( p^2 - q^2 \right ) =\alpha$
\underline{or} $e^{\psi} p = \g$, where $c$ and $\alpha$ are constants. This leads to
four different alternatives for which the Hawking mass is monotonic. 
The uniformly
expanding flow condition (\ref{Flowvector}) with $|c| <\g$ corresponds  to the
case $q e^{\psi}  = c$, and $p e^{\psi} = \g$. This follows easily from the expressions
$\vec{H} = -q \vec{n} + p \vec{m}$ and its dual 
$\vec{H}^{\star} = -q \vec{m} + p \vec{n}$. How restrictive are any of these four
alternatives on a given spacetime remains an open problem.

While the Hawking mass is intrinsically a functional on surfaces in spacetime (although it obviously admits
a rewriting in 3+1 language), the Geroch mass
is directly a functional on surfaces in a spacelike slice $(\Sigma,\gamma)$.
Its definition is
\begin{eqnarray}
\label{GerochMass}
M_G(S) = \sqrt{\frac{|S|}{16\pi}} \left ( \frac{\chi(S)}{2} - \frac{1}{16 \pi}
\int_S \left (p^2  + \frac{4}{3} \C \right ) \bm{\eta_S}  \right ).
\end{eqnarray}
The Geroch mass is fully insensitive to the second fundamental form on the slice (in this
sense, it is a purely Riemannian object) and always satisfies $M_G (S) \leq M_H(S)$.
Its variation can be obtained formally from expression (\ref{dMH2}) simply by putting
all terms depending on the second fundamental form equal to zero. In particular,
this requires substituting $8 \pi \rho \rightarrow 1/2 R(\gamma)$, which follows from the 
Gauss equation for $\Sigma$ when the second fundamental form vanishes. Explicitly
\begin{eqnarray}
\left . 
\frac{d M_{G} (S_{\lambda})}{d \lambda}  \right |_{\lambda=0}   & =  &  
\frac{1}{8 \pi} \sqrt{ \frac{ \left | S \right |}{16 \pi}}  \int_{S}
\left [ e^{\psi} p \left ( \frac{1}{2} R(\gamma) - \C \right )
+ \frac{1}{2} e^{\psi} p \; \Pi^{\vec{m}}_{AB} \Pi^{\vec{m} \, AB}
+ e^{\psi} p D_A \psi D^a \psi + \right . \nonumber \\
& + & \! \! \! \left . \frac{}{} \left ( \Delta_S \psi - \frac{1}{2} R(h)  +
\frac{p^2}{4} \right ) \left (p e^{\psi} - \g \right ) 
\right ]  \label{dMG}
\bm{\eta_{S}}.
\end{eqnarray}
We emphasize that this result is valid for any spacelike hypersurface of the spacetime,
not only for time-symmetric ones.
For vanishing $\C$ this expression was first obtained by Geroch \cite{Geroch1973}.
In the time-symmetric case, for arbitrary $\C$ and imposing  IMCF, this
expression was derived in \cite{BoucherGibbonsHorowitz1984}.

\subsection{Asymptotic flatness}
\label{asymptoticflatness}

%A second basic ingredient of the Penrose inequality is the concept of total mass of 
%a spacetime or of an initial data set. 

An initial data set $(\Sigma,\gamma_{ij},A_{ij})$ is {\it asymptotically euclidean} provided $\Sigma$ is the disjoint
union of a compact set $K$ and a finite union of asymptotic ends $\Sigma^{\infty}_i$,
each of which is  diffeomorphic to $\mathbb{R}^3 \setminus
B$ where $B$ is a closed ball.  Moreover, in Cartesian coordinates $x^{i}$ 
in $\Sigma^{\infty}_i$
induced by this diffeomorphism, the metric and second fundamental forms have the
following asymptotic behaviour (we restrict to 3+1 dimensions for definiteness)
\begin{eqnarray}
\gamma_{ij} = \delta_{ij} + O\left (\frac{1}{r}\right ), \quad 
\partial \gamma_{ij} = O \left (\frac{1}{r^2} \right ), \quad  
\partial \partial \gamma_{ij} = O \left (\frac{1}{r^3} \right ),
\nonumber \\
A_{ij} = O \left (\frac{1}{r^2} \right ), \quad \partial A_{ij} = O \left (\frac{1}{r^3} \right ), \label{decay}
\end{eqnarray}
where $r  = \sqrt{\delta_{ij} x^i x^j}$. 
The Penrose inequality involves the total mass of a spacetime. In the asymptotically euclidean setting,
the total ADM energy-momentum vector \cite{ADM} is defined through the coordinate expressions
\begin{eqnarray*}
E _{ADM}= \lim_{r \rightarrow \infty}
\frac{1}{16 \pi} \int_{S_r} \left ( \partial_j \gamma_{ij} - \partial_i \gamma_{jj} \right)
dS^i,  \\
P_{ADM \, i} = \lim_{r \rightarrow \infty}
\frac{1}{8\pi} \int_{S_r} \left ( A_{ij} - \gamma_{ij} \tr_{\gamma} A \right ) dS^j,
\end{eqnarray*}
where $S_r$ is the surface at constant $r$ and 
$dS^i = n^i dS$ with $\vec{n}$ being the outward unit normal to $S_r$ and $dS$
the surface element. This definition depends a priori on the choice of coordinates $x^i$.
However, they can be shown to define geometric quantities on $(\Sigma,\gamma_{ij},A_{ij})$
provided the constraints satisfy the additional decay properties
\begin{eqnarray}
R(\gamma) = O\left ( \frac{1}{r^4}\right ), \quad \nabla_{i} A^{i}_{j} - \nabla_j A = O \left (\frac{1}{r^4} \right ).
\label{RDA}
\end{eqnarray}
In fact, the ADM energy-momentum is shown in \cite{Bartnik2005}, generalizing
results in \cite{Bartnik1986},
to be well-defined under much weaker conditions, where the metric and second fundamental forms
belong to appropriate weighted Sobolev spaces involving just two derivatives
for $\gamma$ and one for $A_{ij}$, and that the left hand sides of (\ref{RDA}) are integrable
on $\Sigma$. The total ADM mass of an asymptotically euclidean initial data set is defined
as $M_{ADM} = \sqrt{E_{ADM}^2 - \delta^{ij} P_{ADM \, i} P_{ADM \, j}}$.

The positive mass theorem of Schoen and Yau \cite{SchoenYau1979, SchoenYau1981}
states that, provided the initial data satisfies the dominant energy condition,
$\rho \geq |\vec{J}|$, then $M_{ADM}$ is real and in fact 
strictly positive except whenever $(\Sigma,\gamma_{ij},A_{ij})$ corresponds
to a slice of Minkowski spacetime. So far, this theorem has been proven in any space-time
dimension up to 
$n=8$. Furthermore, it also holds for spin manifolds of any dimension, for which 
Witten's spinorial proof \cite{Witten1981,ParkerTaubes1982} of the positive mass theorem
can be applied.

\section{Formulations of the Penrose inequality}
\label{Formulations}

The heuristic derivation of the Penrose inequality for a surface $S$ relies on two fundamental facts.
First and foremost, that the ``establishment viewpoint'' of gravitational collapse holds and second,
that $S$ is known to lie behind the event horizon
 in any situation where a black hole does indeed form. This second aspect makes it useful to 
consider the class of weakly outer trapped surfaces $S$ in a given asymptotically euclidean slice.
In order to have a useful notion of outer direction in this case, one restricts
the surfaces to be boundaries of domains $\D^{+}$
containing the asymptotically euclidean end (if there is more than one end,
one should be selected from the outset). 
The outer normal is chosen to point inside this domain. Such surfaces
will be called {\bf weakly outer trapped boundaries} in the following. 
Given two surfaces $S_1$ and $S_2$ (not necessarily weakly outer trapped)
that bound respectively exterior domains $\D_1^{+}$ and
$\D^{+}_2$, we shall say that $S_2$ encloses $S_1$ provided $\D^{+}_1$ contains $\D^{+}_2$.
The complementary of the outer domain $\D^{+}$ in 
$\Sigma$ is called ``interior region of $S$'' and denoted by $\D^{-}$. It is also useful to define the
outer trapped region $\T^{+}_{\Sigma}$ in $\Sigma$ to be the union of interior regions of all weakly outer trapped
boundaries.  An important fact is that in strongly asymptotically
predictable spacetimes satisfying the null convergence condition (i.e. $\mbox{Ric}(\vec{l},\vec{l}) \geq 0$ 
for each null vector $\vec{l}$) the interior region of any weakly outer trapped boundary
is contained in the black hole region $M \setminus J^{-}(\scri^{+})$ (see Proposition 12.2.4
in \cite{Wald1984}, and Theorem 6.1 in \cite{ChruscielGalloway2008}). Consequently, the same holds
for $\T^{+}_{\Sigma}$. Thus, this type of surfaces
are good candidates for a Penrose inequality. As mentioned in the introduction, the area
of the event horizon cut $\H_{\Sigma} = \H \cap \Sigma$ may be smaller than the area of 
any such surface $S$. However, for spacetime dimensions $n \leq 8$
any surface $S$ that bounds an exterior domain, always has a {\bf
minimal area enclosure}, i.e. the outermost of all surfaces which enclose $S$ and have less
or equal area than any other surface enclosing $S$.  We will denote by $A_{\min} (S)$
the area of the minimal surface enclosure of $S$. For a weakly outer trapped boundary in a black hole
spacetime it follows that $A_{\min} (S) \leq |\H_{\Sigma}|$, because $\H_{\Sigma}$ encloses $S$. Hence the heuristic
argument by Penrose implies
$M_{ADM} \geq \sqrt{A_{\min}(S)/16\pi}$ where $S$ is any weakly outer trapped boundary. Therefore it also
follows
\begin{eqnarray}
M_{ADM} \geq \sup_{S} \sqrt{\frac{A_{\min}(S)}{16 \pi}}, \label{PIHeu}
\end{eqnarray}
where the supremum is taken with respect to all
weakly outer trapped boundaries $S$. This inequality can be rewritten in a much simpler way as follows.

Since a sufficiently large coordinate sphere $S_r$ in the asymptotically euclidean region
has positive outer expansion, Theorem \ref{AnderssonMetzger}
can be applied \cite{AnderssonMetzger2007} to conclude that
$\partial  \T^{+}_{\Sigma}$ is a smooth 
marginally outer trapped boundary, which by construction encloses all other weakly outer 
trapped boundaries. It follows immediately that $A_{\min} (\partial \T^{+}_{\Sigma}) \geq A_{\min} (S)$
for any weakly outer trapped boundary $S$, and hence also
$A_{\min} (\partial \T^{+}_{\Sigma}) \geq \sup_S A_{\min} (S)$. On the other hand
$\partial \T^{+}_{\Sigma}$ is itself
a weakly outer trapped boundary, and hence 
$A_{\min} (\partial \T^{+}_{\Sigma}) \leq \sup_S A_{\min} (S)$. Thus, they necessarily coincide and the Penrose
inequality (\ref{PIHeu}) can be written in the simpler form
\begin{eqnarray}
M_{ADM} \geq \sqrt{\frac{A_{\min}(\partial \T^{+}_{\Sigma})}{16 \pi}}. \label{PIHeuT+}
\end{eqnarray}
Although the Penrose inequality, seen as a consequence of cosmic censorship, should
be expected to hold only for the minimum area enclosure of $\partial \T^{+}_{\Sigma}$,
one can often find in the literature a version of the Penrose inequality 
involving the area of the outermost MOTS,
\begin{eqnarray}
M_{ADM} \geq \sqrt{\frac{|\partial \T^{+}_{\Sigma}|}{16 \pi}}. 
\label{PIT+}
\end{eqnarray}

This is of course a stronger 
and simpler looking inequality. However, it is presently known not to be true. A counterexample has
been found by  I. Ben-Dov \cite{Ben-Dov2004} by considering a spherically symmetric spacetime
composed of four regions. The innermost region is
a portion of a dust-filled closed FLRW spacetime. It is followed by a portion of the Kruskal spacetime of mass $M$
which is then joined to another region made of dust, closed FLRW in such a way that the outermost region,
which is again vacuum, has less mass than the intermediate Kruskal region.
In this context, there exists a slice for which the outermost MOTS lies in the intermediate Kruskal portion,
which leads directly to a violation of (\ref{PIT+}). This example is also a
counterexample to a version of the Penrose inequality suggested by Penrose himself in
\cite{Penrose1982}, which goes a follows. Suppose that an asymptotically euclidean initial data set contains
a weakly future trapped boundary $S$ and the spacetime evolves according to weak cosmic censorship.
Then  $S$ is contained in the black hole region that forms. Consider the null hypersurface $\N$ consisting of past directed
null geodesics emanating orthogonally from $S$ towards the outer direction, and let $\vec{k}$ be the
corresponding null tangent vector. 
Since $S$ is weakly future trapped, we have $\theta_k \geq 0$ and the area of $S$ does not decrease initially
along $\N$, and this remains to be true as long as $\theta_k$ stays non-negative along the null hypersurface.
Since the collapse is taking place in the future and $\N$ extends towards the past and exterior 
region, one expects from physical grounds that $\theta_k$ does not change sign (this is because $\N$
approaches weaker gravitational fields and hence  outer past directed 
expansions should be positive). Under these circumstances,
the intersection of $\N$ with the event horizon has at least the same area than $S$, and hence the usual
heuristic argument implies $M_{ADM} \geq \sqrt{|S|/16 \pi}$. The counterexample by Ben-Dov shows that this inequality
is not generally true either. The difficulty with the argument is double. First of all,
the assumption that $\theta_k$ never becomes negative along $\N$ need not be true (notice that such 
a property is {\it not} implied by the  Raychaudhuri equation (\ref{lthetal})).
In the example by Ben-Dov, this property is not fulfilled
if the starting surface $S$ is any of the spherically symmetric future trapped surfaces
lying either in the intermediate Kruskal region or in 
the innermost closed dust FLRW region (see Figure 6 in \cite{Ben-Dov2004}). Secondly, the argument above also relays
on the  implicit assumption that $\N$ intersects the black hole event horizon. In the example
in \cite{Ben-Dov2004} these hypersurfaces end up in the
white hole singularity and do not intersect the event horizon anywhere.

Weakly outer trapped surfaces on a spacelike hypersurface satisfy $\theta_+ = p + q \leq0$ and the outermost
such surface $\partial \T^{+}_{\Sigma}$ is a MOTS (i.e. $\theta_+ = p +q =0$).
Given an initial data set $(\Sigma,\gamma_{ij}, A_{ij})$, a new one
$(\Sigma,\gamma^{\prime}_{ij}, A^{\prime}_{ij})$
can be obtained simply by changing the time orientation, i.e. by setting $\gamma^{\prime}_{ij} =
\gamma_{ij}$ and  $A^{\prime}_{ij} = -A_{ij}$. 
The heuristic argument to prove the Penrose inequality obviously applies equally well to this new initial data set.
For surfaces which are boundaries of exterior domains, the future outer null direction is now $\vec{l}^{\prime}_{+} =
- \vec{l}_{-}$, and therefore the weakly outer trapped surfaces satisfy $\theta^{\prime}_{+}= -\theta_{-} \leq 0$, 
or equivalently $p  - q \leq 0$ (this is, in fact, obvious since the change in time orientation
changes $q \rightarrow -q$). Similarly as before, one can construct the {\it past trapped region}
$\T^{-}_{\Sigma}$ as the union of interiors of all such boundaries.
Andersson and Metzger's result (Theorem \ref{AnderssonMetzger}) implies again the existence of a
unique  outermost surface $\partial \T^{-}_{\Sigma}$ which satisfies $p -q =0$ (i.e. a 
past marginally outer trapped surface). Thus, Penrose's heuristic argument also supports
the inequality
\begin{eqnarray}
M_{ADM} \geq \sqrt{\frac{A_{\min}(\partial \T^{-}_{\Sigma})}{16 \pi}}. \label{PIHeuT-}
\end{eqnarray}
In a given initial data set, neither $\T^{+}_{\Sigma}$ contains $\T^{-}_{\Sigma }$ nor viceversa, in general.
Although not supported
by any heuristic argument, a version of the Penrose inequality along the lines of (\ref{PIT+}) which has
also been proposed is
\begin{eqnarray}
M_{ADM} \geq \sqrt{\frac{|\partial (\T^{+}_{\Sigma} \cup \T^{-}_{\Sigma})|}{16 \pi}}. 
\label{PIT}
\end{eqnarray} 
Since $\T^{+}_{\Sigma} \cup \T^{-}_{\Sigma}$ contains both trapped
sets, the area of its boundary is not smaller than the minimal area enclosures $A_{\min} (\partial
\T^{+}_{\Sigma})$
or $A_{\min} (\partial \T^{-}_{\Sigma})$. Thus, (\ref{PIT}) is in general
a stronger inequality than (\ref{PIHeuT+}) or
(\ref{PIHeuT-}). So far no counterexample of (\ref{PIT}) has been found and it is not clear
whether the inequality should hold or not. In the time-symmetric case 
this inequality reduces to either of (\ref{PIHeuT+}) or (\ref{PIHeuT-}) and its validity has been
proven in general, as discussed in Sect. \ref{Riemannian} below. Inequality (\ref{PIT}) also reduces
to either (\ref{PIHeuT+}) or (\ref{PIHeuT-}) in any situation where $q$ is known to be identically
zero on the outermost future and past MOTS. An example of this behaviour
is given by a class of axially symmetric and conformally
flat initial data sets presented in \cite{Dain_et_al_2002}. For a subclass thereof, the validity of the inequality
(\ref{PIT}) has been verified numerically in \cite{Jaramillo_Vasset2007}.

In the spherically symmetric case, both
$\T^{+}_{\Sigma}$ and $\T^{-}_{\Sigma}$ are obviously spherically symmetric and one of them
necessarily contains
the other. The Penrose inequality (\ref{PIT}) is known to be true in that case, as we discuss in the following
section. Since the example by Ben-Dov deals with a spherically symmetric spacetime and a spherically symmetric
slice, it cannot provide a  counterexample to (\ref{PIT}). 
On the other hand, very few explicit examples are known where (\ref{PIT}) has been verified in
situations where it does not reduce
to either (\ref{PIHeuT+}) or (\ref{PIHeuT-}).
The only cases outside spherical symmetry 
that I am aware of involve a
numerical analysis of three special classes of  initial data sets \cite{Karkowski-Malec2005}.
In all these examples the inequality (\ref{PIT}) was numerically confirmed.

\begin{figure}[h!]
\begin{center}
\includegraphics[width=8cm]{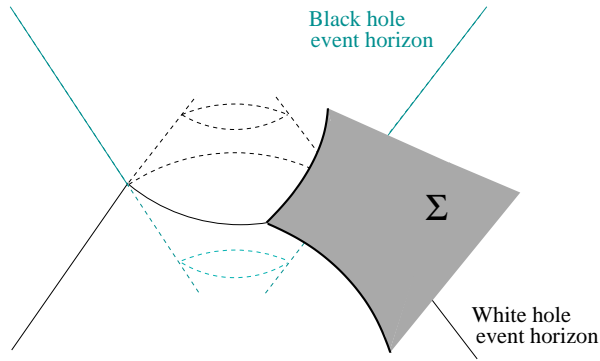}
\caption{Spacelike hypersurface $\Sigma$ in the Kruskal spacetime which intersects both the black hole and white
hole event horizons in such a way that the two surfaces defined by these intersections meet transversaly. 
In the figure, only the portion of $\Sigma$ lying outside the black hole and the white hole regions is
shown. However, $\Sigma$ is a Cauchy hypersurface, so the intersection with the
black hole event horizon is a compact surface without boundary, and the same holds
for the intersection with the white hole event horizon. Since both surfaces are generalized
trapped surfaces, the boundary $\partial \T_{\Sigma}$ must enter the shaded region somewhere.}
\end{center}
\label{Nonsmooth}
\end{figure}

Still another version of the Penrose inequality has been very recently proposed by H. Bray and M. Khuri. 
The fundamental idea behind this proposal is to use generalized trapped surfaces instead of
weakly outer trapped surfaces. Recall that, on an initial data set, a surface
that bounds an exterior domain is a generalized trapped surface provided $p \leq  |q|$. This class
of surfaces has two main advantages over weakly outer trapped surfaces: first, its definition is 
insensitive to time reversals, so that one can get rid of the complications of dealing with 
several sets, like $\T^{+}_{\Sigma}$, $\T^{-}_{\Sigma}$ or their union. There is only one generalized
trapped set $\T_{\Sigma}$ (in a given initial data set) and this contains both
$\T^{+}_{\Sigma}$ and $\T^{-}_{\Sigma}$ (and, obviously, its union). Moreover, the boundary $\partial \T_{\Sigma}$
is a smooth generalized
apparent horizon, due to Eichmair's result (Theorem \ref{Eichmair} above).  The second
advantage is that $\partial \T_{\Sigma}$ is area outer minimizing, i.e.  $|\partial \T_{\Sigma}| = A_{\min} (
\partial \T_{\Sigma} )$. Consequently, a Penrose inequality involving generalized trapped surfaces can be 
stated in a much simpler form in terms of the area of $\partial \T_{\Sigma}$, without the need of 
invoking minimal area enclosures. The Penrose inequality proposed by Bray and Khuri thus reads
\begin{eqnarray*}
M_{ADM} \geq \sqrt{\frac{|\partial \T_{\Sigma}|}{16 \pi}}. \label{PIKhuriBray}
\end{eqnarray*}
This version is automatically stronger than (\ref{PIHeuT+}) or (\ref{PIHeuT-}) because $\T_{\Sigma}$
encloses the other trapped sets. Arguments in favor of this version will be given below, where we discuss
Bray and Khuri's approach in more detail. We should emphasize, however, that the heuristic argument
by Penrose {\it does not} support this version. The reason is that one of the key ingredients for
the heuristics to go through is that the surfaces under consideration lie behind the event horizon provided
a black hole does indeed form. However, this property is not generally true for generalized trapped surfaces. An example
can be constructed as follows. Consider 
a Cauchy slice $\Sigma$  of the Kruskal spacetime.
The portion of $\Sigma$ lying in the domain of outer communications (i.e. in the exterior Schwarzschild region) has
a boundary which necessarily lies in the union of the black hole event horizon, the bifurcation surface and
the white hole event horizons. Select $\Sigma$ such that this boundary has a non-empty intersection with 
both the black hole and the white hole event horizons. Since we are considering the
Kruskal spacetime, the intersection of $\Sigma$ with the black hole event horizon 
is a smooth MOTS and the intersection with the while hole event horizon is a smooth past MOTS,
it follows that the outermost generalized apparent horizon $\partial \T_{\Sigma}$ must contain both of them. However,
$\Sigma$ can be easily chosen so that the former intersect transversally (see Fig \ref{Nonsmooth}).
Consequently, some portion of
$\partial \T_{\Sigma}$ (which must be $C^2$) must lie strictly inside
the domain of outer communications, i.e. outside the black hole region.

\section{The Penrose inequality in spherical symmetry}
\label{Spher}

The Penrose inequality has been proven to hold 
under the assumption of spherical symmetry, i.e. when $(\Sigma,\gamma_{ij},A_{ij})$
is invariant under an $SO(3)$ action with (generically) $S^2$ orbits. This inequality was first established by
Malec and \'O Murchadha \cite{MalecMurchadha1994} assuming that the initial data is maximal, i.e. 
$\tr_{\gamma} A =0$ and in full generality by Hayward \cite{Hayward1996}. In either case the inequality
bounds the ADM energy (not the ADM mass) in terms of the area of the outermost surface among 
$\partial \T^{+}_{\Sigma}$ (the outermost MOTS) and $\partial \T^{-}_{\Sigma}$ (the outermost past MOTS).

Both arguments are ultimately based on properties of the Misner-Sharp quasi-local energy
\cite{MisnerSharp1964}, although
this is made fully explicit only in \cite{Hayward1996}. This quasi-local energy is exactly the Hawking
mass specialized to spherical symmetry. Thus, the inequality in this case can be easily
derived from the general expressions in subsection \ref{UEF}.

For a spherically symmetric asymptotically euclidean initial data set, 
the outer trapped region $\T^{+}_{\Sigma}$, being a geometrically defined set, is necessarily spherically symmetric.
Its boundary $\partial \T^{+}_{\Sigma}$ is therefore a metric sphere (if non-empty, which we assume from now on). 
Now, two different possibilities arise. Either the region outside $\partial \T^{+}_{\Sigma}$ contains
minimal surfaces (case (i))  or not (case (ii)).
In case (i) there is an  outermost minimal surface $S_m$ lying outside
$\partial \T^{+}_{\Sigma}$. This surface is necessarily a two-sphere and therefore encloses
$\partial \T^{+}_{\Sigma}$.
Let us consider the region outside $S_m$ in case (i) and the region outside $\partial \T^{+}_{\Sigma}$
in case (ii). We denote this region by $\Sigma_{ext}$ and
the corresponding first and second fundamental forms by
$\gamma_{ext}$ and $A_{ext}$ respectively. This region is free of minimal
surfaces, so that the metric can be written as $\gamma_{ext}=
\frac{dr^2}{1- \frac{2 m(r)}{r}} + r^2 \left (d\theta^2 + \sin^2 \theta d \phi^2 \right )$,
with $2m(r) > r$. At infinity 
$\lim_{r \rightarrow \infty} m(r) = E_{ADM}$. 
From spherical symmetry, the second fundamental form can be written as
$A_{ext} = W(r) dr^2 + Z(r) \left (d\theta^2 + \sin^2 \theta d \phi^2 \right ).$

A direct calculation shows that the mean curvature $p$ and the trace $q$ of the second fundamental
forms on the
surfaces $\{r=\mbox{const}\}$ are $p=\frac{2}{r}\sqrt{1-\frac{2m}{r}} > 0 $, $q= 2 Z/r^2$. 
Consequently, the Hawking mass (\ref{HawMass})
(with $\C=0$) of these
spheres reads
$M_H(r) \equiv m(r) + Z^2/(2r)$. Its radial derivative can be obtained immediately
from (\ref{dMH2}), using the fact that 
$\partial_r = \sqrt{1 - \frac{2 m(r)}{r}} \, \vec{m}$, which implies $e^{\psi} p = 
r /2$. This gives
\begin{eqnarray}
\frac{d M_H}{dr} = 4 \pi r^2 \left (\rho + \frac{Z}{r} J_r \right ),
\label{DerM_h}
\end{eqnarray}
where $J_r = (\vec{J} \cdot \partial_r)$. The dominant energy
condition $\rho \geq | \vec{J} |$ becomes in this case $\rho \geq \left |
J_r \sqrt{1 - 2 m(r)/r} \right |$. Furthermore,
asymptotically euclidean demands
$W(r) = O(1/r^2)$, $Z(r)=O(1)$ so that 
$\lim_{r \rightarrow \infty} M_H(r) = E_{ADM}$.  

Let us first deal with case (i): since $S_m$ is a minimal surface and cannot be
weakly outer trapped (it lies outside $\partial \T^{+}_{\Sigma}$), it must have $q >0$
and hence it is past weakly outer trapped ($\theta_{-} >0$).  The spheres
in the asymptotic region have $\theta_{-}=q - p <0$, and consequently
there  must exist an outermost sphere $\partial \T^{-}_{\Sigma}$ with vanishing
$\theta_{-}$ (i.e. an outermost past MOTS). In the exterior 
of $\partial \T^{-}_{\Sigma}$ we have $\theta_+ = p +q >0$ and $\theta{^-} = - p + q <0$. This implies
$|q| < p$ and thus the bound  $| Z r^{-1} ( 1 - 2m /r )^{-1/2} | <1$ outside
$\partial \T^{-}_{\Sigma}$. This, together with
the dominant energy condition implies that $M_H(r)$ is non-decreasing
outside $\partial \T^{-}_{\Sigma}$
(this is just a particular case of the monotonicity properties of $M_H$ discussed above).
Being $\partial \T^{-}_{\Sigma}$ a past MOTS, we have
$M_H(\partial \T^{-}_{\Sigma}) = \frac{r}{2} = \sqrt{|\partial \T^{-}_{\Sigma}|/16 \pi}$. The monotonicity
of $M_H(r)$ from this surface to infinity and the fact that $S_{m}$ is area
outer minimizing establishes the Penrose inequality
\begin{eqnarray*}
E_{ADM} \geq \sqrt{\frac{|\partial \T^{-}_{\Sigma}|}{16 \pi}} \geq 
\sqrt{\frac{A_{\min} (\partial \T^{+}_{\Sigma} )}{16 \pi}} 
\end{eqnarray*}
for case (i). For case (ii) we have that $\partial \T^{+}$ is automatically area outer minimizing and
a similar argument applies: if there is an outermost  past MOTS $\partial \T^{-}_{\Sigma}$
outside $\partial \T^{+}_{\Sigma}$, apply the monotonicity of $M_H$
from $\partial \T^{-}_{\Sigma}$ to infinity. If there is none, apply monotonicity of $M_H$
from $\partial \T^{+}_{\Sigma}$ to infinity. In either case one concludes (since
$\partial \T^{+}_{\Sigma}$ is area outer minimizing),
\begin{eqnarray}
E_{ADM} \geq \sqrt{\frac{|\partial \T^{+}_{\Sigma}|}{16 \pi}}. \label{partialT} 
\end{eqnarray}

Notice that the Penrose inequality {\it does not} state (\ref{partialT})
in  case (i). There, the minimum area needed to enclose $\partial \T^{+}_{\Sigma}$ must be used.
This agrees with the discussion in Section \ref{Formulations}. 
As already mentioned there, Ben-Dov \cite{Ben-Dov2004} 
has found an explicit example in
spherical symmetry where the inequality (\ref{partialT}) is violated. On the other hand,
the argument above in fact proves the Penrose inequality for the outermost
of the two surfaces $\partial \T^{+}_{\Sigma}$ and $\partial \T^{-}_{\Sigma}$ and thus
also for $\partial ( \T^{+}_{\Sigma} \cup \T^{-}_{\Sigma} )$, due to spherical symmetry.

Notice also that the spherically symmetric Penrose inequality above
involves the total energy of the slice. This is weaker than the expected Penrose inequality in
terms of the total ADM mass. If the slice $(\Sigma,\gamma_{ij},A_{ij})$ is
such that one can generate a piece of spacetime which admits another slice with vanishing
total momentum, then the Penrose inequality in terms of the ADM mass also follows.  However, the existence
of this piece of spacetime is not always obvious. In any case, it would be of interest
to find a proof of the Penrose inequality in terms of the total ADM mass in spherical
symmetry directly in terms of the given data. This might give some new clues
on how the general Penrose inequality can be addressed.

\section{Riemannian Penrose inequality}
\label{Riemannian}

As already mentioned, the field has experienced a fundamental breakthrough
in the last decade or so with the complete proof of the 
Penrose inequality in the time symmetric case, first by 
Huisken and Ilmanen \cite{HuiskenIlmanen2001} 
for a connected horizon and then by Bray 
\cite{Bray2001} for an arbitrary horizon. Both papers dealt with the
four-dimensional case. However, while Huisken and Ilmanen's proof is
very specific to four dimensions (because of the use of the Geroch mass),
Bray's approach can be generalized to any spacetime dimension not bigger than eight,
as recently shown by Bray and Lee \cite{BrayLee2007}.

By definition, an initial data set is called time-symmetric whenever $A_{ij}=0$. This has
two immediate consequences, namely that the ADM three-momentum vanishes identically, so that 
$M_{ADM}=  E_{ADM}$, and that only one constraint equation remains,
\begin{eqnarray*}
R(\gamma) = 16 \pi \rho
\end{eqnarray*}
which gives $R(\gamma)\geq 0$ provided the dominant energy condition holds. In fact,
the weak energy condition (defined as $\mbox{Ein} (\vec{u}, \vec{u} ) \geq 0$ for all 
causal vectors) suffices in this case.

Another immediate consequence is that $\theta_{+} = - \theta_{-} = p$ and, hence, the trapped region $\T^{+}_{\Sigma}$,
the past trapped region $\T^{-}_{\Sigma}$ and the generalized trapped
region $\T_{\Sigma}$ all coincide in this case. Its boundary $S_m$ is the outermost
minimal surface, which is non-empty as soon as there is a bounding surface
with negative mean curvature (with respect to the normal pointing into the
chosen asymptotically euclidean end). This is a corollary of Theorems \ref{AnderssonMetzger} 
and \ref{Eichmair} above. However, in this context this result is known
to hold even in more generality. As discussed in \cite{HuiskenIlmanen2001}, following
classic results on minimal surfaces \cite{MeeksIIISimonYau1982}, it is sufficient
to define the trapped region $K$ as the 
image of all immersed minimal surfaces in $\Sigma$ together with 
the bounded connected components of its complementary. It follows that the boundary of this set is a
collection of smooth, embedded, minimal 2-spheres and that any connected component of $\Sigma \setminus K$
is an ``exterior'' region, i.e. an asymptotically euclidean manifold free of minimal surfaces (even
immersed) and  with compact and minimal boundary composed of a finite union of 
2-spheres.  The Penrose inequality therefore becomes an inequality relating the total mass
and the area of the outermost minimal surface $S_m$ in $(\Sigma,\gamma)$ with respect
to the chosen asymptotically euclidean end. The corresponding inequality
\begin{eqnarray}
M_{ADM} \geq \sqrt{\frac{|S_m|}{16\pi}}
\label{PIRiem}
\end{eqnarray}
is usually termed  ``Riemannian Penrose inequality'' since it involves
directly Riemannian manifolds, with no further structure coming from the
ambient Lorentzian manifold. 

As already mentioned, the Penrose inequality has a rigidity part, namely that equality is achieved
only for slices of the Kruskal extension of the Schwarzschild metric. The time-symmetric
slices of this spacetime define a manifold
$(\Sigma_{\mbox{\tiny{Sch}}} = \mathbb{R}^3 \setminus {0}, \gamma_{\mbox{\tiny{Sch}}})$ with induced
metric
\begin{eqnarray}
\gamma_{\mbox{\tiny{Sch}}} = \left ( 1 + \frac{m}{2r} \right )^4 \left (dr^2 + r^2 d\theta^2 + r^2 \sin^2 \theta d \phi^2
\right ). \label{gsch}
\end{eqnarray}
The surface $r= m/2$ is minimal and separates the manifold into two isometric
pieces (corresponding to the two asymptotic ends of the Kruskal metric). The rigidity part
of the Riemannian Penrose inequality states that if equality is achieved in (\ref{PIRiem})
then the region in $(\Sigma,\gamma)$ outside its outermost minimal
surface $S_m$ is isometric to the domain $r > m/2$ of $(\Sigma_{\mbox{\tiny{Sch}}},\gamma_{\mbox{\tiny{Sch}}})$ with $m= M_{ADM}$.

Before discussing the breakthroughs of Huisken \& Ilmanen and Bray, let us
discuss some previous attempts to address the Riemannian Penrose inequality.

\subsection{Spinor methods}
\label{spinors}

Shortly after Schoen and Yau proved the positive mass theorem, Witten \cite{Witten1981}
proposed a completely different method using spinors (see \cite{ParkerTaubes1982} for a rigorous 
version of Witten's ideas). It is natural to ask whether spinorial techniques can be also 
applied to the Penrose inequality. After all, the spinor techniques had been 
successfully extended to prove the positive mass theorem in the presence of black holes (more precisely,
marginally trapped surfaces) in the initial data \cite{Gibbons_Hawking_1983} (see \cite{Herzlich1998} 
for a rigorous proof). The main difficulty lies in finding suitable boundary conditions
on Witten's equation on the boundary of the black holes so that the boundary
term arising by integrating the Schr\"odinger-Lichnerowicz \cite{Schrodinger1932,Lichnerowicz1963}
identity can be related to the area of the black hole. An interesting attempt to achieve this 
in the Riemannian case 
is due to M. Herzlich \cite{Herzlich1997}, who obtained a Penrose-like inequality involving
not just the total mass and area of the minimal surface but also a 
Sobolev type constant of the manifold. Depending on the space under consideration, Herzlich's
inequality may turn out  to be stronger or weaker than the Penrose inequality (see
\cite{Malec-Roszkowski1998} for a limiting case where this inequality reduces simply to a positive mass statement).
Nevertheless, the inequality is still optimal in the sense that equality is achieved only for the 
Schwarzschild manifold.

The class of manifolds considered in \cite{Herzlich1997} consists of asymptotically euclidean
3-dimensional, orientable Riemannian manifolds $(\Sigma,\gamma)$ having an inner boundary
$\partial \Sigma$ which  is topologically an $S^2$ and geometrically a minimal surface.
No assumption is made on whether this
surface is the  outermost minimal
surface in $(\Sigma,\gamma)$ or not. This already indicates that the inequality to be proven cannot be
the standard Penrose inequality because minimal surfaces with large area 
can be shielded from an asymptotically euclidean region with small mass by 
an outermost minimal surface with sufficiently small area, 
while maintaining non-negative Ricci curvature everywhere. This is often called ``shielding effect''
and explicit examples are easily constructed by cutting the Schwarzschild manifold (\ref{gsch})
on a sphere at $r < m/2$ and attaching a piece of $S^3$ of large radius with a cap 
near the north pole removed (this manifold belongs to the class of initial data leading to the 
so-called Oppenheimer-Snyder spherical dust collapse \cite{Oppenheimer-Snyder1939})

The inequality proven by Herzlich reads \cite{Herzlich1997}
\begin{eqnarray}
M_{ADM} \geq \frac{\sigma}{2(1 + \sigma)} \sqrt{\frac{|\partial \Sigma|}{\pi}},
\label{HerzlichPI}
\end{eqnarray}
where $\sigma$ is a geometric, scale invariant, quantity on $(\Sigma,\gamma)$ defined
as 
\begin{eqnarray*}
\sigma = \sqrt{\frac{|\partial \Sigma|}{\pi}} 
\inf_{f \in C^{\infty}_c, f \not\equiv 0}
\frac{ \int_{\Sigma} \left (df, df \right )_{\gamma} {\bm{\eta_{\gamma}}} }{\int_{\partial \Sigma}
f^2 \bm{\eta_{\partial \Sigma}}},
\end{eqnarray*}
where $C^{\infty}_c$ denotes, as usual, the collection of smooth functions with compact support.
Equality in (\ref{HerzlichPI}) occurs if and only if $(\Sigma,\gamma)$ is isometric to the exterior
of the Schwarzschild manifold (\ref{gsch}) outside the minimal surface $r= m/2$.

The proof is based on an improvement of the  positive mass theorem (also proven in 
\cite{Herzlich1997}) valid for asymptotically euclidean Riemannian
manifolds $(\overline{\Sigma},\overline{g})$ of non-negative scalar curvature 
having an inner boundary $\partial \overline{\Sigma}$ which is topologically a sphere 
and which may have positive, but not too large, mean curvature (with respect to the  normal pointing
towards infinity). More precisely, 
the mean curvature $p$ must satisfy the upper bound 
\begin{eqnarray}
p \leq 4 \sqrt{\frac{\pi}{|\partial \overline{\Sigma}|}}.
\label{InequalityMeanCurvature}
\end{eqnarray}
This positive mass theorem also has a rigidity part that states that equality is achieved if and only if
$(\overline{\Sigma},\overline{g})$ is the exterior of a ball in Euclidean space (this implies, 
in particular, that the only surface $S$ in $\mathbb{R}^3$ which is topologically an $S^2$
and which satisfies $p \leq 4 \sqrt{\pi / |S|}$ is in fact a sphere). The proof of this
theorem involves finding appropriate boundary conditions
for the Witten spinor on $\partial \overline{\Sigma}$ so that the boundary term in the Schr\"odinger-Lichnerowicz
identity on the inner boundary gives a non-negative contribution.

The Penrose-like inequality (\ref{HerzlichPI}) is then proven by finding a conformally
rescaled metric $\overline{g} = f^4 g$, with $f \rightarrow 1$ at infinity in such 
a way that $\overline{g}$ has vanishing scalar curvature and 
the mean curvature of $\partial \Sigma$ with respect to the metric $\overline{g}$ saturates
the inequality (\ref{InequalityMeanCurvature}). Moreover, the conformal rescaling is shown to
decrease the mass at least by the amount given in the right-hand side of (\ref{HerzlichPI}). 
Here is where the quantity $\sigma$ arises. Since the mass after the rescaling is
non-negative due to the positive mass theorem above,  the Penrose-like inequality follows.
The rigidity part holds because, in the case of equality in (\ref{HerzlichPI}),  $(\Sigma,\gamma)$
is conformal to the flat metric outside a ball and, moreover, its curvature scalar vanishes (if this was non-zero, then
the decrease in mass due to the conformal
rescaling would be {\it larger} than the  right-hand side of (\ref{HerzlichPI}), which cannot occur in the equality case).
Since the only conformally flat, scalar flat and  asymptotically euclidean Riemannian manifold with a minimal 
surface is the Schwarzschild space (\ref{gsch}), the rigidity part follows.

This Penrose-like inequality has been generalized in three different ways. First, Herzlich 
\cite{Herzlich2002} extended the results to spin manifolds of arbitrary dimension $n$. In this case the
boundary $\partial \Sigma$ is assumed to be a compact and connected minimal surface
of positive Yamabe type (i.e. such that it admits a metric of positive constant scalar curvature $R_0$).
Denoting by $\yy$ the Yamabe constant (i.e. $\yy = R_0 |\partial \Sigma|^{\frac{2}{n-1}}$ where 
$|\partial \Sigma|$ is the $(n-1)$-dimensional volume of the boundary), the Penrose-like inequality reads
\begin{eqnarray}
M_{ADM} \geq \frac{1}{8 \pi} \sqrt{ \frac{ \yy (n-1)}{n-2} }
\frac{\sigma }{\sigma + 1} | \partial \Sigma|^{\frac{n-2}{n-1}}
\label{PIHerzlichDimn}
\end{eqnarray}
where $\sigma$ is the analogous scale invariant quantity in higher dimensions
\begin{eqnarray*}
\sigma = \sqrt{\frac{4 (n-1)}{(n-2) \yy}} 
|\partial \Sigma|^{\frac{1}{n-1}} \inf_{f \in C^{\infty}_c, f \not\equiv 0}
\frac{ \int_{\Sigma} \left (df, df \right )_{\gamma} \bm{\eta_{\gamma}} }{\int_{\partial \Sigma}
f^2 \bm{\eta_{\partial \Sigma}}}.
\end{eqnarray*}
The idea of the proof is similar to the three-dimensional case.

The second  generalization \cite{Maerten2007}
involves {\it maximal} initial
data sets $(\Sigma,\gamma_{ij},A_{ij})$. Although this result
is not a Riemannian Penrose inequality, the methods used are very similar to the
previous ones and it is therefore natural to include it here. The idea of the proof
involves, again, a positive mass theorem
for manifolds with suitable boundary and 
finding an appropriate conformal factor which transforms the data so that 
the previous mass theorem can be applied while decreasing the mass by a certain amount. 

More precisely, the class of manifolds $(\Sigma,\gamma_{ij},A_{ij})$ under consideration
are asymptotically euclidean spin manifolds of arbitrary dimension $n$ satisfying the dominant energy
condition $\rho \geq |\vec{J}|$.
As before, the boundary $\partial \Sigma$ is connected, compact and of positive Yamabe type.
The condition of being minimal is replaced by three conditions, namely  
(i) the mean curvature $p$ (with respect to the
direction pointing towards infinity) is non-positive (ii) $q \geq |p|$ (or alternatively $-q \geq |p|$)
everywhere, and 
\begin{eqnarray*}
\mbox{(iii)} \quad \tt \equiv \sqrt{\frac{n-2}{(n-1) \yy}} \,\, |\partial \Sigma|^{\frac{1}{n-1}} 
\sup_{\partial \Sigma} \left ( p + \sqrt{q^2 + S_AS^A} \right ) < 1.
\end{eqnarray*}
($S_A$ is defined in(\ref{DefS_A})). Under these circumstances, the total energy satisfies the bound \cite{Maerten2007}
\begin{eqnarray}
E_{ADM} \geq \frac{1}{8 \pi} \sqrt{ \frac{ \yy (n-1)}{n-2} }
\frac{\sigma \left ( 1- \tt \right ) }{\sigma + 1 - \tt} | \partial \Sigma|^{\frac{n-2}{n-1}}.
\label{MaertenPI}
\end{eqnarray}
The constant $\tt$ is always non-negative under assumptions (i)-(ii). Thus, inequality (\ref{MaertenPI})
is weaker than the corresponding one in the time-symmetric case (\ref{PIHerzlichDimn}). The conditions on the boundary
$\partial \Sigma$ are somewhat surprising. Conditions (i) and (ii) state that the mean curvature
vector of the surface points inward (in
the sense that its product with the outer normal $\vec{m}$ is non-positive) and is 
causal past (future) everywhere. Thus, the boundary is indeed weakly past (future) trapped. However, these type of
surfaces are necessarily not area outer minimizing (except in the very special case
that their mean curvature vector vanishes identically).
Consequently, the minimal area enclosure of the boundary lies, in general, inside $\Sigma$. 
Thus, this Penrose-like inequality is obtained for a surface for which the original Penrose
inequality is not expected to hold. It is therefore interesting that such an inequality exists.

The third generalization is due to M. Khuri \cite{Khuri2009} and again involves a non-vanishing second
fundamental form. Since the method uses the Jang equation in a fundamental way, we postpone its discussion to
the end of Subsection \ref{Jang1}.

\subsection{Isoperimetric surfaces}
\label{isoperimetricprofile}

An interesting attempt to proof the Riemannian Penrose inequality is discussed in H.L. Bray's
Ph.D thesis \cite{BrayThesis}. The idea is to consider a special class of surfaces which interpolate
between the outermost minimal surface $S_m$, tends to large round spheres at infinity and for which the Geroch
mass is non-increasing. Although this sounds familiar with the inverse curvature flow argument of Geroch,
the idea exploited by Bray is in fact very different. Indeed, instead of using
flows of surfaces, Bray considers, for any given volume,
an area minimization problem (i.e. an isoperimetric problem).
Assume that the outermost minimal surface $S_m$ is connected and consider the class of surfaces in the
same homology class as $S_m$. One can associate to each surface $S$ in this class the volume bounded between
$S_m$ and $S$ (counted negatively in the portion where $S$ lies inside $S_m$ and positively where it lies
outside). For a given value of $V \geq 0$, consider the collection of surfaces which bound,
together with $S_m$, precisely a volume $V$ and define $A(V)$ as the infimum of the corresponding areas.
If the infimum is attained on a surface $S_V$, then this surface is obviously of constant mean
curvature $p(V)$ because it is the solution of  an isoperimetric problem. Bray's idea is to show that
the Geroch mass is a non-increasing function of $V$.  

Let us start by assuming that $S_V$ exists and
that the  function $A(V)$ is twice differentiable. Then, the first variation of area together with
the fact that the area is the first variation of  volume, gives
$A'(V) = p(V)$ (prime denotes derivative with respect to $V$) and the Geroch mass (\ref{GerochMass})
 becomes simply 
\begin{eqnarray}
  M_G (V) = \sqrt{\frac{A(V)}{16 \pi}} \left ( 1 - \frac{1}{16 \pi} A (V) A'(V)^2 \right),
\label{GerochMassIsoperimetric}
\end{eqnarray}
provided the surface $S_V$ is connected and of spherical topology. Consider now a variation of $S_V$
along its outer unit normal $\vec{m}$ with unit speed. If we denote by $S_V(t)$ the corresponding
flow of surfaces (with $S_V(t=0)=S_V$), the
second variation of area gives
\begin{eqnarray*}
\left . \frac{d^2 |S_V(t)|}{dt^2} \right |_{t=0}
= \int_{S_V} \left ( \pounds_{\vec{m}} p + p ^2 
\right ) \bm{\eta_{S_V}} =
\int_{S_V}  \left ( - \frac{1}{2} \Pi^{\vec{m}}_{AB} \Pi^{\vec{m} \, AB} + \frac{1}{4}  p^2
- \frac{1}{2} R(\gamma) + \frac{1}{2} R (h) \right ) 
\bm{\eta_{S_V}} 
\end{eqnarray*}
where we have used (\ref{mp}) in the second equality. Using now the 
Gauss-Bonnet theorem and the non-negativity of the curvature scalar of $(\Sigma,\gamma)$
we conclude $ \frac{d^2 |S_V(t)|}{dt^2} |_{t=0} \leq 4 \pi +  \frac{1}{4} p(V)^2 A(V)$. Denoting
by $V(t)$ the volume bounded by $S_V(t)$, it follows that $A(V(t)) \leq |S_V(t)|$ because 
$A(V)$ is the infimum of all areas bounding a volume $V$. Since these two functions touch at $t=0$, it 
follows $\frac{d^2 A(V(t))}{dt^2} |_{t=0}\leq \frac{d^2 |S_V(t)|}{dt^2} |_{t=0}$. Performing
a change of variables $t \rightarrow V$, one concludes
\begin{eqnarray*}
A''(V) \leq \frac{4\pi}{A(V)^2} - \frac{3 A'(V)^2}{4 A(V)} \quad \Longrightarrow \quad M_G(V)' \geq 0.
\end{eqnarray*}
Although the argument just described assumes that $A(V)$ is differentiable,
the conclusion is still valid if a suitable weak (distributional) derivative
is taken \cite{BrayThesis}. 
The other two assumptions that enter into the argument are (i) that $S_V$ is connected
and of spherical topology and (ii) that the infimum of $A(V)$ is in fact attained i.e. that
the surface $S_V$ exists. The first condition turns out to be crucial and needs to be imposed
as an assumption (in fact, the hypothesis can be relaxed to demand just that for each $V>0$
if one or more minimizers of $A(V)$ exists then at least one of them is connected; this
is termed ``Condition 1'' in \cite{BrayThesis}). Most of the technical work in
\cite{BrayThesis} consists in showing that
condition (ii) (i.e. the existence of a minimizer for all $V \geq 0$) imposes no extra
restriction. To that aim, Bray first argues that the class of metrics $(\Sigma,\gamma)$
can be chosen to be exactly Schwarzschild at infinity (a similar, but weaker, reduction is
also used in Bray's full proof of the Riemannian Penrose inequality summarized in subsection \ref{Braysproof}
and will be discussed in  more detail there). The heart of the proof of existence of the minimizer
$S_V$ consists in proving that the isoperimetric surfaces in the Schwarzschild spacetime
are given by the spherical orbits of the $SO(3)$ isometry group. Contrarily to what one may expect, proving
this fact is not a trivial matter. The fundamental idea behind the construction is to use a 
comparison metric (in this case the flat metric on $\mathbb{R}^3$)
for which one knows that the spheres are the solutions of the isometric problem. This argument,
which was tailored for the Schwarzschild metric, has been simplified and substantially extended in 
\cite{BrayMorgan2001}, where simple conditions are found on a given spherically symmetric metric which 
ensure that the spheres are the minimizers of the isoperimetric problem. 
Summarizing, Bray proves in \cite{BrayThesis} that the Penrose inequality (\ref{PIRiem})
holds for each asymptotically
euclidean Riemannian  manifold which has a connected  outermost minimal surface $S_m$ and satisfies ``Condition 1''.

%The case where
%where $S_m$ is not conneted is also discussed in \cite{BrayThesis}, under a suitable restriction 
%that replaces ``condition 1''. However, the inequality obtained involves  

\subsection{Huisken and Ilmanen's proof}
\label{HuiskenIlmanen}

The heuristic idea behind the proof of Huisken and Ilmanen was first proposed by Geroch 
\cite{Geroch1973} and is based on the observation  that the
Geroch mass (\ref{GerochMass}) (with $\C=0$)
is monotonically increasing if the surfaces are moved
by inverse mean curvature (i.e. $p e^{\psi}=1$) provided the scalar curvature of 
$(\Sigma,\gamma)$ is non-negative.  This fact is clear from 
(\ref{dMG}) since $p e^{\psi} = a=1$ and the derivative of $M_G(S_{\lambda})$ is then
a sum of non-negative
terms. Another immediate property of $M_G$ is that its value on any
connected, topologically $S^2$ minimal surface is $M_G = \sqrt{|S|/(16\pi)}$. 
For surfaces $S_r = {r = \mbox{const}}$ in the asymptotically euclidean end $\Sigma^{\infty}$,
the asymptotic decay of $\gamma$ implies $\lim_{r \rightarrow \infty} M_G(S_r) = M_{ADM}$. Thus,
$M_G$ indeed interpolates between the left and right-hand sides of the Penrose inequality.

Geroch original idea was to prove the positive mass theorem by starting the inverse mean
curvature flow from a point so that $M_G=0$ initially (the point can be 
approached as the limit of very small coordinated spheres). The positivity of mass
would follow provided the inverse mean curvature flow remained smooth all the way to infinity
and the flow approached large coordinate spheres in the asymptotically euclidean end. 
Jang and Wald \cite{JangWald1977} realized later on that the same argument could be used
to prove the Penrose inequality by starting from the outermost minimal surface. In fact, $p$
vanishes on this minimal surface, and hence the velocity $e^{\psi}$ diverges there. However, by
starting the flow from surfaces which approximate the outermost minimal surface from outside, the
Penrose inequality would follow.  However, it was immediately
realized that the flow will not remain smooth in general and that singularities
will develop. The first variation of area implies that, as long as the flow remains smooth,
the area of the surfaces increases exponentially. Consider now, 
as a very simple example, two disjoint balls
in Euclidean space. By symmetry, each surface will evolve under IMCF to a larger sphere with 
exponentially increasing radius. Thus, the two surfaces will necessarily touch
in finite time and the flow cannot remain smooth forever. A less trivial example, discussed
in \cite{HuiskenIlmanen1997}, consists of a thin torus in Euclidean space. The differential
equation satisfied by $p$ under IMCF is of parabolic type and the maximum principle implies that
$p$ stays bounded above in terms of its initial value (and the background geometry) as long as the flow remains
smooth. Thus, the torus will flow outwards at positive speed bounded away from zero and it will thicken.
However, sufficiently thick torus have vanishing
mean curvature at points on their inner rim. The speed becomes infinity there
and the flows necessarily stops being smooth.

The presence of singularities made this idea dormant for decades. The only case were the
method was made rigorous involved a particular case of metrics called {\it quasi-spherical
metrics with divergence-free shear} \cite{Bartnik1995}, where the flow was seen to remain
smooth all the way to infinity. Huisken and Ilmanen's 
fundamental contribution was to define the flow in a suitably weak sense so that
the singularities could be treated (and, in fact, basically avoided). 

An important ingredient in Huisken and Ilmanen's approach is the use of a level set
formulation for the IMCF (which is a
geometric parabolic flow). This means describing 
the leaves of the flow as the level sets of a real function $u$ on $\Sigma$. The IMCF 
condition translates directly into the following 
degenerate elliptic equation for $u$
\begin{equation}
\label{degel}
\mbox{div}_{\gamma} \left(\frac{{\nabla} u}{|{\nabla} u|} \right) = |{\nabla} u|,
\end{equation}
where $\mbox{div}_{\gamma} V$ is
the divergence of $V$ in $(\Sigma,\gamma)$. 
In principle it is possible that  $u$
remains constant on open sets, which has the immediate consequence that 
the flow may jump across regions with positive measure. The fundamental idea is 
to use these jumps precisely to avoid the singularities that the smooth IMCF would
otherwise have. In order to achieve this, Huisken and Ilmanen find a variational
formulation for (\ref{degel}). This equation is not the Euler-Lagrange equation of any
functional. However, by freezing the right-hand side to $|{\nabla} u|$, the authors
write down the functional (which now depends on $u$)
\begin{equation}
\label{funct1} 
J_u(v) = \int_{\Sigma}\left( |{\nabla} v| + v |{\nabla} u|\right) \etagamma.
\end{equation}
The critical points of this functional with respect to compactly supported variations
of $v$ gives (\ref{degel}) with $u$ replaced by $v$ on the left-hand side. One then looks for
functions $u$ which minimize their own functional, thus giving (\ref{degel}). 

This variational formulation has a geometric counterpart which, in rough terms, implies that
each of the level sets of $u$ (defined as $\partial \{u < t\}$ for positive $t$) is area outer minimizing. 
Thus, if we start from a smooth surface $S_0$ which is area outer minimizing and with positive
mean curvature, the IMCF (which enjoys short time existence thanks to its
geometric parabolic character) will evolve 
the surface smoothly for some ``time'' $\lambda$. For small enough values of $\lambda$, each level set
will be outer area minimizing. However, there may well exist a value for which $S_{\lambda}$
ceases to be area outer minimizing. Take the smallest of such values, $\lambda_1$. This means that 
there exists a surface $S'_{\lambda_1}$ which encloses $S_{\lambda_1}$ and has less or equal area. 
In fact, it must have equal area because if if had 
strictly less area, it would also have less area than some close enough previous leaf (the flow
is smooth up to $\lambda_1$ and the area changes smoothly).  The surface $S'_{\lambda_1}$ may 
have common points with $S_{\lambda_1}$. However, the surface $S'_{\lambda_1}
\setminus S_{\lambda_1}$
must be a minimal surface because otherwise it would admit a compactly supported
variation that would reduce its area while still remaining 
outside $S_{\lambda_1}$ and enclosing it. However, this varied surface would
(i) enclose all surfaces in the flow from $S_0$ up to $S_{\lambda_1}$, (ii) have larger area than
$S_0$ (because the latter is area outer minimizing) and (iii) have less area that $S_{\lambda_1}$.
So, for some value of the flow parameter smaller than $\lambda_1$ it would have the same area, and
it would enclose it. This would contradict the definition of $\lambda_1$ as being the smallest 
value with such property. Notice that, a priori, this outer minimizing prescription
does not exclude that the flow may become singular already in the interval $0 < \lambda < \lambda_1$,
where all the leaves remain area outer minimizing.
That this cannot happen 
is one of the several statements that Huisken and Ilmanen had to show, and which makes their proof
technically difficult.

In the torus example before, the smooth flow would thicken the torus (in an axially
symmetric way) until a surface is obtained which has exactly the same area than the surface
obtained by closing its hole by two horizontal planes. It is clear in this example
that the new pieces have vanishing mean curvature.

Since the level set function $u$ is such that each level set is outer area minimizing, this means
that the IMCF evolves smoothly for as long as the leaves remains area outer minimizing until
a surface which is not area outer minimizing is eventually reached.
At this point, the surface jumps (meaning that $u$ remains constant between
the two surfaces). The flow should be continued from the new surface outwards. Since
the new surface has pieces with $p=0$, the IMCF cannot be defined there in a classical
sense. However, the weak formulation in terms of level sets also takes care of this.

A crucial condition for the monotonicity argument to go through in this formulation
is that  the Geroch mass does not increase in the jumps. We know that the total area does not change. 
Moreover, the new pieces of the surface
have $p=0$, while the deleted pieces have $p>0$ because the flow was smooth up to and including
$S_{\lambda}$. Thus, across the jump the Geroch mass is non-decreasing provided the Euler
characteristic of the surface does not decrease. The example with the torus 
shows that in general the topology of the surface may change across the jump, in that case
changing from a surface with genus one to a topological sphere. In this particular case,
the Euler characteristic increases and the Geroch mass would be monotonically increasing. The 
question is therefore whether this is a general property or not. The example with two 
spheres in Euclidean space shows that this cannot be true in general. There, the value of the Euler 
characteristic is initially four and after the jump a topological sphere forms, for which
$\chi=2$. A general result by Huisken and Ilmanen \cite{HuiskenIlmanen2001} is that a
topologically $S^2$
surface in an asymptotically euclidean three-manifold $(\Sigma,\gamma)$ outside
the outermost minimal surface cannot jump, via the weak formulation of the IMCF, to
another surface with smaller Euler characteristic, i.e. no holes can appear on the 
surface after the jump. Nevertheless, the outermost minimal surface $S$ in $(\Sigma,\gamma)$
needs not be connected. Consequently, for the Geroch mass to remain 
monotonic under the weak IMCF, a connected component $S_i$ of 
the outermost minimal surface must be chosen as initial surface. The Penrose inequality proven by
Huisken and Ilmanen is, therefore,
\begin{eqnarray}
M_{ADM} \geq \max_i \sqrt{\frac{|S_i|}{16 \pi}}, \label{PIHI}
\end{eqnarray}
where $i$ runs over the connected components of the outermost minimal surface.

As mentioned before, the Riemannian Penrose inequality also has a rigidity part. In this setting, this states
that equality is achieved in (\ref{PIHI}) if and only if 
the region in $(\Sigma,\gamma)$ outside its outermost minimal
surface $S_m$ is isometric to the domain $r > m/2$ of $(\Sigma_{\mbox{\tiny{Sch}}},\gamma_{\mbox{\tiny{Sch}}})$
(\ref{gsch})  with $m= M_{ADM}$.
Heuristically, it is clear that some rigidity is to be expected already
from the monotonicity property of the Geroch mass.
If equality is achieved, then the derivative of the Geroch mass is everywhere zero. In particular, in
the smooth part of the flow,  $R(\gamma)=0$, $\psi= \mbox{const}$ and each leaf is totally
umbilical (i.e. $\Pi^{\vec{m}}_{AB}=0$) at each point. The IMCF condition gives $p = \mbox{const.}$ on 
each leaf. Since $\partial_{\lambda} = e^{\psi} \vec{m} = \frac{1}{p} \vec{m}$, the
variation formula (\ref{mp}) implies $\frac{d p}{d \lambda} = - \frac{3}{4} p + \frac{R(h)}{2p}$. 
This implies that each leaf is a metric sphere (because $R(h) = \mbox{const.}$). The exponential grow of
the area gives $|S_{\lambda}| = |S_0 | e^{\lambda}$. Defining a new variable $\hat{r}= \sqrt{|S_{\lambda}|/4\pi}$, 
the metric $\gamma$ takes the form $\gamma = \frac{4}{p^2 \hat{r}^2} d\hat{r}^2 + \hat{r}^2 d \Omega$.
The expression for the Geroch
mass and the fact that it takes a value $m$ independent of $\lambda$  gives $\hat{r}^2 p^2 = 4 (1 -
\frac{2 m}{\hat{r}} )$, and
hence $\gamma$ is the Schwarzschild metric outside the minimal surface $\hat{r} = 2 m$. This is, of course,
a heuristic derivation. Huisken and Ilmanen are able show the same results using only the weak flow.

Huisken and Ilmanen's proof has interesting side consequences. First of all, the argument does not use
the positive mass theorem anywhere and therefore the method gives an independent proof of 
this important result. In the presence of an outermost minimal surface, positivity of $M_{ADM}$
is an obvious consequence of the theorem. If $\Sigma$ is free of minimal surfaces, it suffices to start
the weak inverse mean curvature flow at one point, making rigorous the original idea of Geroch.

Another interesting consequence deals with the so-called Bartnik mass (or capacity)
\cite{Bartnik1989}, defined as follows. A 3-dimensional Riemannian manifold $(\Omega,\gamma)$ with no boundary
and compact metric closure is called 
admissible if it  can be isometrically embedded into a complete and connected
asymptotically euclidean three-dimensional manifold $(\Sigma,\gamma)$ satisfying the following
three properties: (a) the curvature scalar is non-negative, (b) the boundary of $\Sigma$
is either empty or a minimal surface, and (c) $\Sigma$ is free of any other minimal
surfaces (the original definition excludes minimal surfaces even at the boundary of $\Sigma$, this 
modification is due to Huisken and Ilmanen \cite{HuiskenIlmanen2001}). $(\Sigma,\gamma)$ is called an admissible extension.
The Bartnik capacity is the infimum of the ADM masses of all possible extensions of 
$(\Omega,\gamma)$. By the positive mass theorem, it is immediately non-negative and it was conjectured
in \cite{Bartnik1989-2, Bartnik1995, Bartnik1997} to be positive if $(\Omega,\gamma)$ is not a subset of Euclidean space.
Huisken and Ilmanen used the weak IMCF to reach a slightly weaker conclusion, namely
that if the Bartnik capacity is zero, then $(\Omega,\gamma)$ is locally isometric to Euclidean 
space. The idea of the proof is to take a point $p$ in $\Omega$ where the metric
is non-flat and choose any admissible extension 
$(\Sigma,\gamma)$. Consider the weak IMCF starting at $p$. For some small enough value of the
flow parameter the corresponding leaf $S_{\lambda_0}$ 
can be proven to have a positive Geroch mass, and to lie within $\Omega$ independently of the extension (this means in
particular that $S_{\lambda_0}$ is area outer minimizing independently of the extension). Thus, monotonicity
of the IMCF gives the lower bound $M_{ADM} \geq M_G (S_{\lambda_0})$. Since this is independent of the extension, the result
is established.

Another consequence of Huisken and Ilmanen's construction is the so-called ``exhaustion'' property of the Bartnik mass.
This can be formulated as follows. Consider an asymptotically euclidean Riemannian manifold $(\Sigma, \gamma)$ and any
increasing collection of bounded sets $\Omega_i \subset \Sigma$ satisfying $\cup_i \Omega_i = \Sigma$. Then
the Bartnik mass $m_B(\Omega_i)$ tends to the ADM mass when $i \rightarrow \infty$. This is proven
in \cite{HuiskenIlmanen2001} by taking suitable coordinate balls within each $\Omega_i$ in such a way that the coordinate
radius $R_i$ tends to $\infty$. It is shown, by using the IMCF, that the Bartnik mass of each $\Omega_i$ is bounded below
by the Geroch mass of the corresponding coordinate balls. The Geroch mass of the coordinate balls tends to the ADM 
mass when $R_i \rightarrow \infty$. Since the Bartnik mass of $\Omega_i$ is obviously bounded above by $M_{ADM}$ of $\Sigma$,
the exhaustion property $m_B (\Omega_i) \rightarrow M_{ADM}$ follows.

\subsection{Bray's proof}
\label{Braysproof}

H.L. Bray was able \cite{Bray2001} to prove the Riemannian Penrose inequality for arbitrary outermost
minimal surfaces, with no restriction of connectedness. Bray's proof also uses geometric
flows in an essential way, but instead of flowing surfaces in a fixed Riemannian 
manifold, as in Huisken and Ilmanen's proof, it uses a flow of metrics.
The idea is to modify the metric with a one-parameter family of conformal factors  so that
the curvature scalar remains non-negative,  the new horizon with respect to the conformal metric
has the same area as the original one and the total ADM mass does not increase during the process.
If, moreover, the flow can be defined in such a way that, as $t\rightarrow \infty$, the Riemannian
manifold outside its horizon converges to the ${r > m/2}$ portion of the Schwarzschild
manifold (\ref{gsch}) in a suitable sense, then the Penrose inequality for the original manifold
follows because the inequality (in fact equality) holds in the limit. This conformal flow also allows
Bray to show the rigidity part of the Penrose conjecture, namely that equality 
holds if and only if the starting Riemannian manifold is isometric to the exterior region of
Schwarzschild.

Bray's statement of the Penrose inequality deals with three-dimensional 
asymptotically euclidean, Riemannian manifolds $(\Sigma,\gamma)$ with one or several
asymptotic ends (in the latter case, one of them is selected and the mass and the concept of ``outer''
are taken with respect to this end) and which contain 
an area outer minimizing horizon $S_0$. By definition,
a horizon is a smooth, compact, minimal
surface which is a boundary of an open set $\Omega^{-}$ containing all
possible asymptotic ends on $(\Sigma,\gamma)$ except the one that has been selected.
In the case of several ends the existence of such a horizon is
guaranteed, otherwise $(\Sigma, \gamma)$ is assumed to have one. An area outer minimizing horizon need not be
outermost, although the outermost one (which
always exists) will have at least the same area as $S_0$ (by the 
area outer minimizing property of the former). As already said, $S_0$
need not be connected and  Bray's theorem states $M_{ADM} \geq \sqrt{\frac{|S_0|}{16 \pi}}$ with 
equality if and only if $(\Sigma,\gamma)$ is isometric to the exterior region of the Schwarzschild manifold.

The first step in Bray's proof is to reduce the class of metrics under consideration. 
To that aim, Bray uses an interesting result due to Schoen and Yau \cite{SchoenYau1981-1}
(c.f. Theorem 7 in \cite{BrayThesis}) which states that
given an asymptotically euclidean, three-dimensional Riemannian
manifold $(\Sigma,\gamma)$  (see Proposition 4.1 in \cite{Schoen1989} for a similar statement
in any dimension)
with $R(\gamma) \geq 0$, and any positive number $\epsilon$, there always exists an asymptotically
euclidean metric ${\gamma}_{\epsilon}$ on $\Sigma$, with $R(\gamma_{\epsilon}) \geq 0$ everywhere
and identically zero outside a compact set $K$, which is conformally flat outside $K$, i.e.
${\gamma}_{\epsilon} = s_{\epsilon}^4 \delta$ on $\Sigma^{\infty} \equiv \Sigma \setminus K$, where
$s_{\epsilon}: \Sigma^{\infty} \rightarrow \mathbb{R}$ is a positive function
approaching a positive constant at infinity (on each one of the asymptotic ends). Moreover,
the following conditions are fulfilled,
\begin{eqnarray}
\left | M_{ADM} (\gamma) - M_{ADM} ({\gamma}_{\epsilon}) \right | & \leq & \epsilon, \nonumber \\
\left | \frac{{\gamma}_{\epsilon} ( \vec{X}, \vec{X} )}{\gamma (\vec{X},
\vec{X} )} - 1 \right | & \leq & \epsilon, \quad \forall \vec{X} \in T_x \Sigma
\mbox{\,\,\,\, and  \,\,\,\,} \forall x \in \Sigma, \label{epsilonchange}
\end{eqnarray}
i.e. the total mass changes within $\epsilon$ and the metric itself changes pointwise
at most by $\epsilon$ (in the sense that unit vectors change their norm at most by $\epsilon$).
The condition that ${\gamma}_{\epsilon}$
has vanishing curvature scalar outside $K$ implies that $s_{\epsilon}$ is harmonic (with respect
to the flat metric). Therefore, the metric ${\gamma}_{\epsilon}$ 
is called {\it harmonically flat}. This perturbation result has been strengthened in
Bray's Ph.D. thesis \cite{BrayThesis} to show that $s_{\epsilon}$ can in fact be taken to be
spherically symmetric, in which case $\gamma_{\epsilon}$ is the Schwarzschild metric outside $K$.
Such a metric is called {\it Schwarzschild at infinity}. This stronger version was needed to apply the
isoperimetric methods described in Sect. \ref{isoperimetricprofile}.

Schoen and Yau's perturbation result implies that, in order to prove the Riemannian Penrose inequality,
it is sufficient
to consider harmonically flat metrics. Indeed, if there existed a counterexample to 
the Penrose inequality, i.e. a metric with total ADM mass strictly smaller than
the Geroch mass of its outermost minimal surface, then we would be able to find
a harmonically flat metric ${\gamma}_{\epsilon}$ with ADM mass within $\epsilon$ of the original one. 
Since the area of the outermost minimal surface also depends continuously on
$\epsilon$, it is clear that $\epsilon$ can be arranged so that 
the Penrose inequality would also 
be violated for the harmonically flat metric ${\gamma}_{\epsilon}$. 
%Thus, it is sufficient to prove
%the Penrose inequality for harmonically flat metrics.

For this  class of manifolds,
Bray defines a flow of metrics by a conformal rescaling ${\gamma}_{t} =  u_{t}^4 \gamma$ which depends
on a parameter $t \geq 0$.  The function $u_t$ is defined via an elliptic equation for
$v_t \equiv \frac{d u_t}{dt}$
together with the initial value $u_0 = 1$ (so that ${\gamma}_0 = \gamma$). $v_t$ is defined by solving the 
Dirichlet problem
\begin{eqnarray}
\left . \begin{array}{rcl}
\Delta_{\gamma} v_{t} & = &  0 \mbox{\,\,\, \,\,\,\,\, on \,\,   } \Sigma_{t}, \\
v_t & = & 0 \mbox{\,\,\, \,\,\,\,\, on \,\,   } \Sigma \setminus \Sigma_{t}, \\
\left . v_{t}  \right |_{S_{t}} & = &  0, \\
v_{t} & \rightarrow & -e^{-t} \mbox{\, \,
at \, \,} \infty 
\end{array} \right \},  \label{BraysFlow}
\end{eqnarray} 
where $S_t$ is the outermost minimal area surface enclosing $S_0$ with respect to 
the metric ${\gamma}_t$ and $\Sigma_t$ is the exterior of $S_t$. Since $S_t$
is defined using the metric one is trying to construct, it is not obvious a 
priori that such problem admits a solution. Bray uses a discretization argument whereby
$u_t(x)$ as a function of $t$ (i.e. for fixed value of $x$) is discretized
as a continuous piecewise linear function $u^{\epsilon}_t(x)$ (in $t$), where the jumps in the derivatives occur
at fixed intervals of length $\epsilon$. The slope of $v^{\epsilon}_t = \frac{d u^{\epsilon}_t}{dt}$ on 
the $k$-th interval, i.e. on $t \in [(k-1) \epsilon, k \epsilon)$, $k \in \mathbb{N}$ 
is defined inductively in $k$. At $k=1$, $v^{\epsilon}_t$ is zero inside 
$S_0$ and solves the Laplace equation $\Delta_{{\gamma}_0} v^{\epsilon}_t =0$ with boundary data
$v^{\epsilon}_t =0$ on $S_0$ and $v^{\epsilon}_t \rightarrow -1$ at infinity. $S^{\epsilon}_1$ is then defined
as the outermost minimal area enclosure of $S_0$ with respect to the metric $(u^{\epsilon}_{t=\epsilon})^4 {\gamma}_0$
(the function $u^{\epsilon}_{t=\epsilon}(x)$ is known by continuity). 
Notice that since $S_0$ was not assumed to be the outermost minimal surface, $S_1$ may  be far away from
$S_0$ even for arbitrarily small $\epsilon$. Assume that the construction has been carried out for
$k \in \mathbb{N}$, then the slope $v^{\epsilon}_t$ on the interval $t \in [k \epsilon, (k+1) \epsilon )$ is defined
as being zero inside $S_k$ and solving the Laplace equation $\Delta_{{\gamma}_0} v^{\epsilon}_t =0$ which
approaches the constant $- (1- \epsilon)^{k}$ at infinity and vanishes on $S_{k}$.
The surface $S_{k+1}$ is defined as the 
outermost minimal area enclosure of $S_k$ with respect to the metric $(u^{\epsilon}_{t=(k+1)\epsilon})^4 {\gamma}_0$
(again, the function $u^{\epsilon}_{t=(k+1)\epsilon}(x)$ is known by continuity).

Thus, while the metric is changed continuously in $t$, the boundaries $S^{\epsilon}_{k}$ 
change only at discrete values. Existence of $u^{\epsilon}_t$ follows from the inductive construction.
The conformal flow of metrics depends on $\epsilon$.
It is an important ingredient of Bray's proof to show that a suitable limit $\epsilon \rightarrow 0$
exists, thus defining the flow of metrics in (\ref{BraysFlow}).
The limit is performed in two ways, first the conformal factors are seen to converge
when $\epsilon \rightarrow 0$ to a locally Lipschitz function $u_t$. This
already defines the metric ${\gamma}_t$ and the surfaces $S_t$ as before. Extending the discrete
collection of $S^{\epsilon}_k$ to a one-parameter family via $S^{\epsilon}(t) \equiv
S^{\epsilon}_k$ for $t \in [k \epsilon, (k+1)\epsilon)$ the convergence of these surfaces when $\epsilon \rightarrow 0$ 
(for fixed $t$) is then considered. It turns out that the limit may depend
on the subsequence $\epsilon_i \rightarrow 0$. Bray denotes the collection of all limit surfaces
as $\{ S_{\alpha} (t) \}$, where $\alpha$ identifies the element in the collection. Convergence is seen to be in the
Hausdorff distance sense and each element in $\{S_{\alpha}\}$ is proven to be a smooth surface.
The surfaces $S_t$ and $\{ S_{\alpha} (t) \}$ need not always coincide. However, they are
closely related: for any $t_2 > t_1 \geq 0$, $S_{t_2}$ encloses $S_{\alpha_1}(t_1)$ (for any 
$\alpha_1$) and $S_{\alpha_2} (t_2)$ encloses $S_{t_1}$ (for any $\alpha_2$). Thus, they must coincide for any value where
$S_t$ depends continuously on $t$. Bray then proves that $S_t$ is continuous except
for $t$ in a countable set $J$ (which may be empty). Furthermore, $S_t$ is everywhere upper-semicontinuous
and the right limit always encloses the left limit. Moreover, for $t_2 > t_1$,
$S_{t_2}$ encloses $S_{t_1}$. Thus, the flow of surfaces $S_{t}$ is everywhere outward
and may jump, also outwards, at a countable number of places. 

In addition, and this is crucial for the proof of the Penrose inequality, the area of $S_t$
is constant for all $t$, even at the jumps. Outside the jumps, this constancy can be understood heuristically
because $S_t$ is a closed minimal surface and hence its area does not change to first order with
respect to any variation, in particular the variation which transforms $S_t$ into
$S_{t + \delta}$ ($\delta$ small) in the metric ${\gamma}_t$.
The area also changes because the metric depends on $t$.
However, since $v_t =0$ on $S_t$ by construction,
the metric ${\gamma}_{t+ \delta}$ coincides with ${\gamma}_t$ on $S_t$  to first order  
and therefore
the area of $S_t$ with the metric ${\gamma}_t$ coincides to first order with its area
with respect to ${\gamma}_{t+\delta}$. This implies that $|S_t|$ is constant. At the jumps, the statement
requires a careful analysis of the various limiting procedures involved.

The second crucial ingredient in Bray's proof is the fact that the total mass $m(t)$
of the conformally rescaled metric ${\gamma}_t$ is a non-increasing function of $t$. The
proof relies on two facts. First of all, the definition of $v_t$ in
$(\ref{BraysFlow})$ seems to depend on $\gamma$ and on $t$. However, as a simple consequence
of how the Laplacian changes under conformal rescalings, 
it follows that $v_t$ depends only on ${\gamma}_t$ (i.e. $v_t$
solves a suitable Dirichlet problem with respect to the metric ${\gamma}_t$,
with no reference to the original metric ${\gamma}_0$). Therefore, proving that
$\frac{d m(t)}{dt} \leq 0$ is equivalent to showing 
$\left . \frac{dm}{dt} \right |_{t=0} \leq 0$ because there is nothing that distinguishes ${\gamma}_0$
from any other metric ${\gamma}_t$ in the flow. 
For $t=0$, the function $v_0$ restricted to
$\Sigma_0$ (the exterior of $S_0$ in $\Sigma$) is just minus the Green function of ${\gamma}_0$, defined as
\begin{eqnarray*}
\left . \begin{array}{rcl}
\Delta_{{\gamma}_0} G & = &  0 \mbox{\,\,\, \,\,\,\,\, on \,\,   } \Sigma_0, \\
\left . G  \right |_{S_0} & = &  0, \\
G  & \rightarrow & 1 \mbox{\, \, at \, \,} \infty 
\end{array} \right \}.
\end{eqnarray*} 
The total mass for a harmonically flat metric $u^4 \delta$, can be computed directly from the 
behaviour of $u$ at infinity. Namely, if $u = a +  b /(2 r) + O(1/r^2)$
then the ADM mass is $m= ab$. The metric ${\gamma}_t$ is of the form $\gamma_t
= (1 - t G + O(t^2) )^4 {\gamma}_0 = (1- t G + O(t^2))^4 {\cal U}^4 \delta$,
with ${\cal U} = 1 - \frac{M_{ADM}}{2r} + O(r^{-2})$, where $M_{ADM}$ is the ADM mass of ${\gamma}_0$.
Expanding the Green function $G$ in the asymptotic region
as $G = 1 - c/(2r) + O(1/r^2)$, it follows from a direct calculation that 
$m(t)= M_{ADM} + t (c - 2 M_{ADM}) + o(t)$. Thus, the mass
will not increase provided $c \leq 2 M_{ADM}$. This turns out to be a general
property of the Green function $G$ of any asymptotically flat metric on a manifold with
compact minimal boundary $S_0$. This result, also due to Bray \cite{Bray2001},
has independent interest and has been already used in another context by P. Miao \cite{Miao2005}
to characterize the Schwarzschild initial data as the only static initial data
with a non-empty minimal boundary. 

The proof of $c \leq 2M_{ADM}$ relies on the positive mass theorem and uses a technique first introduced by
Bunting and Masood-ul-Alam \cite{BuntingMassood-ul-Alam1987}
to prove uniqueness of the Schwarzschild black hole. The idea is to double
$(\Sigma_0,{\gamma}_0)$ across its boundary $S_0$ (i.e. take two copies and identify the boundaries)
to define a new manifold $\hat{\Sigma}$.
Define also a function $\Phi(x) = \frac{1+G(x)}{2}$
on one of the copies and $\Phi(x) = \frac{1-G(x)}{2}$ on the other copy. This defines a 
function which is harmonic away from $S_0$ (which remains a minimal surface) and which approaches
$1$ on one asymptotic end and zero on the other asymptotic end.  This function is also $C^1$
everywhere. Consider the conformally rescaled metric $\hat{\gamma} = \Phi^4 {\gamma}_0$ on
$\hat{\Sigma}$. The asymptotic behaviour of $G$ near the infinity where it vanishes
shows that $(\hat{\Sigma},\hat{\gamma})$ admits a one-point compactification there, thus
defining a complete asymptotically euclidean Riemannian manifold. Furthermore, the behaviour
of the curvature scalar under conformal rescalings and the fact that $\Phi$ is harmonic
implies that ${\gamma}_0$ has non-negative curvature scalar away from $S_0$. If $(\hat{\Sigma}, \hat{\gamma})$
were smooth, the positive mass theorem could be invoked to conclude that the
mass of $\hat{\gamma}$ is non-negative, and zero if and only if $(\hat{\Sigma},\hat{\gamma})$ is 
Euclidean space. However, the metric
$\hat{\gamma}$ is not smooth across $S_0$ and Bray needs to use an approximation argument 
to reach the same conclusion. Consequently, the ADM mass of $(\hat{\Sigma},\hat{\gamma})$ is non-negative.
The mass of this manifold is just $M_{ADM} - c/2$, from which $c \leq 2M_{ADM}$ follows.

%Notice that this smoothing argument is not needed in the black hole uniqueness
%theorem by Bunting and Masood-ul-Alam \cite{BuntingMassood-ul-Alam1987}
%because there the boundary $S_0$ is not only minimal but also totally geodesic, and $(\hat{\Sigma},\hat{g})$
%is automatically $C^{1,1}$.

Using related techniques, Miao \cite{Miao2002} has extended these results and
has proven that the positive mass theorem holds for complete, asymptotically euclidean Riemannian manifolds 
$(\Omega,\gamma)$ of dimension $n \leq 7$ and non-negative curvature scalar, with a metric
which is allowed to be non-smooth across  a compact 
codimension one hypersurface $S$ provided (i) this hypersurface separates $\Omega$ into two
pieces $\Omega^{\pm}$, with $\Omega^{+}$ containing the chosen asymptotically euclidean end, (ii) 
the induced metric on $S$ from both sides coincides and (iii) the mean curvature 
with respect to the outer direction satisfies the inequality 
\begin{eqnarray}
p^+(S) \geq p^- (S), \label{IneqH}
\end{eqnarray}
where $p^{\pm}(S)$ is computed with the geometry of $(\Omega^{\pm},\gamma|_{\Omega^{\pm}} )$.
In the case considered by Bray, the metrics are $\hat{\gamma}_{\pm} = [(1\pm G)/2]^4 {\gamma}_0$. Under
a conformal rescaling $\hat{\gamma} = u^4 \gamma$,
the mean curvature $p$ of a hypersurface changes according to
$\hat{p} = (p + 4 \vec{n} (u))/u^2$, where $\vec{n}$ is the unit normal along
which $p$ is calculated. Applying this to the metrics $\hat{\gamma}_{\pm}$ and using that
$S_0$ is minimal, it follows $p^{-}(S_0) = p^{+}(S_0)$ and the positive mass theorem 
proven by Bray is recovered.  It is worth mentioning that a related positive mass theorem has been proven
using spinor techniques by Shi and Tam \cite{ShiTam2002} 
on spin manifolds in any dimension (this 
entails no restriction in three dimensions, provided the manifold is orientable)
for metrics which are smooth on $\overline{\Omega^{+}}$ and
on $\overline{\Omega^{-}}$, only Lipschitz across $S$ and such that equality 
in (\ref{IneqH}) holds.

Returning to Bray's proof of the Penrose inequality, once the area of the horizons $S_t$ is known
to be constant and that the mass does not increase, the final step is to 
show that the metric outside $S_t$ approaches the Schwarzschild metric in a suitable sense. The
surfaces $S_t$ flow outwards in $\Sigma$. In fact, Bray proves that, for 
sufficiently large $t$, $S_t$ encloses any bounded set of $\Sigma_0$.
Thus, the shrinking manifold $\Sigma_t$ behaves in such a way that any given point $p \in \Sigma_0$
is left out for a sufficiently large $t$. It may seem, therefore, that this manifold
disappears in the limit. Here is where
the asymptotic condition $u_t \rightarrow e^{-t}$ at infinity comes into play.
For very large $t$, the asymptotic value of the conformal factor $u_t$ also tends to zero. Thus, vectors
that were unit at large distances in the original metric ${\gamma}_0$ have increasingly smaller
lengths in the metric ${\gamma}_t$. However, this metric is also asymptotically euclidean, so for a suitable choice 
of coordinates, the metric must approach the flat metric in Cartesian coordinates. In these coordinates,
the  region $\Sigma_t$ shrinks more slowly (or it may even expand). In order to see why is this so, consider
the metric 
defined by
\begin{eqnarray}
g^{\mbox{\tiny{Euc}}}_t = e^{-4t} \left (dr^2 + r^2 d\theta^2 + r^2 \sin^2 \theta d\phi^2 \right)
\label{euclidean}
\end{eqnarray}
on the domain $r > R_0(t)$ with $R_0(t) \rightarrow \infty$ as $t \rightarrow \infty$. The domain
is shrinking in the coordinates $\{r,\theta,\phi\}$. However, (\ref{euclidean}) is 
the Euclidean metric for any $t$. The transformation to the standard spherical coordinates $\{ r^{\prime},
\theta, \phi \}$ is given by $r^{\prime} = e^{-2t} r$. In the new coordinates, the domain is
$r^{\prime} > e^{-2t} R_0(t)$ which is not shrinking if $R_0(t)$ grows at most as $e^{2t}$. 
This is the type of behaviour that occurs to the conformal flow of metrics introduced by Bray. This can be seen explicitly
by considering the flow in the particular case when the starting metric is the Schwarzschild manifold (\ref{gsch}). Rotational
symmetry allows one to integrate the equations easily. The flow of metrics ${\gamma}_t$ in the exterior domain $r \geq m/2$
are, for $t \geq 0$, \cite{Bray2001}.
\begin{eqnarray}
{\gamma}_t = \left \{  \begin{array}{lll}
\frac{4m^2}{r^2} \left ( dr^2 + r^2 d\theta^2 + r^2 \sin^2 \theta d \phi^2 \right )
& \frac{m}{2} \leq r \leq \frac{m}{2} e^{2t} & \mbox{interior domain,} \\
\left ( \frac{m}{2 r} e^t + e^{-t} \right )^4 \left ( dr^2 + r^2 d\theta^2 + r^2 \sin^2 \theta d \phi^2 \right )
& r \geq \frac{m}{2} e^{2t} & \mbox{exterior domain.} \\
\end{array} \right .  \label{FlowSchw}
\end{eqnarray} 
The exterior domain indeed shrinks in these coordinates. However, performing the
coordinate transformation (or the diffeomorphism if an active point of view is preferred) $r^{\prime} = r e^{-2t}$,
with $\theta$ and $\phi$ unchanged, the exterior region is transformed back into the original exterior
part of the Schwarzschild manifold. The metric (\ref{FlowSchw}) 
in the region already swapped by the surfaces $\{S_t\}$ is cylinder-like in the
sense that each surface $r= \mbox{const.}$ has the same area. This is a general property of Bray's flow, since each 
of the surfaces $S_t$ has the same area as the starting horizon.
The behaviour in the exterior part is also general. The idea is that since the surface $S_t$ grows unboundedly,
eventually it swallows all the interior geometric features of $(\Sigma_0,{\gamma}_0)$ and leaves only the asymptotic
geometry. Since the mass does not increase along the flow of metrics, its limit exists and is non-negative due to the 
positive mass theorem. Bray proves that the limit mass is, in fact, positive and hence the geometry in the
exterior approximates Schwarzschild with increasing accuracy. The asymptotic
behaviour of the conformal factor $u_t \rightarrow e^{-t}$ has the effect that this exterior remains unbounded,
in the sense that after performing the appropriate diffeomorphism, all the surfaces $S_t$ stay within a bounded set. 
More precisely, Bray shows that for each $\epsilon >0$
there exists a $T$ so that for $t > T$ 
there exists a diffeomorphism $\Phi_t$ between
$(\Sigma_t, {\gamma}_t)$  and the exterior region $r \geq m_F/2$ of a fixed Schwarzschild manifold 
$(\Sigma_{\mbox{\tiny{Sch}}},\gamma_{\mbox{\tiny{Sch}}})$ of mass $m_F$,
such that both metrics are $\epsilon$-close to each other
(in the sense that the length of any unit vector of the metric ${\gamma}_t$ has length 
in the interval $(1-\epsilon,1+\epsilon)$ for the metric $\Phi^{\star}_t ({\gamma}_{sch})$). Moreover,
$m(t)$ and $m_F$ also differ at most by $\epsilon$. This concludes the proof
of the inequality
\begin{eqnarray}
M_{ADM} \geq \sqrt{\frac{|S_0|}{16 \pi}} \label{BrayPI}
\end{eqnarray}
for the original metric. As Bray points out, no property from $(\Sigma,{\gamma}_0)$ inside
$S_0$ is used in the proof, therefore establishing (\ref{BrayPI}) also 
for manifolds with boundary $S_0$,  provided this is an area outer minimizing horizon.

While Huisken and Ilmanen's proof is strongly dependent on the dimension of the manifold, Bray's
approach can be generalized to any dimension $n \leq 7$,
as proven by Bray and Lee in \cite{BrayLee2007}. The limitation $n\leq 7$ comes basically from two facts. First,
regularity of minimal surfaces holds only for  dimensions $n \leq 7$, and the outermost
minimal surface enclosure of $S_0$, which is a crucial ingredient in the method, needs to be smooth
for the argument to go through. Secondly, the positive mass theorem for general asymptotically euclidean manifolds
(not spin) is only known to hold so far for dimensions $n \leq 7$ (the underlying reason is again regularity
of minimal surfaces, as Schoen and Yau's proof uses this type of surfaces in a fundamental way). The statement of the 
Riemannian Penrose inequality in higher dimensions is, of course, different to (\ref{BrayPI}), since the physical 
dimension of mass is not length in higher dimensions. The proper statement in (space) dimension $n$ is
\begin{eqnarray}
M_{ADM} \geq \frac{1}{2} \left ( \frac{|S_0|}{\omega_{n-1}} \right )^{\frac{n-2}{n-1}}
\label{RiemPIHiguerDim}
\end{eqnarray}
where $\omega_{n-1}$ is the area of an $(n-1)$-dimensional sphere of unit radius.
The rigidity statement says that
equality in (\ref{RiemPIHiguerDim}) occurs only for the higher dimensional Schwarzschild metric (first 
discussed by Tangherlini \cite{Tangherlini1963})
\begin{eqnarray}
\gamma^{(n)}_{\mbox{\tiny{Sch}}} = \left ( 1 + \frac{m}{2 |x|^{n-2}} \right )^{\frac{4}{n-2}} \delta_{ij} dx^i dx^j.
\label{Tanger}
\end{eqnarray}
As discussed in \cite{BrayLee2007} several of the steps in Bray's proof extend easily to
higher dimensions: the definition and existence of the conformal flow,
the proof that the area of $S_t$ remains constant under the flow, and the fact that the ADM mass $m(t)$
does not increase. However, the arguments required to show that
$S_t$ eventually encloses all bounded sets,
and that the flow of metrics converges to a Schwarzschild
metric after a suitable $t$-dependent diffeomorphism, do in fact depend strongly on the dimension. This is because
the
original argument is based on
the Gauss-Bonnet theorem and a Harnack type inequality which is valid only in three-dimensions. 
Moreover, the convergence to Schwarzschild requires 
a refinement of the rigidity part of the positive mass theorem, which roughly speaking gives a quantitative description
of how the metric approaches the flat metric when the mass tends to zero. In Bray \cite{Bray2001}, this
is proven using spinor techniques. This result is extended
by Lee \cite{Lee2007} 
to arbitrary higher dimensional manifolds for which the positive mass theorem holds (in particular for dimensions
$n \leq 7$).   

The main technical work to establish (\ref{RiemPIHiguerDim}) in \cite{BrayLee2007} is therefore devoted to prove
that the conformal flow converges in a suitable sense to 
the metric (\ref{Tanger}). Besides being more general, the argument presented in \cite{BrayLee2007} 
in fact simplifies and streamlines
some of the original steps in the three-dimensional case.

\section{Penrose inequality for asymptotically hyperbolic Riemannian manifolds}
\label{hyperbolicSect}

Despite its difficulty, the Riemannian Penrose inequality is considerably simpler than
the general Penrose inequality because all the complications arising from the
second fundamental form disappear. 
The time-symmetric case, however, is not the only one where this kind of simplification
occurs. Another example is given by
initial data sets $(\Sigma, \gamma_{ij}, A_{ij})$ which are umbilical, i.e.
when the second fundamental form is proportional to the metric with a constant
proportionality factor. Writing $A_{ij} =
\lambda \gamma_{ij}$, the Hamiltonian constraint (\ref{HamiltonianConst}) 
becomes $R(\gamma) = - 6 \lambda^2 + 16 \pi \rho$ (or
$R(\gamma) = - n (n-1) \lambda^2 + 16 \pi \rho$ if $\Sigma$ is $n$-dimensional).
If the energy condition $\rho \geq 0$ holds, then $R(\gamma) > - 6 \lambda^2$

In an asymptotically flat spacetime, slices with this property and $\lambda
\neq 0$  cannot reach
spacelike infinity because the decay (\ref{decay})
is obviously not satisfied. This type of initial data are usually called ``asymptotically
hyperbolic'' (also ``hyperboloidal'') since the simplest example is given by 
the hyperboloid $t = \sqrt{r^2 + \lambda^{-2} }$ in Minkowski spacetime. 
As a Riemannian
manifold, this hypersurface is just the standard three-dimensional 
hyperbolic space of radius $1/\lambda$ (i.e. $\mathbb{R}^3$ endowed
with the metric of constant negative
curvature $-\lambda^2$). Asymptotically hyperbolic initial data sets approach null infinity
provided $\gamma_{ij}$ satisfies suitable 
asymptotic conditions. In the context of the Penrose inequality, the model example 
consists of the  spherically symmetric slices $\Sigma$ of the Kruskal spacetime
satisfying $A_{ij} = \lambda \gamma_{ij}$. It turns out that for $\lambda>0$, $\Sigma$
is fully contained in the advanced Eddington-Finkelstein portion of the spacetime
and satisfies the equation (in advanced coordinates $(v,r,\theta,\phi)$)
%\begin{eqnarray*}
%ds^2 = - \left ( 1- \frac{2m}{r} \right ) dt^2 -2 2 dv dr + r^2 d \Omega^2, \quad
%(v, r, \theta, \phi) \in \mathbb{R} \times \mathbb{R}^{+} \times S^2
%\end{eqnarray*} and 
\begin{eqnarray*}
\frac{dv}{dr} = \frac{\lambda r - \sqrt{1 - \frac{2m}{r} + \lambda^2 r^2} }{\left (1- \frac{2 m}{r}
\right ) \sqrt{1 -\frac{2m}{r} + \lambda^2 r^2 }}.
\end{eqnarray*}
These hypersurfaces approach future null infinity
for $r \rightarrow \infty$ and 
intersects the white hole event horizon on the surface $S_1 \subset \{r=2m\}$, which has
mean curvature $ p =
2 \lambda$ (because $q = 2 \lambda$ on any surface of $\Sigma$ and $\theta_{-} = - p + q$ vanishes
on the white hole event horizon). The induced metric on $\Sigma$ is 
\begin{eqnarray}
dl^2 = \frac{dr^2}{1- \frac{2m}{r} + \lambda ^2 r^2} + r^2 d \Omega^2
\label{hyperbolic}
\end{eqnarray}
and obviously satisfies $R(\gamma) = -6 \lambda^2$. This model space arises not only as an umbilic hypersurface
of Kruskal, but also as time-symmetric hypersurfaces in spherically symmetric solutions
of the vacuum Einstein field equations with cosmological
constant, i.e. satisfying $\mbox{Ein}( g ) = - \Lambda  g$.
The general spacetime solution is given by the so-called Kottler metric 
\cite{Kottler1918} (discovered independently by Weyl \cite{Weyl1919}) and reads
\begin{eqnarray}
ds^2= - \left ( 1- \frac{2m}{r} - \frac{\Lambda r^2}{3} \right ) dt^2
+ \frac{dr^2}{ 1- \frac{2m}{r} - \frac{\Lambda r^2}{3}} + r^2 d \Omega^2
\label{Kottler}
\end{eqnarray}
This metric is often called 
Schwarzschild-de Sitter ($\Lambda >0$) or Schwarzschild-anti de Sitter ($\Lambda <0$) since it
contains both the Schwarzschild (when $\Lambda=0$) and the de Sitter metrics (when $m=0$). The Kottler metric admits
an interesting generalization (still satisfying the vacuum Einstein equations with cosmological
constant) whereby the spheres  $t=\mbox{const}$, $r=\mbox{const}$ are replaced  
by any compact connected manifold $M^2$ without boundary and the metric is
\begin{eqnarray}
ds^2= - \left ( k - \frac{2m}{r} - \frac{\Lambda r^2}{3} \right ) dt^2
+ \frac{dr^2}{ k - \frac{2m}{r} - \frac{\Lambda r^2}{3}} + r^2 d \Omega_k^2
\label{GeneralizedKottler}
\end{eqnarray}
where $k = {1,0,-1}$ depending, respectively, on whether the genus of $M^2$ is zero, one or higher.
Here $d \Omega_k$ stands for the two-dimensional metric of constant curvature $k$.
The Kottler metric obviously corresponds to the case $k=1$.
For $\Lambda <0$ and $m >0$, let $r_+$ be the only positive root of $k - 2m/r - \Lambda r^2 /3 =0$. 
The hypersurface $\{t=0\}$ of (\ref{GeneralizedKottler}) 
restricted to the range $r \geq r_{+} >0$ is a manifold with boundary which admits a smooth 
conformal compactification (see \cite{ChruscielSimon2001}, \cite{ChruscielHerzlich2003})
to a manifold with a boundary consisting of two
copies of $M^2$. One of the copies corresponds to the inner boundary $r=r_+$
and the other one to the ``surface at infinity'', usually denoted by $\partial^{\infty} \Sigma$. 

In the case $k=1$ the induced metric on $\{t=0\}$ is exactly 
(\ref{hyperbolic}) with $\Lambda = -3 \lambda^2$. 
In this context, the surface $S_1 = \{r = 2 m\}$ plays no special
role regarding the Penrose inequality (because the surface is outer untrapped) and instead one has to look for minimal
surfaces (similarly as in any time-symmetric case). Assuming again $m >0$, the outermost minimal surface of (\ref{hyperbolic}) 
is given by $S_2 = \{r = r_{+}\} $. Since $r_{+} < 2m$ we have that $S_1$ encloses $S_2$. 

It follows from this discussion that the  Penrose inequality in this  context
has two flavors, one for which the surface of interest has $p=2 |\lambda| $ and another
one where the surface to consider is minimal. The Riemannian manifolds $(\Sigma,\gamma)$ 
relevant for this setting 
satisfy $R(\gamma) > - 6 \lambda^2$,
which is a consequence of (\ref{HamiltonianConst}) both when $A_{ij} = \lambda \gamma_{ij}$ and $\rho \geq 0$ 
or when the spacetime satisfies $\mbox{Ein} ( g) = - \Lambda  g + 8 \pi T$ provided $\Lambda \equiv - 3 \lambda^2$
and the energy-momentum $T$
satisfies the week energy condition.

The metric $\gamma$ has to satisfy appropriate asymptotics so that a useful concept of mass can be defined.
Both the appropriate asymptotic behaviour and the definition of mass is considerably 
more difficult than the corresponding definition in the
asymptotically euclidean case. In the case when the boundary at infinity $\partial^{\infty} \Sigma$ is a two-sphere,
the definition was first given in \cite{Wang2001}. The definition for other topologies at infinity (and
admitting weaker asymptotic conditions) appears in \cite{ChruscielHerzlich2003} (see also 
\cite{ChruscielNagy2001} for a definition of mass from a spacetime viewpoint). For the $\{t=0\}$ slice of
(\ref{GeneralizedKottler}) the mass according to this definition is $m$. 

The statement of the Penrose inequality in this setting should be expected to be different 
when it involves outermost minimal surfaces or outermost surfaces with mean curvature $p = 2 |\lambda|$.
In the case of boundaries with several
connected components or when the topology of $\partial^{\infty} \Sigma$ is not 
spherical, it is not clear how the precise statement of the inequality should read.
However, when $\partial^{\infty} \Sigma= S^2$ and the inner boundary $S$ is connected,
a natural version of the inequality reads
\begin{eqnarray}
M \geq (1- g_{S} ) \sqrt{\frac{|S|}{16 \pi}} + \frac{\lambda^2}{2} (1- \epsilon) \left ( \frac{|S|}{4 \pi}
\right )^{\frac{3}{2}},
\label{PIhyperbolic}
\end{eqnarray}
where $M$ is the total mass, $g_{S}$ is the genus of $S$
and $\epsilon =0,1$ depending, respectively, on whether $S$ is outermost
minimal or outermost with mean curvature $p = 2 |\lambda|$. In the minimal case, this inequality has been proposed in 
\cite{Gibbons1999} 
for the case $g_S=0$ and in \cite{ChruscielSimon2001} for arbitrary genus (in this reference, 
the Penrose inequality in terms of the asymptotic value of the Geroch mass under smooth inverse mean curvature flow is 
also discussed for $\partial^{\infty} \Sigma$ of arbitrary genus). 
In \cite{Gibbons1999}, an inequality for $\partial^{\infty} \Sigma$
of arbitrary genus also appears. However, as noted in 
\cite{ChruscielSimon2001}, this version fails for the slice $\{t=0\}$ of the generalized Kottler metric when $M^2$ has at
least genus three. In the case 
of surfaces with $p = 2 |\lambda|$ the inequality (\ref{PIhyperbolic}) was conjectured in \cite{Wang2001}.
Support for the validity of (\ref{PIhyperbolic}) comes from the fact that the $\{t=0\}$ slice 
of the Kottler metric gives equality. Moreover, by choosing the
value $C = - 3 \lambda^2$, the Geroch mass 
(\ref{GerochMass}) evaluated on a surface $S$ with $p = 2 |\lambda| \epsilon$ gives precisely the 
right-hand side of (\ref{PIhyperbolic}). This value of the constant $C$ is adapted to the inequality $R(\gamma) 
\geq  -6 \lambda^2$ because then the Geroch mass is monotonic under smooth inverse mean curvature flow
\cite{BoucherGibbonsHorowitz1984, Gibbons1999}. This can be seen explicitly from the general formula (\ref{dMG}).

Thus, it is tempting to try and adapt Huisken and Ilmanen's proof to the hyperbolic case, at least when the
boundary $S$ is connected. However, a recent result by Neves \cite{Neves2007} shows that this is not possible
in general. The difficulty lies in the fact that the Geroch mass of a flow of surfaces
moved by inverse mean curvature does not necessarily approach  
the total mass of the asymptotically hyperbolic manifold. This is proven in \cite{Neves2007} by showing that
in the manifold $\{r \geq 2m\} \times S^2$ with metric 
(\ref{hyperbolic}) there exist surfaces 
with Geroch mass larger than $m$ and which can be flowed smoothly by inverse
mean curvature all the way to infinity. Consequently,
the limit of the Geroch mass under the flow is still larger than $m$ due to the monotonicity of $M_G$. 
In  this  example the inner boundary does not satisfy $p = 2 |\lambda|$. 
However, the author is able to modify the construction so that the 
flow starts on a horizon ($p=2 |\lambda| $) and remains smooth for all values of the
parameter in such a way that the leaves of the flow do {\it not} become
rounder (in a precise sense) at infinity. Thus, according to the author, it becomes impossible 
to compare the limit value of  Geroch mass with the value of the total mass of the manifold, which makes the 
inverse mean curvature flow inconclusive for the Penrose inequality.
Despite this failure of the inverse mean curvature method to prove the
inequality (\ref{PIhyperbolic}), its validity is still open. 

%The construction of this example is based
%on the inverse mean curvature flow in the metric (\ref{hyperbolic}) discussed above and proceeds first
%by adapting coordinates to the flow $\{S_u \}$, i.e. such that the metric is written as
%$dl^2 = \frac{du^2}{p^2(u)} + g_u$, where $S_{u} =  \{u = \mbox{const.}\}$ ($p(u)$ is then the mean curvature of
%this surface and $g_u$ its induced metric).  The next step follows a previous idea of
%Shi and Tam \cite{ShiTam2002}, and consists in rescaling $d u^2 $ factor in this metric so
%in such a way that the new metric has constant curvature $\hat{R} = -6 \lambda^2$ and the initial surface
%$\partial \Sigma = S_0$  has mean curvature $\hat{p} = -2 \lambda$. The surfaces $\{S_u\}$ still define an inverse mean
%curvature flow for the new metric. Since the induced geometry on each $S_u$ remains unchanged, this surfaces do not become
%round al infinity, since the original ones did not have this property (basically because their Geroch mass  in the original
%metric (\ref{hyperbolic}) did not tend to  $m$).

Using 
isoperimeric surface methods \cite{Corvino_et_al2007} (see 
Subsect. \ref{isoperimetricprofile}), the inequality (\ref{PIhyperbolic}) (with 
$\epsilon=0$) has been proven in the special case that $(\Sigma,\gamma)$ 
outside a compact set is isometric to the Kottler metric (\ref{hyperbolic}) outside a
sphere $r = \mbox{const}$ and the following two conditions are satisfied: (i) 
$(\Sigma,\gamma)$ contains a unique closed and connected surface $S_m$ with $p = 2 |\lambda|$ and (ii) 
the isoperimetric surfaces $S_V$ (with respect to $S_m$) are connected and coincide with the spheres
$r=\mbox{const.}$ in the asymptotic region for $V$ large enough.

It is also interesting to note that the Penrose inequality in the hyperbolic setting 
may be a powerful tool to address the uniqueness problem of asymptotically hyperbolic
static initial data sets \cite{ChruscielSimon2001}. More precisely, 
denoting by $U$
the square norm of the static Killing vector and by $M_G(U)$ the Geroch mass of the 
level sets of $U$, it is proven in \cite{ChruscielSimon2001} that the validity of the
Penrose inequality
\begin{eqnarray}
M_G (U) \geq (1- g_{S} ) \sqrt{\frac{|S|}{16 \pi}} +
\frac{\lambda^2}{2} \left ( \frac{|S|}{4 \pi}
\right )^{\frac{3}{2}} \label{uniq}
\end{eqnarray}
implies a uniqueness theorem for the 
generalized Kottler metric in the case that $\partial^{\infty} \Sigma$ if of genus larger than one and $S$
is connected. The reason is that in the static setting, the Geroch mass $M_{G}(U)$ can be bounded
{\it above} in terms of the mass of the generalized Kottler solution 
with the same surface gravity $\kappa$ (provided this satisfies the inequality $0 < \kappa \leq | \lambda|$).
The area $|S|$ is also bounded {\it below}  by the area of the Killing horizon of the corresponding Kottler
solution. Combining these inequalities with (\ref{uniq}), it follows that 
equality in (\ref{uniq}) is in fact the only possibility. However, 
the only static initial data set which saturates (\ref{uniq}) 
turns out to be the $\{t=0\}$ slice of the Kottler metric.
Thus a proof of the Penrose inequality for $M_G(U)$
would establish a uniqueness result for static initial data sets in the hyperbolic setting.

%The generalization to
%arbitrary dimensions $n\geq 3$ is
%\begin{eqnarray*}
%ds^2= - \left ( 1- \frac{2m}{r^{n-3}} - \frac{\Lambda r^2}{3} \right ) dt^2
%+ \frac{dr^2}{( 1- \frac{2m}{r^{n-3}} - \frac{\Lambda r^2}{3}} + r^2 d \Omega^2
%\end{eqnarray*}
%and was first considered by

\section{On the general Penrose inequality}
\label{GeneralPI}

The validity of the
Penrose inequality for arbitrary initial data sets (with an arbitrary second fundamental form
and without the assumption of spherical symmetry)
is still open. The proofs by Huisken and Ilmanen and Bray of the Riemannian Penrose
inequality involve manifolds $\Sigma$ with a positive definite metric of non-negative curvature scalar 
admitting an outermost minimal surface. Although primarily intended to 
cover the time-symmetric case, the proofs 
only require the presence of a minimal surface and a Riemannian metric with non-negative
scalar of curvature. Thus, the method also gives useful results in the case 
of  maximal hypersurfaces, $\tr_{\gamma} A=0$ (assuming  the energy density $\rho$
in (\ref{HamiltonianConst}) to be non-negative), provided
the Riemannian manifold contains at least a bounding minimal surface. The Riemannian arguments
in the previous section, however, do
not settle the general Penrose inequality 
in the maximal hypersurface case for two reasons: firstly because they would give a lower bound 
for the ADM energy instead of the ADM mass
and secondly because the outermost minimal surface does not coincide, in general, with
any of the minimal area enclosures arising in any of the  
versions of the Penrose inequalities discussed in Sect. \ref{Formulations} (except in the 
time-symmetric case, of course).

%In the time symmetric case, all the
%proposed versions of the Penrose inequality coincide among themselves and with 
%\ref{BrayPI}. 
Although the general Penrose inequality remains open, several methods have been proposed to address it. We devote this section to discuss them.

\subsection{Null shells of dust}

As described in the Introduction, 
Penrose's original setup \cite{Penrose1973}
to test the validity of cosmic censorship consisted of a shell of matter moving inwards
at the speed of light in a flat spacetime. The shell is assumed to have closed 
and connected cross sections and the matter within the shell is made of
null dust (meaning that the particles 
defining the shell are massless and that 
all pressures vanish). 
In order to have a flat metric inside the imploding shell, all points in their interior
must be causally disconnected (to the past) with all points on the shell. Choosing a
Minkowskian time $t$ inside the shell, this demands that, to the past of some $t=t_0$, the null
hypersurface $\N$ swept by the incoming shell of matter
must be free of self-intersections. Since the matter within the shell
is moving at the speed of light, the cross section
$S_t \equiv \N \cap \{ t = \mbox{const} \}$ can be viewed, after 
the natural identification of points in different instants of time, as the 
surface  lying  at distance $t_0-t$ from $S_{t_0}$, where distance is positive
to the exterior and negative to the interior (the fact that $S_{t_0}$ separates 
Euclidean space into an interior and an exterior is always true  
as a consequence of the 
Jordan-Brower separation theorem for smooth hypersurfaces in $\mathbb{R}^n$, see e.g.
\cite{Lima1988}).
The distance level function from a given closed surface
in Euclidean space is smooth everywhere in its exterior if and only if the surface is convex (i.e.
it has non-negative principal curvatures). The setup, therefore, requires that the null hypersurface
has one cross section $S_{t_0}$ which is convex. This property is then true for all $t \leq t_0$. 
Towards the future, 
$\N$ will become singular
at the first focal point of the incoming null geodesics. A spacetime singularity 
will form there.  Outside the null shell, the metric is no longer flat, in particular because
gravitational waves may be emitted by the collapsing dust.

Penrose devised this physical process as a potential counterexample to the inequality
(\ref{PI1}). The fundamental simplification of this problem is that the inequality can be
translated into an inequality directly in Minkowski space, as follows 
\cite{Penrose1973,Tod1985,Tod1992,Gibbons1997}:

Let $\vec{l}^{-}$ be the future directed null tangent 
to $\N$ normalized to satisfy $\vec{l}^{-} (t)=1$,  
where, as above, $t$ is a Minkowskian time in the interior part of the shell. Take any closed, spacelike surface
embedded in $\N$ and let $\vec{l}^{+}$ be its future null normal satisfying $(\vec{l}^{+} \cdot \vec{l}^{-})=-2$.
The energy momentum of the spacetime is a distribution supported on $\N$ which reads
$T_{\alpha\beta} =  8 \pi \mu l^{-}_{\alpha} l^{-}_{\beta} {\bm \delta}$, where $\mu$ is the energy
density of the shell and the Dirac $\bm{\delta}$ is defined with
respect to  the volume form $d\sigma$ induced by the normal $l^{-}_{\alpha}$ to $\N$, i.e.
$l^{-}_{\mu} d \sigma = \eta_{\mu\alpha\beta\gamma} e^{\alpha}_1 e^{\beta}_2 e^{\gamma}_3$ where
$e^{\alpha}_i = \frac{\partial x^{\alpha}}{\partial y^i}$ and $y^i \rightarrow x^{\alpha}(y^i)$ is a
coordinate expression for the embedding of $\N$ (see e.g. \cite{MarsSenovilla1993} for details). The 
null expansion $\theta_{+}$ jumps across $\N$. The jump can be determined using the Raychaudhuri
equation (\ref{lthetal}). One way
of doing this is by extending the null vector $\vec{l}_{+}$ to a geodesic null congruence and taking its divergence
on each side of the shell. This defines the null expansion $\theta_{+}$ as a discontinuous function
on the spacetime.  A distribution can then be introduced as
$\bm{\theta_{+}} = \theta_{+}^{E} \bm{\Theta} + \theta_{+}^{I} ( \bm{1} - \bm{\Theta})$, where
$\bm{\Theta}$ is th standard Heaviside distribution (it acts on tests functions by integration on the domain outside
the shell) and the superscript $I (E)$ stands
for interior (exterior) of the shell.
Since $\partial_{\mu} \bm{\Theta} = - l^{-}_{\mu} \bm{\delta}$, the derivative
of $\bm{\theta_{+}}$ along $\vec{l}_{+}$ gives 
\begin{eqnarray*}
l^{\mu}_{+} \partial_{\mu}\bm{\theta_{+}} = l^{\mu}_{+} \partial_{\mu}\theta_{+}^E \left (
\bm{1} - \bm{\Theta} \right )
+ l^{\mu}_{+} \partial_{\mu}\theta_{+}^I \bm{\Theta} + 2 \left [ \theta_{+} \right ] \bm{\delta}, 
\end{eqnarray*}
where the jump $[ \theta_{+}] \equiv (\theta^{E}_{+} - \theta^{I}_{+}) |_{\N}$. The Raychaudhuri equation (\ref{lthetal}), which 
in this case is a distributional equation, has a singular part supported on $\N$ only through the term
$ - \mbox{Ric} (\vec{l}_{+},  \vec{l}_{+} ) = - 32 \pi \mu \bm{\delta}$. The jump must therefore satisfy
$\theta^{E}_{+} |_{\cal N} = \theta^{I}_{+} |_{\cal N} - 16 \pi \mu$.

For an arbitrary surface embedded within a null hypersurface, the expansion along the null direction tangent
to the hypersurface
coincides with the null expansion of the hypersurface. This means, in particular, that 
it only depends on the point where it is calculated but not on the specific surface passing through that point. 
This has as immediate consequence that the incoming null
expansion $\theta_{-}$  of the shell is continuous
across the shell. On the surface $S_{t_0}$, the null expansion $\theta_{-}$ coincides with the mean curvature 
of $S_{t_0}$  as a surface
of Euclidean space with respect to the inner normal and it is therefore non-positive
(since $S_{t_0}$ is convex) and not everywhere zero (since
$S_{t_0}$ is closed). It follows from the Raychaudhuri
equation that $\theta_{-} \leq 0$ everywhere on $\N$. Consequently, 
right after the shell has passed, a spacelike surface $S \subset \N$ is
marginally outer trapped 
(i.e. $\theta^{E}_{+} =0$) if and only if it is  
marginally future trapped.

Assume such an $S$ exists along the shell. We want to show that the Penrose heuristic argument based on cosmic censorship
then implies 
$M_{B} \geq \sqrt{\frac{|S|}{16 \pi}}$ 
where $M_B$ is the Bondi mass (see e.g. \cite{Wald1984} for its definition)
of past null infinity at the cut defined 
by $\N$. Indeed, under cosmic censorship the singularity that necessarily forms in the future
is shielded from infinity by an event horizon. Since $S$ must be contained in the
black hole region, the intersection $\H$ of the event horizon with $\N$ must lie completely in the causal past of $S$. 
Using the fact that the null expansion $\theta_ {-}$ is non-positive, this implies $|\H| \geq |S|$. Since the
standard heuristic argument gives $M_{B} \geq \sqrt{\frac{|\H|}{16 \pi}}$, the claim above follows.
Notice that, in this case,
the inequality is expected to hold for $S$ irrespectively of whether 
or not this surface is area outer minimizing with respect
to any spacelike slice. Using now the conservation equation $\nablafour_{\alpha} T^{\alpha\beta}=0$ (which 
holds in distributional sense, see e.g. \cite{MarsSenovilla1993})), it follows that 
the integral $\int_{\hat{S}} \mu \bm{\eta_{\hat{S}}}$ does not depend on the cut $\hat{S}$ of $\N$.
Evaluating this integral at past null infinity 
gives precisely the Bondi mass (this can be easily shown for instance using the Hawking mass introduced in Subsect. \ref{UEF}).
Using now that $S$ is marginally outer trapped from the exterior, we have, after defining $\theta_{+}
\equiv \theta_+^I$,
\begin{eqnarray*}
\int_{S} \theta_{+} \bm{\eta_S} = 
\int_{S} \theta^{I}_{+} \bm{\eta_S} = 
\int_{S} 16 \pi \mu \bm{\eta_S} = 
16 \pi M_B \geq \sqrt{16 \pi S},
\end{eqnarray*}
where the last step is precisely the Penrose inequality in this
setting. Thus, the Penrose inequality for incoming shells can  be rewritten as 
\begin{eqnarray}
\int_{S} \theta_{+} \bm{\eta_S} \geq \sqrt{16 \pi |S|}, \label{Shells}
\end{eqnarray}
which  has the remarkable property of making no reference to the exterior geometry at all. Since the density 
$\mu$ is freely specifiable, this inequality
should hold for any closed spacelike surfaces $S$ in Minkowski spacetime for which 
the null hypersurface generated by past directed and outer null geodesics orthogonal to $S$ remains regular
everywhere. A similar inequality can be derived in any spacetime dimension \cite{Gibbons1997}.

In the particular case of a surface lying on the past null cone of a point $p$, the surface $S$ can be described
by a single positive function $r$ which measures the distance to $p$ (after projection to a constant time hypersurface).
A straightforward calculation gives the outer null expansion $\theta_{+}$ and  
the inequality (\ref{Shells}) becomes
\begin{eqnarray}
\int_{S^2} \left ( r + \frac{|d r |^2}{r} \right )  \bm{{\eta}_{S^2}} \geq
\sqrt{ 4 \pi \int_{S^2} r^2  \bm{{\eta}_{S^2}}},
\label{PIShellSpher}
\end{eqnarray}
where all geometric objects refer to the standard metric of unit radius on the sphere.
This inequality already appears in \cite{Penrose1973} and a more detailed derivation can be
found in \cite{BarrabesIsrael1991}. Its validity (in fact of a stronger version)
was proven by Tod \cite{Tod1985}
as a consequence of the Sobolev inequality applied to a 
suitable class of functions on $\mathbb{R}^4$ (see
\cite{Tod1986} for a different proof which gives an even stronger inequality).

\begin{figure}[h!]
\begin{center}
\psfrag{S}{$S$}
\psfrag{hatS}{$\hat{S}$}
\psfrag{P}{$p$}
\psfrag{T}{Hyperplane $t=0$}
\psfrag{sm}{$s(\theta)$}
\includegraphics[width=10cm]{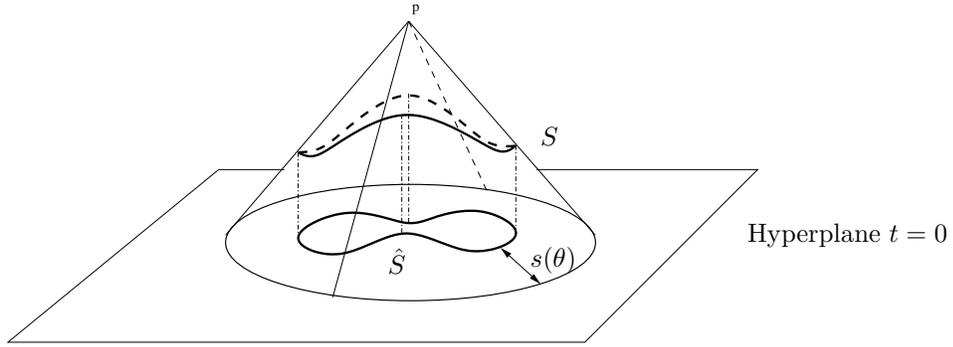}
\caption{Surface $\hat{S}$ in Euclidean space which is non-mean convex and which 
can be obtained by orthogonal projection of a surface $S$ lying in the past null cone of a point $p$
in Minkowski spacetime.}
\label{Projection}
\end{center}
\end{figure}

If  $S$ lies in a spacelike hyperplane in Minkowski, then $\theta_{+}$ is the mean curvature
$p$ of $S$ as a surface of Euclidean space and the inequality becomes
\begin{eqnarray}
\int_{S} p \bm{\eta_{S}} \geq \sqrt{16 \pi |S|}. \label{MinkowskiIneq}
\end{eqnarray}
As first noticed by Gibbons in his Ph.D. thesis, this inequality is exactly the Minkowski inequality for convex
bodies in Euclidean space, see e.g. \cite{BuragoZalgaller}. This settles the Penrose inequality when $S$
lies on a constant time hyperplane \cite{Gibbons1997}. The range of validity of the
Minkowski inequality  (\ref{MinkowskiIneq})  has been extended by Trudinger \cite{Trudinger1994},
to cover all mean convex bodies in Euclidean space, i.e. all closed surfaces 
with $p \geq 0$. Gibbons has used this result to claim the validity of the Penrose inequality
(\ref{Shells}) in the general case. The idea of the argument was to project $S$ orthogonally onto
a constant time hyperplane. The projection $\hat{S}$ can be seen to have at least the same area as $S$,
i.e. $|\hat{S}| \geq |S|$. Furthermore, by direct calculation, the author finds that 
the mean curvature $\hat{p}$ of the projected surface is non-negative and that 
$\int_{S}{\theta_{+}} \bm{\eta_{S}}  = \int_{\hat{S}} \hat{p} \bm{\eta_{\hat{S}}}$. Thus, the Penrose inequality would follow
from Trudinger's version of the Minkowski inequality. Unfortunately, the calculation leading
to $\hat{p} \geq 0$ and $\int_{S}{\theta_{+}} \bm{\eta_{S}}  = \int_{\hat{S}} \hat{p} \bm{\eta_{\hat{S}}}$ contains
an error which invalidates the argument. Instead of going into the details of the derivation, it is simpler
to just notice that one can easily construct a surface on a null hypersurface $\N$ in Minkowski spacetime
which has a projection $\hat{S}$ which is not mean convex. Consider the past null cone of a point at $t=1$, 
and consider the sphere obtained as the cross section of $\N$ with $\{t=0\}$. A function $s$ on $S^2$
taking values in $[0,1)$ defines a surface $\hat{S}$ on the hyperplane $\{t=0\}$ simply by moving
radially inwards each point of the sphere a distance $s$. It is easy to construct  surfaces $\hat{S}$ which are
not mean convex. Consider, for instance,
%The idea is to consider surfaces which have a wide but steep neck.
%More precisely, take 
a surface of revolution defined by $s(\theta)$ 
with equatorial symmetry 
and with a neck on the equator (i.e. such that $s(\theta)$ is symmetric under
$\theta \rightarrow \pi - \theta$ and $s_0= s(\theta = \pi/2)$ is a local maximum).
The principal curvatures on a point on the equator are simply 
$1/(1-s_0)$ and $1/(1-s_0) + s''_0/(1-s_0)^2$, where $s''_0$ is the second derivative of $s(\theta)$
at the equator. Thus, if $s''_0 < -2 (1-s_0)$ (i.e. the second derivative is
negative and sufficiently large in absolute value on the equator )
then the surface is not mean convex.  But $\hat{S}$ is obviously the orthogonal projection
on $t=0$ of the surface on $\N$ constructed by lifting each of its points a temporal amount $s$
 to the future, see Fig. \ref{Projection}. 
This example shows that the projection performed in \cite{Gibbons1997} is not correct. 
The validity of the Penrose inequality for null shells is therefore still open (in any spacetime
dimension).

Analogously as in (\ref{PIShellSpher}), the inequality (\ref{Shells}) can be rewritten in terms of the 
geometry of an arbitrary closed and convex surface $S_0$ in $\mathbb{R}^3$
and a function $s$ defined on $S_0$, as 
\begin{eqnarray}
\int_{S_0} \left ( P_0 - 2 s K_0 \right )
\left ( 1 + g_0^{AB} \partial_A s \partial_B s 
\frac{1 - 2 s P_0 + s^2 \left ( P_0^2 - K_0 \right )}{
\left ( 1- s P_0 + s^2 K_0 \right )^2 }
+ \kappa_0^{AB} \partial_A s \partial_B s 
\frac{s \left ( 2- s P_0 \right )}{ 
\left ( 1- s P_0 + s^2 K_0 \right )^2 } \right ) \bm{\eta_{S_0}} \geq 
\nonumber \\
\geq  \sqrt{ 16 \pi \int_{S_0} \left (1 - s P_0 + s^2 K_0 \right ) \bm{\eta_{S_0}}},
\hspace{5cm}
\label{PIshells_rewritten}
\end{eqnarray}
where $g_0^{AB}$ and $\kappa_{0}^{AB}$ are, respectively, the induced metric and second fundamental form of $S_0$
respect to the outer normal, $P_0$ and $K_0$ are the mean curvature and the Gauss curvature of $S_0$  or, 
in terms of the principal curvatures $\lambda_1,\lambda_2$ 
of the surface, $P_0 = \lambda_1 + \lambda_2$ and $K_0 = \lambda_1 \lambda_2$.
The smooth function $s$ must satisfy the inequalities
$s < 1/ \lambda_1$ and $s < 1 /\lambda_2$, but is otherwise arbitrary. It defines the surface $\hat{S}$ (and hence $S$)
in a similar way as before.
Details of the derivation of (\ref{PIshells_rewritten}) and some of its consequences will be given elsewhere
\cite{Mars2009}.

\subsection{Spinor techniques on null hypersurfaces: Ludvigsen-Vickers and Berg\-qvist approaches}

Witten's proof of the positive mass theorem is based on the properties of spinors satisfying a
suitable elliptic equation and which approach a constant spinor at spatial infinity. The same ideas
have been applied to prove positivity of the Bondi mass using asymptotically hyperbolic initial data sets (see
\cite{Chrusciel_et_al_2004} and references therein). Another possibility to
approach the positivity of the Bondi mass is to use null hypersurfaces. This allowed
Ludvigsen and Vickers \cite{LudvigsenVickers1982}
to replace the elliptic equations for the spinor by much simpler transport equations.
Similar ideas (with different transport equations) allowed the same authors \cite{LudvigsenVickers1983} 
to argue 
that, in spacetimes satisfying the dominant energy condition,
the inequality 
$M_B \geq \sqrt{|S|/16 \pi}$ holds for any weakly future trapped (in particular marginally future
trapped) surface $S$ which has the property that
one of the two null hypersurfaces generated by past directed null geodesics normal to $S$ can be extended
to past null infinity while remaining smooth everywhere. Although this is a global assumption on the 
spacetime, it makes no reference to the {\it future} evolution of the spacetime and hence it is logically
independent of cosmic censorship. For instance, this assumption is automatically satisfied for the incoming
null shell of dust discussed in the previous section. However,  Bergqvist \cite{Bergqvist1997} found 
a gap in the proof and the range of validity of the argument remains, at present, open. We discuss this next.

Bergqvist  reformulated  Ludvigsen and Vickers' argument so that all spinors could
be completely dispensed of. The idea
is, in some sense, analogous to the inverse mean curvature flow for the Geroch mass
and is based on using a
functional on spacelike closed surfaces $S_{\mu}$ constructed as follows: start with 
a closed spacelike surface $S$ and let $\vec{l}$ and $\vec{k}$ be {\it past} directed
null normals to $S$ partially fixed by the normalization $(\vec{l} \cdot \vec{k}) = -2$. Assume also
that $\vec{l}$ points outwards of $S$ in the sense that the null geodesics starting on $S$ 
with tangent vector $\vec{l}$ extend to infinite values of the affine parameter and intersect
$\scri^{-}$. Let $\mu$ be the affine parameter of this geodesic with $\mu = \mu_0$ on $S$ where $\mu_0$
is a constant to be chosen later and assume that the surfaces $\{\mu = \mbox{const}\}$ are smooth for
all $\mu \geq \mu_0$. The collection of all these surfaces defines a null hypersurface
$\N$ (the ``outer past null cone'' of $S$) which is assumed to intersect $\scri^{-}$ on a smooth cut.
The Bergqvist mass  is defined on each leave $\{S_{\mu}\}$ of this foliation as 
\begin{eqnarray*}
M_{b} (S_{\mu}) = \frac{1}{16 \pi}
\int_{S_{\mu}} \theta_k \bm{{\eta}_{S_{\mu}}} + 4 \pi \chi(S_{\mu}) \mu. 
\end{eqnarray*}
where $\theta_k$ is the null expansion along $\vec{k}$. 
Using the expressions (\ref{FirstVarArea}) and (\ref{lthetak2}), it is easy to obtain 
\begin{eqnarray*}
\vec{l} \left ( \int_{S_{\mu}} \theta_k \bm{\eta_{S_{\mu}}} \right ) =
\int_{S_{\mu}} \left ( \mbox{Ein}(\vec{l},\vec{k}) + 2 S_A S^A  \right ) \bm{\eta_{S_{\mu}}} - 4 \pi \chi(S_{\mu}),
\end{eqnarray*}
where the Gauss-Bonnet theorem $\int_S R(h) \bm{\eta_S} = 4 \pi \chi(S)$ has been used. We therefore get
\begin{eqnarray*}
\frac{ d M_b (S_{\mu})}{d \mu} = \frac{1}{16 \pi}  
\int_{S_{\mu}} \left ( \mbox{Ein} (\vec{l}, \vec{k})  + 2 S_A S^A \right )  \bm{\eta_{S_{\mu}}} \geq 0,
\end{eqnarray*}
provided the dominant energy condition is satisfied. Assuming that $\mu_0$
can be chosen so that asymptotically $\theta_{k} = -2/\mu + O(1/\mu^3)$, i.e. without a term in $\mu^{-2}$,
then the Bergqvist mass can be seen to approach the Bondi mass when $\mu \rightarrow \infty$.
Assuming now that the initial surface $S$ is a marginally trapped surface, if the value of $M_B$
could be somehow related to the area of $S$, a Penrose inequality
would follow. Bergqvist \cite{Bergqvist1997} shows that this can indeed be done but only under the
assumption that the induced metric $g^{S_{\mu}}_{AB}$ of the surface $S_{\mu}$ becomes round at infinity 
in the sense that $\lim_{\mu \rightarrow \infty} \mu^{-2} g^{S_{\mu}}_{AB} \rightarrow g^{S_2}_{AB}$, where
$g^{S_2}_{AB}$ is the standard unit metric on the sphere. At present, it is not clear how restrictive 
is this requirement.
The idea however remains interesting and deserves further investigation.

%
%A natural situation where the Ludvigsen-Vickers-Bergqvist
%argument can be tested is the imploding null shell in Minkowski discussed in the previous section. Investigating
%the limit behaviour of the surfaces $S_{\mu}$ in that case shows that 

\subsection{Uniformly expanding flows}

The success of proving the Riemannian Penrose inequality in the connected
horizon case using the monotonicity of the Geroch mass under the (weak) inverse mean curvature flow
suggests a possible scenario for approaching the Penrose inequality in the general case. As discussed
in Subsect. \ref{UEF}, the Hawking mass (\ref{HawMass}) is a functional on surfaces which coincides
with the Geroch mass in the time symmetric context. Moreover, the Hawking mass (with $\C=0$) takes the value
$\sqrt{|S|/(16\pi)}$ on any topological 2-sphere which is either a marginally outer trapped surface,
a past marginally outer trapped surface, or a generalized
apparent horizon (all of which have null mean curvature vector). Moreover, as discussed in Subsect. \ref{UEF},
under suitable spacetime variations of 
a given surface, the Hawking mass is monotonically increasing \cite{BrayHayward2007}.
Since its numerical value for large coordinate spheres
in the asymptotically euclidean region approaches the ADM energy, a flow which interpolates between the horizon and infinity
and falls into any of the four monotonicity cases discussed in Sect (\ref{UEF}) would imply the general Penrose
inequality between the ADM energy and the area of the horizon. Among the four cases, the closest one to the Riemannian inverse mean
curvature flow is the so-called ``uniformly expanding flow'', defined by (\ref{Flowvector}). 

With these monotonicity properties, the situation regarding the general Penrose inequality can be compared
to the status of the Riemannian Penrose inequality after Geroch's heuristic argument. It is conceivable that the
uniformly expanding flows might be useful for the proof of the general Penrose inequality. However, many issues
remain open. For instance, in a spacetime formulation, the jumps that occurred in the Riemannian setting will remain,
but it is unclear how and when the jumps should take place, even from a purely heuristic point of
view. Moreover, the inverse mean curvature flow is a parabolic equation in the Riemannian setting, so that
local existence is guaranteed, but the situation is quite different for the uniformly expanding flows.

Local existence in this case has been proven only for null flows, i.e. $|c| = \g$.
A null flow will obviously not reach spacelike infinity. Nevertheless, assuming that
a sufficiently large portion of the spacetime is at hand, this null flow could be used to study the Penrose inequality 
involving the Bondi mass. The difficulty, however, is that the uniformly expanding null flow seems to have the tendency
to deform the surface in such a way that the mean curvature vector becomes causal at some places, even if one starts with a 
surface with spacelike mean curvature everywhere. This behaviour is observed in explicit examples in Minkowski spacetime
(provided the surfaces do not lie on a constant time hyperplane and in fact
cover a sufficiently large time interval).
One alternative that may still give interesting results is to use
a flow which is null and future directed for some interval and then continue with a null {\it past directed}
uniformly expanding flow, then with a {\it future} directed null flow, and so on. This flow can be constructed in a tubular neighbourhood of the
given initial data set, and therefore does not need strong global assumptions on the spacetime. Moreover, it is conceivable
that this broken flow can approach spacelike infinity. However, it is not clear which criterion should be used
to stop the future flow and continue with the past one (and viceversa).
Moreover, the construction must be such that the mean curvature vector remains spacelike
everywhere
in order to ensure monotonicity of the Hawking mass.

Regarding the non-null case, $|c| < \g$,  the flow equations  form
a so-called forward-backward parabolic system, for which no local existence theory is known (this was noted by
Huisken and Ilmanen \cite{HuiskenIlmanen2001} in the case $c=0$ and was extended to arbitrary $c$ in \cite{BrayHayward2007}).
This is, of course, a major difficulty and addressing it would require a much better understanding of this type
of partial differential equations. In Huisken and Ilmanen's work, a fundamental part of the analysis dealt with the
level set formulation, which gives a degenerate elliptic equation. More specifically, the existence of a variational
formulation turned out to be a fundamental ingredient to solve the problem of existence and to study the jumps. Remarkably, the
uniformly expanding flows also admit a variational formulation \cite{BrayHayward2007}. The new basic ingredient
is that the field to be varied is not just the level set function $v$ (see (\ref{funct1})) as
in the Riemannian case, but also the
spacelike hypersurface $\Sigma$ where this function is defined. Whether this variational formulation      
can give useful hints on how to define the jumps remains an open and difficult problem. 

It should be remarked in this context that the simplest spacelike uniformly expanding flow corresponds
to $c=0$, i.e. $\vec{\xi} = \frac{1}{(\vec{H} \cdot \vec{H})} \vec{H}$ ($\g=1$ can be chosen without loss of generality). The flow vector is
therefore the inverse mean curvature vector.
In terms of initial data information $(\Sigma,\gamma_{ij},
A_{ij})$ this corresponds to the case $q=0$ (this condition has been often termed ``polar gauge'' in the literature).
Assuming this gauge condition and a second restriction which turns out to coincide with the inverse
mean curvature flow condition $p e^{\psi} = \mbox{const.}$ Jezierski was able to prove \cite{Jezierski1994.1,Jezierski1994.2}
the Penrose
inequality for small (but non-linear) electrovacuum perturbations of the Reisner-Nordstr\"om time-symmetric initial data 
outside the black hole event horizon. His method was based on writing the Hamiltonian
constraint as a divergence term plus a non-positive reminder. The Penrose inequality is
established by integrating this equation between the event horizon and infinity
after using the Gauss theorem to transform the divergence into a surface integral at infinity and on the horizon (the
former gives the ADM energy and the latter the area term in the inequality).
Although Jezierski's argument does not use the monotonicity of the Hawking mass, the calculation does in fact
correspond to the general identity (\ref{dMH2}) specialized to the case at hand. With hindsight, one can therefore
conclude that this monotonicity property of the Hawking mass is the 
underlying reason why the argument works. Jezierski supports the plausibility of the gauge conditions $q=0$ and $p e^{\psi} = 1$
by studying linear perturbations of Reissner-Nordstr\"om, where he finds that the two equations decouple, one giving a
parabolic equation that needs to be integrated radially outwards and another one
also parabolic but which needs
to be integrated radially inwards. This is, of course, a manifestation of the forward-backward parabolic nature of the full  
system. A similar existence result for linear, axial (i.e. odd) perturbations of maximal slices of the Schwarzschild spacetime
has been obtained in \cite{RoszkowskiMalec2005}.

Another observation worth mentioning regarding the inverse mean curvature vector flow
(i.e. $c=0$) is that
the monotonicity formula turns out to be  insensitive to the value of the 
energy flux $\vec{J}$. If follows that the weak energy condition is already sufficient to ensure monotonicity
of the Hawking mass in this context. In principle this opens up the possibility that the full Penrose
inequality might be true for spacetimes satisfying just the weak energy condition. This is not so, however.
An explicit counterexample for scalar field initial data has been constructed by V. Husain \cite{Husain1999}.
%The existence of this counterexample implies that the inverse mean curvature flow {\it vector} ($c=0$) cannot be capable
%of proving the general Penrose inequality. This example implies nothing for uniformly
%expanding flows with $c\neq 0$ because then the monotonicity formula does require the dominant
%energy condition.

\subsection{Jang equation}
\label{Jang1}

Schoen and Yau's proof of the positive mass theorem proceeded in two steps. First, the purely Riemannian case
(i.e. vanishing second fundamental form) was solved by using minimal surface techniques. In a second step,
the general case $(\Sigma,\gamma_{ij},A_{ij})$ was treated by modifying the metric $\gamma_{ij}$ with a transformation
introduced by Jang \cite{Jang1978}, namely 
\begin{eqnarray}
\hat{\gamma}_{ij} = \gamma_{ij} + \partial_i f \partial_j f,
\label{JangTrans}
\end{eqnarray}
where $f$ solves the so-called {\it Jang equation}
\begin{eqnarray}
\label{JangEq}
\left ( \gamma^{ij} - \frac{\nabla^i f \nabla^j f}{
1+ |df|^2_{\gamma}} \right ) \left (
\frac{\nabla_{i} \nabla_j f }{\sqrt{ 1 + |df|^2_{\gamma}}} - A_{ij} \right )=0.
\end{eqnarray}
This transformation has the property that
the curvature scalar of $\hat{\gamma}$ has a lower bound that allows to prove existence of a conformal
factor $\Omega >0$ such that the conformally rescaled metric $\Omega^2 \hat{\gamma}_{ij}$ has
vanishing scalar curvature. This metric is still asymptotically euclidean and has at most the same mass as
the original metric.  The Riemannian positive mass theorem then gives the desired result. 
The proof is however involved because the Jang equation does not
admit regular solutions when the initial data set $(\Sigma,\gamma_{ij},A_{ij})$ contains marginally trapped surfaces,
but the idea can nevertheless be made to work \cite{SchoenYau1981}.

A natural question, already asked in \cite{Bray2001}, is whether a similar idea can be applied to prove
the general Penrose inequality. The Penrose inequality has already been proven in full generality in the
Riemannian setting so the present status is similar to the situation of the proof of the positive mass theorem after its
Riemannian proof. This idea has been analyzed by Malec and \'O Murchadha 
\cite{Malec-Murchadha2004}. Their argument is based on the observation that, 
if the Jang equation could be used to prove the general Penrose inequality, it should be able to do
so in the particular case of spherical symmetry. Restricting to spherically symmetric functions $f$,
the Jang equation becomes a simple ODE and existence of regular solutions in the exterior region outside
the full trapped region $\T^{+}_{\Sigma} \cup \T^{-}_{\Sigma}$ 
%(i.e. the connected component 
%containing the asymptotic end 
%of the region foliated by spheres satisfying $p > |q|$
can be easily shown.
The difficulty of the method is that along the
process (Jang's transformation and subsequent conformal rescaling) the mass of the manifold should not
increase and the area
of the outermost horizon $S = \partial (\T^{+}_{\Sigma} \cup \T^{-}_{\Sigma} )$ should not decrease.
This is because one wants to use the Riemannian Penrose inequality for the final manifold and 
conclude that the same inequality holds in the original initial data. It is also clear that $S$ must be transformed
into a minimal surface after the conformal rescaling. The simplest situation where this can be achieved is by 
demanding that $S$ becomes minimal already for the Jang transformed metric $\hat{\gamma}_{ij}$. This requires that
the outer normal derivative of $f$ diverges to either $+\infty$ or to $-\infty$ on $S$. In the first case, it follows
that the metric $\hat{\gamma}$ has a cylindrical end near $S$. The conformal transformation is expected to compactify
this end with one point (this behaviour was found in \cite{SchoenYau1981}, and this was important in order
to apply the Riemannian positive mass theorem). Hence, the area of $S$ decreases in the process (it vanishes in the final
manifold) and nothing can be concluded. In the second case (outer normal derivative of $f$ diverging to 
$- \infty$) the conformal factor $\Omega$ 
%with boundary data $\Omega |_S = 1$ and $\Omega \rightarrow 1$ at infinity
is expected to have an interior local minimum. This would imply that the outer normal derivative of $\Omega$
 is negative on $S$, and hence that this surface is not the outermost minimal surface in the final manifold.
Thus, the Riemannian Penrose
inequality applied to the final manifold is again inconclusive for the original one.
Although these arguments are not definitive
in discarding the Jang equation method for the Penrose inequality, they indeed show that difficulties should be 
expected for the method to work, at least when $S$ is required to transform to a minimal surface by the 
Jang transformation. The situation where $S$ is minimal only after the final conformal transformation is not considered
in \cite{Malec-Murchadha2004}  and should be further investigated.

Very recently, the Jang equation has been successfully applied to prove a Penrose-like inequality in the spirit
of Herzlich's inequality discussed in Subsect. \ref{spinors}, but allowing non-vanishing
second fundamental form. The class of initial data 
$(\Sigma,\gamma_{ij},A_{ij})$ under consideration are such that  $\Sigma = K \cup \Sigma^{\infty}$ with $K$ 
compact and $\Sigma^{\infty}$ an asymptotically euclidean end. The boundary $\partial \Sigma$ is compact and
consisting of a finite collection of future MOTS (with respect to the normal pointing towards $\Sigma$).
Moreover, no weakly future or past trapped surface strictly enclosing $\partial \Sigma$ is allowed
to exist in $\Sigma$. In other words, $\partial \Sigma$ is the outermost MOTS in $\Sigma$
and, moreover, no past weakly outer trapped boundary is allowed to exist in $\Sigma$ except possibly $\partial \Sigma$ itself.
Assuming also 
that $(\Sigma,\gamma_{ij},A_{ij})$ satisfies the dominant energy condition $\rho \geq |\vec{J}|$, Khuri
has recently proven \cite{Khuri2009} that the Penrose-like inequality
\begin{eqnarray}
E_{ADM} \geq \frac{\hat{\sigma}}{2 ( 1 + \hat{\sigma})} \sum_{a=1}^{k} \sqrt{\frac{|\partial_a \Sigma|}{\pi}}
\label{KhuriIneq}
\end{eqnarray}
holds, where $k$ is the number of connected components of $\partial \Sigma$ and
the scale invariant quantity $\hat{\sigma}$ is defined as
\begin{eqnarray*}
\hat{\sigma} = \frac{1}{\sum_{a=1}^{k} \sqrt{4 \pi |\partial_a \Sigma|}} \inf_{v \in C^{\infty}} 
\int_{\Sigma} \left (dv, dv \right )_{\ggamma} \bm{\eta_{\ggamma}}. 
\end{eqnarray*}
The infimum is taken with respect to functions $v$ with approach one at infinity and zero on $\partial \Sigma$.
The metric $\ggamma$ is constructed using the Jang transformation (\ref{JangTrans})  and $f$ is a solution
of the Jang equation approaching zero at infinity and blowing up to $+ \infty$ on $\partial \Sigma$. The existence of 
such an $f$ has been established by Metzger in \cite{Metzger2008}. The asymptotic behaviour of $f$ at infinity is such that 
$(\Sigma,\ggamma)$ is asymptotically euclidean and $E_{ADM}(\gamma) = E_{ADM} (\ggamma)$.
Since $f \rightarrow \infty$ on $\partial \Sigma$, the level sets $S_T \equiv \{f =T\}$, for $T$ large enough, 
form a foliation near $\partial \Sigma$ which converge to $\partial \Sigma$. The idea is to conformally transform
$\ggamma$ outside $S_T$ in such a way that the conformally rescaled metric $\gamma^T_{ij} = u_T^4 \ggamma_{ij}$ 
is still asymptotically euclidean, with vanishing
curvature scalar and such that each connected component $S_{T,a}$ of $S_T$
satisfies $p^T_a = 4 \sqrt{\pi/|S_{T,a}|_{\gamma^T}}$, where $p^T_a$ is the mean curvature of $S_{T,a}$
with respect to $\gamma^T_{ij}$ (and the area is also calculated with this metric). Existence
of $u_T$ is established as a consequence of the positivity properties of $R(\ggamma)$  and the
blowing up behavior of $f$ at $\partial \Sigma$.
The conformal rescaling is such that the ADM energy decreases by an amount which is at least
equal to the right-hand side of (\ref{KhuriIneq}) except for terms that vanish in the limit $T\rightarrow \infty$.
The only remaining step to conclude (\ref{KhuriIneq})  is that the ADM energy of 
$\gamma^T$ is non-negative. But this is precisely the content of the positive mass theorem
proven by Herzlich \cite{Herzlich1997} for asymptotically euclidean manifolds with a connected boundary of spherical
topology and satisfying (\ref{InequalityMeanCurvature}). Khuri notices that this positive mass theorem still holds
if the boundary has a finite number of connected components of spherical topology, each one satisfying the bound 
(\ref{InequalityMeanCurvature}). In the case at hand, this inequality is satisfied (in fact, saturated) by construction and 
the spherical topology is guaranteed by Galloway and Schoen's results
\cite{SchoenGalloway2005, Galloway2007} on the topology of outermost MOTS.

Comparing this Penrose-like inequality with the difficulties
described by Malec and \'O Murchadha to use the Jang transformation to prove the
general Penrose inequality in spherical symmetry, the main difference is that this method ultimately relies
on a positive mass theorem, instead of on the Riemannian Penrose inequality. Thus, there is no need to obtain
an outermost minimal boundary after the metric is modified by the Jang transformation and the subsequent 
conformal rescaling.

\subsection{Bray and Khuri approach}

Very recently Bray and Khuri have made an important step forward towards establishing the 
general Penrose inequality. As mentioned in Sect. \ref{Formulations}, these authors propose to use generalized
apparent horizons as the appropriate surfaces for which the Penrose inequality should hold.
An important guiding principle that led Bray and Khuri to make this conjecture is related to the fact that,
independently of which method for proving the inequality is used, 
the estimates involved must all
give equality whenever $(\Sigma,\gamma_{ij},A_{ij})$ is any of the slices of
the Kruskal spacetime for which equality holds. Therefore, a preliminary issue of importance is: for which slices of the Kruskal spacetime should equality be expected? Any spacelike Cauchy slice
$\Sigma$ of the Kruskal spacetime must intersect both 
the black hole and the white
hole event horizons. In Kruskal coordinates, the metric reads (c.f. 
\cite{Wald1984})
\begin{eqnarray*}
ds^2 = \frac{32 m^3}{r} e^{-\frac{r}{2m}} dudv + r^2 \left ( d\theta^2 + \sin^2 \theta d \phi^2 \right ),
\end{eqnarray*}
where $uv > -1$ and $r(uv)$ is defined by $uv = e^{\frac{r}{2m}} ( \frac{r}{2m} -1 )$ (our sign
convention for $u$ is different to that in \cite{Wald1984}). The null vectors
$\partial_v$ and $-\partial_u$ are future directed, the black hole event horizon is located at $u=0$, the white
hole event horizon at $v=0$ and the domain of outer communications $\U$ is located at $\{ u >0, v>0 \}$. If the 
boundary of $\Sigma^{\mbox{\tiny{DOC}}} \equiv \Sigma \cap \U$ satisfies $u=0$ everywhere, then 
$\partial \Sigma^{\mbox{\tiny{DOC}}}$ is an area outer minimizing MOTS. Moreover,  since its area is $16 \pi m^2$
and there are no other weakly outer
trapped surfaces (future or past) in $\Sigma^{\mbox{\tiny{DOC}}}$, it follows that
this slice satisfies any of the versions of the Penrose inequality (\ref{PIHeuT+}), (\ref{PIHeuT-})
or (\ref{PIT}). It also satisfies the inequality involving generalized apparent horizons
(\ref{PIKhuriBray}) provided there are no generalized trapped surfaces in $\Sigma^{\mbox{\tiny{DOC}}}$
(this is plausible although not yet proven, as far as I know). Similar things happen if $v=0$ everywhere
on the boundary of $\Sigma^{\mbox{\tiny{DOC}}}$. However, it neither $u$ and $v$ are
identically zero on $\partial
\Sigma^{\mbox{\tiny{DOC}}}$, the situation is quite different. In this case the boundary 
$\partial \Sigma^{\mbox{\tiny{DOC}}}$ is neither a future
or past MOTS and, in most cases, it is not even smooth. Moreover, the intersection of $\Sigma$ with the $\{u=0\}$
hypersurface (which is always a MOTS and has area $16 \pi m^2$) is not
area outer minimizing because its mean curvature $p$ is negative whenever $v < 0 $ (i.e.
on the points lying in the white hole event horizon with respect to the second asymptotically flat 
spacetime region). Thus, its minimal area enclosure has strictly less
area. Consequently, the version (\ref{PIHeuT+}) of the Penrose inequality holds but
{\it not} with equality (this statement assumes
that $\{u=0 \} \cap \Sigma$ is the outermost MOTS of the slice, which is again plausible
but not yet proven, as far as I know). It follows
that not all slices of the Kruskal spacetime are automatically equality cases, at least
for the version (\ref{PIHeuT+}). Obviously, the more slices of Kruskal satisfy equality, the sharper is the version of the Penrose inequality, in the sense of being capable of identifying the
Kruskal spacetime in a larger number of  cases. 
A version that gives equality for any slice of the Kruskal
spacetime is (\ref{PIT}), even when the boundary of $\Sigma^{\mbox{\tiny{DOC}}}$ is non-smooth. Although, as
already mentioned, no counterexample of this version has been found, its validity would however come as
a surprise because the minimal area enclosure of
$\partial (\T^{+}_{\Sigma} \cup \T^{-}_{\Sigma})$ may 
a priori have much smaller area than itself.
The alternative put forward by Bray and Khuri involves
generalized trapped surfaces.
This has the advantage that, as soon as $\Sigma^{\mbox{\tiny{DOC}}}$
has a smooth boundary,
this is
a generalized trapped surface (in fact, a generalized apparent horizon). 
Eichmair's result, see Subsect. \ref{embedded}, implies that
an outermost generalized apparent horizon must exist on $\Sigma$.
It is highly plausible that
$\partial \Sigma^{\mbox{\tiny{DOC}}}$ is in fact the outermost apparent horizon in this case. 
Since its area is $16 \pi m^2$, any slice with smooth
$\partial \Sigma^{\mbox{\tiny{DOC}}}$ would 
belong to the equality case of the Penrose inequality (\ref{PIKhuriBray}). 
If the boundary 
$\partial \Sigma^{\mbox{\tiny{DOC}}}$ is not smooth, then it cannot be its own
minimal area enclosure, and hence it cannot give equality neither in
(\ref{PIHeuT+}) nor in (\ref{PIHeuT-}).

These considerations led Bray and Khuri
to conjecture the following version of the Penrose inequality
(recall that our definition of asymptotically euclidean requires in 
particular that $(\Sigma,\gamma_{ij})$ is complete and that
an initial data set is called {\bf Schwarzschild at infinity} if outside a compact set, the metric 
$\gamma$ is exactly Schwarzchild, c.f. the discussion after (\ref{epsilonchange})).

\begin{conjecture}[Bray and Khuri \cite{BrayKhuri2009}]
\label{BrayKhuriConjecture}
Suppose that the initial data set $(\Sigma,\gamma_{ij},A_{ij})$
is asymptotically euclidean and Schwarzschild at
infinity, with total mass $M_{ADM}$ (in a chosen end) and satisfying the dominant energy condition
$\rho \geq |\vec{J}|$. 
Let $S$ be a closed surface which bounds an open set $\Omega$
containing all the asymptotically euclidean ends
except the chosen one. Assume that $S$  is a generalized trapped surface 
(with respect to the normal
pointing towards the chosen end). Then
\begin{eqnarray*}
M_{ADM} \geq \sqrt{\frac{|S_{\min}|}{16 \pi}},
\end{eqnarray*}
where $S_{\min} = \partial \Omega_{\min}$ is the minimal area enclosure of $S$
(i.e. $\Omega \subset \Omega_{\min}$ and $S_{\min}$ has least area among all surfaces with
this property). Furthermore, equality occurs if
and only if $(\Sigma \setminus \Omega_{\min}, \gamma_{ij},
A_{ij})$ is the induced data of an embedding of $\Sigma \setminus \Omega_{\min}$
into the Kruskal spacetime such that $S_{\min}$ is mapped to a generalized apparent
horizon.
\end{conjecture}
The use of generalized apparent horizons is indeed a completely new idea for the
Penrose inequality. This version is not supported by Penrose's heuristic argument of gravitational collapse
because it is not true that all generalized apparent horizons lie inside the 
event horizon in a black hole spacetime. In fact, little is known in general about this type
of surfaces in general spacetimes. An important question regarding the plausibility of Conjecture
\ref{BrayKhuriConjecture} was posed by R. Wald \cite{Wald2008} who asked whether
there are any generalized apparent horizons in Minkowski spacetime (the question of whether
surfaces with causal mean curvature can exist in Minkowski spacetime was already asked in
\cite{MarsSenovilla2003} in a somewhat different context). The existence of any generalized apparent
horizon embedded in a spacelike Cauchy slice of Minkowski and bounding a compact set would immediately falsify
Conjecture \ref{BrayKhuriConjecture}. Khuri has been able to prove \cite{Khuri2009-2}
that no such surfaces are present
in Minkowski spacetime.
In fact he proves much more by showing that any asymptotically euclidean initial data set satisfying the
dominant energy condition and with compact (non-empty)
boundary consisting of finitely many generalized trapped surfaces 
satisfies a strict positive mass theorem
$E_{ADM} > |\vec{P}_{ADM}|$. The proof is based on Witten's spinorial method for the positive mass.

The strategy that Bray and Khuri propose to address Conjecture \ref{BrayKhuriConjecture}
is related to
the Shoen and Yau's reduction of the general positive mass theorem
to the time-symmetric case. Recall that this was based on 
the Jang transformation  (\ref{JangTrans}) of the metric, where the function $f$
solves the Jang equation. This equation is specifically
tailored so that it always admits a solution for any initial data set
in Minkowski spacetime (the solution is the height function of the slice
over any constant time hyperplane). This is relevant because slices
of Minkowski immediately give equality in the positive mass theorem.
A fundamental observation of Bray and Khuri is that, since
the equality case of the Penrose inequality should correspond to the Schwarzschild metric,
the Jang equation
should be accordingly modified so that it holds identically for slices in this spacetime.
More generally, Bray and Khuri consider static spacetimes
$(\M,g) = ( \mathbb{R} \times \Sigma, ds^2 = - \phi^2 dt^2 + \ggamma)$,
where $\phi >0$ and  $\ggamma$ is a Riemannian metric. The idea
is to consider spacelike graphs $(t=f(x),x)$ for $x \in \Sigma$ and derive
an equation for the graph function which 
holds identically in this case and which still makes sense
for an arbitrary initial data set $(\Sigma,\gamma_{ij},A_{ij})$.

%For the Schwarzschild metric 
%\begin{eqnarray*}
%\phi_{\mbox{\tiny Sch}} = 1 - \frac{2m}{r \left ( 1 + \frac{m}{2r} \right )^2}, \quad r > \frac{m}{2}
%\end{eqnarray*}
%and $\ggamma$ is given by (\ref{gsch}) for $r > m/2$. 

The induced metric on the graph is 
\begin{eqnarray*}
\gamma_{ij} = \ggamma_{ij} - \phi^2 \partial_i f \partial_j f,
\end{eqnarray*}
which is Riemannian provided the gradient of $f$ is not too large with respect
to $\ggamma$,
namely if  $\phi |df|_{\ggamma} < 1$. However, this expression can also be used
to {\it define} $\ggamma_{ij}$
given an initial data set $(\Sigma,\gamma_{ij}, A_{ij})$ and two arbitrary functions $\phi$ and $f$.  This implies that the
Riemannian metric $\gamma$ of any initial data set can be obtained
as the induced metric of a hypersurface in a
suitably constructed static spacetime. With this point of view, the function
$f$ becomes arbitrary (with no restriction on its gradient)
due to the identity
$(1- \phi^2 |df|^2_{\ggamma} ) ( 1 + \phi^2 |df |^2_{\gamma} ) = 1$. 
The inverse metric is 
$\ggamma^{ij} = \gamma^{ij} - v^i v^j$ with
\begin{eqnarray}
v^i \equiv  \frac{\phi \nabla^i f}{\sqrt{1 + \phi^2 |df|^2_{\gamma}}}
\label{vecv}
\end{eqnarray}
where indices are lowered and raised with the metric $\gamma_{ij}$ and its inverse.
The second fundamental form on the graph is \cite{BrayKhuri2009}
\begin{eqnarray}
\hh_{ij} = \frac{\phi \nabla_i \nabla_j f + \nabla_i \phi \nabla_j f + 
\nabla_j \phi \nabla_j f}{\sqrt{ 1 + \phi^2 |df|^2_{\gamma}}}.
\label{NewSecForm}
\end{eqnarray}
Of course, this second fundamental form has nothing to do a priori with the second fundamental form $A_{ij}$ of the given initial data set. Nevertheless,
$(\Sigma,\gamma_{ij},\hh_{ij})$ is the induced geometry
of a spacelike hypersurface in a static spacetime. The
existence of the isometry generated by
$\partial_t$ can be used to relate the geometry of this slice with the geometry of the
corresponding $\{t=0\}$ slice, 
i.e. of $(\Sigma,\ggamma)$. After an involved calculation,
this observation leads to the following remarkable identity
for the curvature scalar $R (\ggamma)$ \cite{BrayKhuri2009},
called {\it generalized Schoen-Yau identity}, 
\begin{eqnarray}
R (\ggamma ) & = &  16 \pi \left ( \rho - J_i v^i \right ) + || \hh - A ||^2_{\ggamma} 
+ 2 |\vec{\zz}\, |^2_{\ggamma} - \frac{2}{\phi} \mbox{div}_{\ggamma} (\phi \vec{\zz} \,) \nonumber \\
& + &  \tr_{\ggamma} (\hh - A)  
\left [ \tr_{\ggamma} (\hh + A ) + 2 A_{ij} v^i v^j \right ] 
+ 2 v^i \partial_{i} \left ( \tr_{\ggamma} (\hh - A) \right ), 
\label{Generalized}
\end{eqnarray}
where $\zz_i = ( A_{ij} - \hh_{ij} ) v^j$ and $\vec{\zz}$ is obtained
after raising its index with the inverse of the metric $\ggamma_{ij}$.
In the case $\phi=1$, this identity was obtained by Schoen and Yau \cite{SchoenYau1981} and was used to
show that any solution $f$ of the Jang equation  defines
a metric $\ggamma_{ij}$ which admits a conformal rescaling with non-negative curvature scalar.

Bray and Khuri's approach aims at finding appropriate functions $\phi$  and $f$ so that the
Penrose inequality for $(\Sigma,\gamma_{ij},A_{ij})$ follows as a consequence
of the  Riemannian Penrose inequality applied to the transformed data $(\Sigma,\ggamma)$.
In order to use
the Riemannian Penrose inequality, it is necessary that $R(\ggamma) \geq 0$.
The dominant energy condition $\rho \geq |\vec{J}|_{\gamma}$ together with the fact that
the vector $\vec{v}$ satisfies $|\vec{v} |_{\gamma} <1$ (see (\ref{vecv}))
implies that the first three
terms in (\ref{Generalized}) are non-negative.  The remaining terms have no sign in general.
Motivated by the structure of 
(\ref{Generalized}), Bray and Khuri
introduce the {\it generalized  Jang equation} $\tr_{\ggamma} ( \hh - A)=0$ or, explicitly,
\begin{eqnarray}
\label{GenJangEq}
\left ( \gamma^{ij} - \frac{\phi^2 \nabla^i f \nabla^j f}{
1+ \phi^2 |df|^2_{\gamma}} \right ) \left (
\frac{\phi \nabla_{i} \nabla_j f + \nabla_{i} \phi \nabla_j f +
\nabla_j \phi \nabla_i f }{\sqrt{ 1 + \phi^2 |df|^2_{\gamma}}} - A_{ij} \right )=0.
\end{eqnarray}
This obviously reduces to the original Jang equation (\ref{JangEq}) when $\phi=1$. In the present
case, however, this equation involves  two unknowns, $\phi$ and $f$. 
The Jang equation is known not to admit regular solutions 
when $\Sigma$ contains a future or past MOTS. In a similar fashion, Bray and Khuri observe
that the generalized Jang equation may blow up on surfaces satisfying $|p|  = |q|$, which
are generalized trapped surfaces.
This, combined with the existence of an outermost 
generalized apparent horizon on $(\Sigma,\gamma_{ij},A_{ij})$, led to authors to study the conjecture
(\ref{BrayKhuriConjecture})  in the particular case that $\partial \Sigma$ is a generalized apparent
horizon and that no further generalized trapped surfaces exist in $\Sigma$.
The conjecture in 
this setting is
\begin{conjecture}[Bray and Khuri \cite{BrayKhuri2009}]
\label{BrayKhuriConjecture-2}
Suppose that the initial data set $(\tilde{\Sigma},\gamma_{ij},A_{ij})$ is asymptotically euclidean 
and Schwarzschild at
infinity, with total mass $M_{ADM}$ (in a chosen end) and satisfying the dominant energy condition $\rho \geq
| \vec{J}|$. 
Let $S$ be a closed surface which bounds 
an open set $\Sigma^{\mbox{\tiny{int}}}$
containing all the asymptotically euclidean ends
except the chosen one and that 
$S$ is  an outermost generalized trapped surface with respect to the normal
pointing towards the chosen end, i.e. that $\Sigma \equiv \tilde{\Sigma} \setminus
\overline{\Sigma^{\mbox{\tiny{int}}}}$ contains no generalized trapped surfaces.
Then 
\begin{eqnarray}
M_{ADM} \geq \sqrt{\frac{|\partial \Sigma|}{16 \pi}}.
\label{Ineq}
\end{eqnarray}
Furthermore, equality occurs if
and only if $(\Sigma, \gamma_{ij}, A_{ij})$ is the induced data of an embedding
of $\Sigma$
into the Kruskal spacetime such that $\partial \Sigma$ is mapped to a generalized apparent horizon.
\end{conjecture}
Under the conditions of this conjecture, 
it is plausible
that (\ref{GenJangEq}) admits regular solutions in  $\Sigma$. The boundary
behaviour is typically singular, as examples in the Kruskal spacetime show. The hope is that this singular
behaviour on the boundary can be adjusted so that the surface $\partial \Sigma$
has non-positive mean curvature with respect to the transformed metric $\ggamma_{ij}$. 
This is useful because the area of any surface never decreases under 
the transformation $\gamma_{ij} \rightarrow \ggamma_{ij}$  (due to the fact that
the volume form of $\ggamma_{ij}$ is $\bm{\eta_{\ggamma}} = ( 1 + \phi^2 |df|^2_{\gamma} ) \bm{\eta_{\gamma}}$).
Consequently, the minimal area enclosure $\overline{S}_{\min}$ of
$\partial \Sigma $ in $(\Sigma,\ggamma)$ 
satisfies
\begin{eqnarray}
| \overline{S}_{\min} |_{\ggamma} \geq |\overline{S}_{\min} |_{\gamma} \geq |\partial \Sigma |_{\gamma},
\label{chain}
\end{eqnarray}
where the subindex denotes which metric is used to calculate the area and the second inequality
holds because $\partial \Sigma$
is area outer minimizing in $(\Sigma,\gamma)$. Thus, 
an upper bound for  $|\overline{S}_{\min}|_{\ggamma}$
(via the Riemannian Penrose inequality) implies an upper bound for
$| \partial \Sigma |_{\gamma}$, which is the type of information relevant for the
full Penrose inequality.

Showing existence of solutions
of the generalized Jang equation (\ref{GenJangEq}) which satisfy this
criterion is a fundamental open issue in this approach.
Nevertheless, by construction there is an interesting particular
case where (\ref{GenJangEq}) admits solutions, namely when the initial data set $(\Sigma,\gamma_{ij},A_{ij})$ is in fact a slice of a static spacetime,
i.e. when there exist functions $\phi>0$ and $f$ such that
$A_{ij} = \hh_{ij}$. In this case not only the generalized
Jang equation holds trivially, but also  $\vec{\zz} =0$. 
Using (\ref{Generalized}), this implies $R (\ggamma) \geq 0$.
If $f$ decays fast enough at infinity (for instance if $f$ is of compact support) then $M_{ADM} (\gamma)
= M_{ADM} (\ggamma)$. Assuming that $\partial \Sigma$
is a generalized apparent horizon, it follows
that $\partial \Sigma$ is a minimal surface in $(\Sigma,\ggamma_{ij})$ provided
$\phi$ and $f$ are smooth up to the boundary and $\phi  |_{\partial \Sigma} =0$ (it should
be remarked that these two conditions
are quite restrictive, for instance they hold for slices of the Kruskal spacetime
only if they intersect
the black hole event horizon precisely at the bifurcation surface $\{u=v=0\}$). The Riemannian Penrose inequality then
gives $M_{ADM}(\ggamma)  \geq \sqrt{ |\overline{S}_{\min} |_{\ggamma}/(16 \pi)}$, and hence
the Penrose inequality $M_{ADM} (\gamma) \geq \sqrt{|\partial \Sigma|_{\gamma} / (16 \pi)}$ follows from
(\ref{chain}). 

Returning to the general case, whenever the generalized Jang equation is satisfied, 
the curvature scalar of $\ggamma_{ij}$ reduces to
\begin{eqnarray}
R (\ggamma ) & = &  16 \pi \left ( \rho - J_i v^i \right ) + || \hh - A ||^2_{\ggamma} 
+ 2 |\vec{\zz}\, |^2_{\ggamma} - \frac{2}{\phi} \mbox{div}_{\ggamma} (\phi \vec{\zz}\, ). 
\label{RicciUnderJang}
\end{eqnarray}
This expression has no sign in general, so the Riemannian Penrose inequality cannot
be applied directly. However, the generalized Jang equation involves two
unknowns, so it must be supplemented by a second condition in order to have
a determined problem. Bray and Khuri discuss
two possibilities. 

The simplest one consists in putting equal to zero
the  last summand in (\ref{RicciUnderJang}), i.e. 
\begin{eqnarray}
\mbox{div}_{\ggamma} (\phi \vec{\zz} \,) =0.
\label{divergence}
\end{eqnarray}
The two equations (\ref{GenJangEq})-(\ref{divergence}) are called the {\it Jang -  zero  divergence
equations} in \cite{BrayKhuri2009}. The equation (\ref{divergence}) is third order in $f$. However,
after substracting suitable derivatives of (\ref{GenJangEq}) it can be converted into a second order
equation for $f$ (with quadratic second derivatives). The resulting system is degenerate
elliptic. Bray and Khuri conjecture that the system
admits solutions with appropriate behaviour at infinity and such that 
$\partial \Sigma$ is a minimal surface with respect to $\ggamma_{ij}$.
Under this conjecture, the 
Penrose inequality as stated in (\ref{BrayKhuriConjecture-2}) follows (modulo a
technical point regarding the equality case, see Conjecture 7 in \cite{BrayKhuri2009}).

The second possibility 
is based on the observation that, while (\ref{GenJangEq})
does not imply $R (\ggamma)\geq 0$, the integrated inequality
$\int_{\Sigma} \phi R(\ggamma) \bm{\eta_{\ggamma}} 
\geq 0$ follows from (\ref{RicciUnderJang})
under suitable decay conditions at infinity and restrictions on $\partial \Sigma$.
Bray and Khuri make the interesting observation that in any situation
(irrespective of whether $R(\ggamma) \geq 0$ or not) where the 
mass $M_{ADM} (\ggamma)$ can be shown to satisfy 
\begin{eqnarray}
M_{ADM} (\ggamma) - \sqrt{\frac{|S_{\min}|_{\ggamma}}{16 \pi}} \geq \int_{\Sigma} Q(x) R (\ggamma) \bm{\eta_{\ggamma}}
\label{GeneralType}
\end{eqnarray}
with some $Q(x) \geq 0$, then the prescription
$\phi = Q$ implies $M_{ADM} (\gamma) \geq \sqrt{|S_{\min}|_{\ggamma}/(16 \pi)}$
and hence the Penrose inequality (\ref{Ineq}) provided $M_{ADM}(\gamma) \geq M_{ADM} (\ggamma)$. 
The question is, therefore, under which circumstances a general type inequality of the form (\ref{GeneralType})
holds. As the authors point out,
any such  inequality would imply the Riemannian Penrose inequality as a particular case.
It is therefore natural to study whether the known proofs of the Riemannian Penrose inequality are capable of 
establishing the validity of (\ref{GeneralType}) for some $Q \geq 0$.
The authors show explicitly that
this is indeed the case for the Huisken and Ilmanen method,
provided the second homology class of $\Sigma$
is trivial and $\partial \Sigma$ is connected. The idea is to integrate (\ref{dMG}) with respect to 
$\lambda$ and convert the double integral (in $\lambda$ and on $S_{\lambda}$) as a volume integral. 
By using the weak formulation in terms of level sets, this can be accomplished even along the
jumps. The result is that $Q = | \overline{\nabla} u |_{\ggamma} \sqrt{e^u
|S_{\min}|_{\ggamma}}/(16 \pi)^{3/2}$
where $u$ is the weak solution of the inverse mean curvature flow equal to
zero on $\partial \Sigma$.
Thus, existence of appropriate solutions for the pair of equations (\ref{GenJangEq}) and $\phi = Q(x)$ implies the general
Penrose inequality (\ref{Ineq}) for connected $\partial \Sigma$ (assuming
that $\Sigma$ has trivial second homology class).
As noted by the authors, existence in this case looks harder 
than for the Jang-zero divergence system 
because $Q$ vanishes identically wherever
the inverse mean curvature flow jumps. 
The equation therefore implies $\phi =0$ on the jumps.
However, if $f$ stays smooth, then $\hh_{ij} =0$  there (see (\ref{NewSecForm})). 
But then, the
generalized Jang equation
(\ref{GenJangEq}) requires $\tr_{\gamma} A=0$ along the jumps, which is a
condition on the initial data  and not an equation.
Thus, existence of classical solutions of the system 
$\phi =Q$ and (\ref{GenJangEq})
should not be expected in general.
It may be, however, that existence can be granted if $f$ is allowed to be unbounded (or even undefined)  in suitable regions.

The other existing method to prove the Riemannian Penrose inequality is the conformal flow of metrics
due to Bray \cite{Bray2001}. As discussed in \cite{BrayKhuri2009},
this method is also capable of producing an inequality of the form (\ref{GeneralType}). In this case $Q$
is expected to be continuous and strictly positive outside $S_{\min}$. On the other hand, the
resulting equation $\phi = Q$
is not local, in the sense that it does not define a local P.D.E. at each point. Existence of the
coupled system with the generalized Jang equations appears to be difficult in this case as well.

\section{Stronger versions of the Penrose inequality}
\label{stronger}

The Penrose inequality can be strengthened under some circumstances,  in particular when suitable
matter fields are present in the spacetime. In order to understand heuristically why this is to be expected,
let us return to the original argument by Penrose based on cosmic censorship. Assume that the 
collapsing matter is electrically charged and that the end-state of the collapse is a black hole in
equilibrium. In this situation, all the matter sources of the electromagnetic field are expected to lie within the
event horizon and the black hole is therefore electrovacuum in its exterior. According to the
black hole uniqueness result, the end-state is therefore a Kerr-Newman black hole. The area
of any cut of the event horizon in this spacetime is given by
\begin{eqnarray}
|S| = 4 \pi \left ( 2 M^2 - Q^2 + 2 M \sqrt{M^2 - L^2/M^2 - Q^2} \right ),
\label{KN}
\end{eqnarray}
where $L$ is the total angular momentum of the final state and $Q$ the total electric charge. 
From (\ref{KN}) it follows immediately $|S| \leq 4 \pi ( M + \sqrt{M^2 - Q^2} )^2$
which makes no reference to the angular momentum. Since the total electric charge of the spacetime is
conserved provided no charged matter escapes to infinity, the Penrose heuristic argument implies that any 
asymptotically euclidean electrovacuum initial data set 
should satisfy the inequality
\begin{eqnarray}
\label{PICharged}
\sqrt{\frac{A_{\min} (\partial \T^{+}_{\Sigma} )}{16 \pi}} \leq 
\frac{1}{2} \left (M_{ADM} + \sqrt{M_{ADM}^2 - Q^2 } \right ).
\end{eqnarray}
In the time  symmetric case, the electrovacuum initial data reduces to the triple $(\Sigma,\gamma_{ij},E_i)$, where 
the electric field $\vec{E}$ satisfies $\mbox{div}_{\gamma} \vec{E} =0$. The total charge is defined
as $4 \pi Q = \int_S (\vec{E} \cdot \vec{m} ) \bm{\eta_{S}}$ where $S$ is homologous to any large sphere in the asymptotically euclidean end. The inequality (\ref{PICharged}) simplifies to
\begin{eqnarray}
\label{PIChargedSym}
\sqrt{\frac{|S_{m}|}{16 \pi}} \leq 
\frac{1}{2} \left (M_{ADM} + \sqrt{M_{ADM}^2 - Q^2 } \right ).
\end{eqnarray}
where $S_{m}$ is the outermost minimal surface. For this inequality to make sense it
is necessary that $M_{ADM} \geq |Q|$ for all electrovacuum initial data. This is a strengthening
of the positive mass theorem in the presence
of electromagnetic
fields and was first proven in \cite{Gibbons_Hawking_1983} (see \cite{GibbonsHull1982}
for a generalization including matter and \cite{Chrusciel_Reall_Tod2006} for a rigorous statement).
The minimum value of the right-hand side of (\ref{PIChargedSym}) is $|Q|/2$. Thus, 
for horizons of small area ($|S_m| < 4 \pi Q^2$) the Penrose inequality (\ref{PIChargedSym})
reduces to the positive mass theorem $M_{ADM} \geq |Q|$, with no reference to the area of the horizon. 
This implies, in particular, that (\ref{PIChargedSym}) does not admit an equality case (i.e.
a rigidity statement) for horizons of small area. On the
other hand, the equality case in the Penrose inequality 
$|S_m| \leq 16 \pi M_{ADM}^2$  has the interesting
consequence of providing 
a variational characterization of the Schwarzschild metric (\ref{gsch}) as the absolute
minimum of total mass among asymptotically euclidean, Riemannian manifolds of non-negative Ricci
scalar and with an outermost minimal surface
of given area $|S_m|$. This is similar to the variational characterization of Euclidean space as the absolute
minimum of total mass among asymptotically euclidean Riemannian manifolds with $R(\gamma) \geq 0$. 

A natural question is whether there exists another version of the charged Riemannian Penrose inequality
which is able to give a variational characterization (among metrics of fixed charge and fixed area of 
the outermost minimal surface) of the Reisner-Nordstr\"om and Papapetrou-Majumdar metrics, which are the only
static and electrovacuum  black holes (see \cite{ChruscielTod2007} and references therein). A strengthening
of (\ref{PIChargedSym}) that has been proposed is
\begin{eqnarray}
M_{ADM} \geq \frac{1}{2} \left (\sqrt{\frac{|S_m|}{4 \pi}}  + Q^2 \sqrt{\frac{4 \pi}{|S_m|}} \right ). 
\label{PIChargedCon}
\end{eqnarray}
This inequality was first discussed and proven by Jang \cite{Jang1979} in the case
of asymptotically euclidean, electrovacuum initial data sets $(\Sigma,\gamma_{ij}, E_{i})$ with a {\it
connected} outermost minimal surface $S_m$, provided the inverse mean curvature flow starting on $S_m$
remains smooth. This last requirement is, in fact, unnecessary thanks to the weak formulation of the flow
introduced by Huisken and Ilmanen. This establishes (\ref{PIChargedCon}) for connected $S_m$.

Inequality (\ref{PIChargedCon}) is, however, not generally true when the outermost minimal 
surface is allowed to have several connected components. A counterexample has been found by
Weinstein and Yamada \cite{Weinstein-Yamada2005}. Their basic idea 
was  to realize that
the Papapetrou-Majumdar spacetime, which represents a static configuration of $N$
black holes of masses $m_i>0$ and charges $Q_i = \epsilon m_i$, with $\epsilon =\pm 1$,  
has the property that the total area $|S|$ of the event horizon violates the inequality (\ref{PIChargedCon}). Indeed, using $Q= M_{ADM}$, it follows
\begin{eqnarray*}
M_{ADM} - \frac{1}{2} \left (\sqrt{\frac{|S|}{4 \pi}}  
+ Q^2 \sqrt{\frac{4 \pi}{|S|}} \right ) 
= - \frac{1}{2} \sqrt{\frac{4 \pi}{|S|}} \left ( M_{ADM} - \sqrt{\frac{|S|}{4 \pi}}  \right )^2 \leq 0,
\end{eqnarray*}
irrespectively of the value of $|S|$. The only way how (\ref{PIChargedCon}) could hold is $|S| = 4 \pi M^2_{ADM}$. 
However, a simple computation gives $|S| = 4 \pi \sum_{i} ( m_i)^2$. Since the ADM mass is
$M_{ADM} = \sum_i m_i$, equality can only happen where there is only one black hole (i.e.
when the spacetime is the extreme Reissner-Nordstr\"om black hole). For any configuration with two
or more black holes, (\ref{PIChargedCon}) is violated for the area of the event horizon.
This argument is however, not a proof that (\ref{PIChargedCon}) fails to hold because the static 
initial data set (i.e. the hypersurface
orthogonal to the static Killing vector) in the Papapetrou-Majumdar spacetime does not contain any minimal
surface. Each connected component of the event horizon corresponds to an asymptotic cylinder. Thus, some engineering
is required to construct an electrovacuum initial data set with a minimal surface and which violates
(\ref{PIChargedCon}). The method followed in \cite{Weinstein-Yamada2005} consists in an adaptation of the 
gluing technique developed in \cite{Isenberg_Mazzeo_Pollack_2002}. More specifically, it consists in
taking two copies of a static initial data set of the Papapetrou-Majumdar spacetime with two black holes of equal
mass $m_1=m_2=m$, such that one of the copies has positive charges and the other one negatives charges. By modifying
the geometry far enough along the cylindrical ends, the two copies can be glued together to construct
an initial data set with two asymptotically euclidean ends and a minimal surface with two connected components. The
final step is to conformally transform the data so that the curvature  scalar vanishes.
By taking $m$
small enough, the resulting manifold violates the inequality (\ref{PIChargedCon}).
As the authors stress,  this is not a counterexample of
(\ref{PIChargedSym}), which is the inequality that follows from Penrose's heuristic argument.

Returning to the question of whether the charged Riemannian Penrose inequality provides a variational characterization
of the electrovacuum static black holes (see also \cite{Gibbons1984} for a related discussion),
we notice that the inequality (\ref{PIChargedSym})  can be written in the
following equivalent way
\begin{eqnarray}
\left \{ \begin{array}{ll}
                  M_{ADM} \geq |Q| & \quad \mbox{if} \quad |S_m| \leq 4 \pi Q^2 \quad \mbox{(case (i))} \\
                  M_{ADM} \geq \frac{1}{2} \left (\sqrt{\frac{|S_m|}{4 \pi}}  + Q^2 \sqrt{\frac{4 \pi}{|S_m|}} \right ) &
\quad \mbox{if} \quad |S_m|\geq  4 \pi Q^2  \quad \mbox{(case (ii))} \\
\end{array} \right . \label{split2}
\end{eqnarray}
The Reissner-Nordstr\"om black holes have event horizons (or equivalently outermost minimal surfaces in their
static initial data) with satisfy $|S_m| \geq 4 \pi Q^2$ (see (\ref{KN}) with $L=0$), with equality only for
the extreme Reisner-Nordstr\"om case ($M=|Q|$). So, these metrics belong to case (ii) above and,
in fact, saturate the corresponding inequality. Similarly, 
the event horizon of the Papapetrou-Majumdar spacetime has area 
$|S| = 4 \pi \sum_i m_i^2 \leq 4 \pi \left ( \sum_i m_i \right )^2 = 4 \pi Q^2$, with equality only if there is only
one black hole (i.e. the metric is extreme Reissner-Nordstr\"om again). So, this case belongs to case (i)
and obviously the inequality is again saturated. Moreover, the Papapetrou-Majumdar static initial data
is the only asymptotically euclidean, electrovacuum initial data $(\Sigma,\gamma_{ij},E_{i})$ satisfying $M=|Q|$ (see
Theorem 1.2 in \cite{Chrusciel_Reall_Tod2006} as well as the related previous work  \cite{GibbonsHull1982},
\cite{Tod1983}). Thus, the formulation (\ref{split2}) would in fact provide a
variational characterization of all static charged black holes provided the inequality can be proven
in case (ii) with equality only for the Reissner-Nordstr\"om initial data.

In the non time-symmetric case, Gibbons conjectured \cite{Gibbons1984}
the inequality (\ref{PIChargedCon}) for connected
and outermost future (or past) marginally outer trapped surfaces $S$.
In the non-connected case, the corresponding conjecture
involves the sum over each connected component of the right-hand side
of (\ref{PIChargedCon}). Although no counterexample is explicitly known,
such an inequality would reduce in vacuum to a stronger statement
than the standard Penrose inequality. As noted by
Weinstein and Yamada \cite{Weinstein-Yamada2005}, it seems that
an initial data representing two Schwarzschild black holes sufficiently far
apart should violate this version of the inequality. 
In the particular case of spherical symmetry (where
$S$ is automatically connected), Gibbons conjecture has been proven by
Malec  and \'O Murchadha \cite{MalecMurchadha1994} for maximal initial data sets and by Hayward 
\cite{Hayward1998} in the general case.

In the discussion above we have dropped the angular momentum term in (\ref{KN}) and have
retained the charge. It is natural to ask what is the situation 
in the reverse case, i.e. when the charge is dropped
and the angular momentum in kept. The inequality that results is
$|S| \leq  8 \pi M ( M + \sqrt{M^2 - L^2/M^2 })$. However, contrarily to the electromagnetic case, the
total angular momentum of the final end-state after the collapse has been completed need not coincide
with the initial one since gravitational waves carry angular momentum and this can be radiated away.
As first discussed in \cite{Friedman_Mayer_1982} (see also \cite{Horowitz1984}), there is one important
situation where angular momentum must be conserved along the evolution, namely in the axially symmetric case.
Under this restriction, the Penrose heuristic argument implies 
\cite{Hawking1972},\cite{Dain_et_al_2002},
\begin{eqnarray}
\label{PIAngularMomentum}
A_{\min} (\partial \T^{+}_{\Sigma} ) 
\leq 
8 \pi M_{ADM} \left (M_{ADM} + \sqrt{M_{ADM}^2 - L^2/M_{ADM}^2 } \right ).
\end{eqnarray}
Similarly as before, this inequality only makes sense provided one can show that any asymptotically euclidean
and axially symmetric initial data set satisfying the dominant energy condition 
$\rho \geq |\vec{J}|$ satisfies the inequality $M_{ADM} \geq \sqrt{|L|}$.
Again, this is a strengthening of the positive mass theorem. This inequality is supported by a heuristic
argument based on cosmic censorship and the conservation of angular momentum in the axially symmetric case
\cite{DainPRL2006} and its validity has been rigorously proven in 
\cite{Dain2008} for any initial data set $(\Sigma,\gamma_{ij},A_{ij})$
which is vacuum, maximal
($\mbox{tr}_{\gamma} A=0$), contains one or more asymptotically euclidean ends 
as well as possibly additional
asymptotically cylindrical ends (which correspond to degenerate horizons) and such that
the outermost MOTS is connected 
(see also \cite{Chrusciel_Li_Weinstein_2008} for an extension which
admits furthermore non-negative energy-density and relaxes some of the technical requirements in
\cite{Dain2008}). Moreover, the case
of equality $M_{ADM} = \sqrt{|L|}$ occurs if and only if
the initial data is a slice of the extreme Kerr black hole.  This provides
a variational characterization of extreme Kerr. The inequality in the
case with an outermost MOTS with several connected components remains still open. Numerical 
evidences for its validity have been recently given in \cite{DainOrtiz2009}.

The situation for the Penrose inequality with angular momentum 
is therefore similar to the charged
case. The inequality (\ref{PIAngularMomentum}) is equivalent to 
\begin{eqnarray}
\left \{ \begin{array}{ll}
                  M^2_{ADM} \geq |L| & \quad \mbox{if} \quad |S| \leq 8 \pi |L| \quad \mbox{(case (i))} \\
                  M^2_{ADM} \geq \frac{|S|}{16 \pi}  + \frac{4 \pi L^2}{|S|} &
\quad \mbox{if} \quad |S|\geq  8 \pi |L|   \quad \mbox{(case (ii))} \\
\end{array} \right . \label{M-L-2}
\end{eqnarray}
where $|S| = A_{\min} (\partial \T^{+}_{\Sigma} ) $. This version of the Penrose conjecture (for axially 
symmetric initial data sets) admits a rigidity case which states that equality in case (ii) can only occur of the 
initial data is a slice of the Kerr black hole. Again, this would provide a variational characterization
of the Kerr metric.

\section{Some applications of the Penrose inequality}
\label{applications}

In this section we briefly mention some recent situations where the Penrose inequality has been 
exploited to derive new results. The list is probably not exhaustive, but it gives a hint on the potential
power of the Penrose inequality as a geometric tool for adressing, a priori, completely unrelated problems.

We have already mentioned in Subsect. \ref{HuiskenIlmanen} 
that the Riemannian Penrose inequality has interesting consequences for the
quasi-local definition of mass due to Bartnik. The Riemannian Penrose inequality has also allowed
for a dual definition of quasi-local mass due to Bray \cite{Bray2001} (see also \cite{BrayChrusciel2004}). 
Here, an asymptotically euclidean domain with non-negative curvature scalar and whose boundary $S$ is area outer minimizing
is kept fixed, and  all possible ``fill in''s (with non-negative curvature scalar) are considered.
The {\it inner mass} is defined as the supremum of $\sqrt{|S|/(16\pi)}$
where $S$ is the minimal area needed to enclose all the asymptotically euclidean ends, except the given one. 
As a consequence of
the Riemannian Penrose inequality, the inner mass is always bounded above by the ADM mass 
of the given region.
The definition can also be extended to the case of non-zero second fundamental form.

We have also mentioned in Section \ref{hyperbolicSect}
another application of the Penrose inequality for the uniqueness problem of
static black holes with negative cosmological constant and topology at infinity of genus larger than one.

The Riemannian Penrose inequality (in fact, its proof using inverse mean curvature flow) has been applied
recently to obtain lower bounds of the so-called Brown-York energy for simply connected, compact, three-dimensional
domains $(\Omega,\gamma)$ with non-negative curvature scalar and with smooth boundary $\partial \Omega$ of positive
Gauss curvature and positive mean curvature  $p$ (with respect to the outer direction). The Brown-York mass
is defined as
\begin{eqnarray*}
M_{BY} (\partial \Omega ) = \frac{1}{8 \pi} \int_{\partial \Omega} \left ( p_0 - p \right ) \bm{\eta_{\partial \Omega}},
\end{eqnarray*}
where $p_0$ is the mean curvature of $\partial \Omega$ when this surface is embedded isometrically
in $\mathbb{R}^3$. This mass is proven to be non-negative in \cite{ShiTam2002}. Using the inverse mean curvature flow,
Shi and Tam prove \cite{ShiTam2007} (among other things)  that the inequality
$M_{BY} (\partial \Omega) \geq M_{G} (\partial \Omega)$ holds with equality if only if $\Omega$ is a
standard ball in $\mathbb{R}^3$. 

Still another
application of the Riemann Penrose inequality is due to J. Corvino \cite{Corvino2005}, who has shown that 
asymptotically euclidean, 3-dimensional, Riemannian manifolds with non-negative curvature scalar and small mass cannot 
contain any minimal surface and must be diffeomorphic to $\mathbb{R}^3$. The required  ``small mass''
condition reads $2 M_{ADM} \sqrt{K} \leq 1$, where $K$ is a positive upper bound for all the sectional
curvatures of the manifold. The proof is based on the fact that
any outermost minimal 
surface $S$ must satisfy $|S| \geq \frac{4 \pi}{K}$ as a consequence
of the Gauss-Bonnet theorem. Therefore, the Penrose inequality implies
that no minimal surface can exist in these circumstances.

\section {Concluding remarks}
\label{concluding}

In this review I have discussed the present status of 
the Penrose inequality. The emphasis has been put on trying to describe the
techniques involved in the various approaches to prove it and trying to place the results into the right context 
so that a clear picture emerges of how impressive the body of work in this field has already been
and what are the open problems that still remain. Although I have tried to cover the main results in this
field, some aspects have been touched upon in less detail. For instance, I have concentrated mostly
on the four dimensional case, although several results in higher dimensions have been
discussed at various places. Some further results in 
higher dimensions can be found in the references 
\cite{IdaMakao}, \cite{BarrabesFrolov}, \cite{GibbonsHolzegel}, \cite{DafermosHolzegel}.

Numerical work has also been important for a better understanding of the Penrose inequality. Outermost
marginally outer trapped surfaces (i.e. the boundary of the outer trapped region $\partial \T^{+}_{\Sigma}$) are
located routinely in numerical black hole evolutions in order to track the boundaries
of the black holes. The numerical routines to do this job are collectively termed
{\it apparent horizon finders} (see \cite{Thornburg2007} for a review) and obviously they can also be
used to test the validity of the Penrose conjecture. They have been used to 
check  whether the Penrose inequality
is fulfilled in explicit numerical examples, as well as for looking for counterexamples to
some of its versions. They can also serve as a test to make sure that the MOTS being located
is, in fact, the outermost one \cite{Jaramillo_Vasset2007}. In this review, numerical results have been
mentioned only very tangentially. Further details can be found in
\cite{Karkowski_Malec_1993}, \cite{Karkoswski_Koc_1994}, \cite{Husain1998}, 
\cite{Dain_et_al_2002}, \cite{Karkowski-Malec2005},  \cite{Karkowski2006}, 
\cite{Jaramillo_Vasset2007},
\cite{Jaramillo_Kroon} 
and \cite{Tippett2009}.

In a 3+1 evolution of a spacetime the outermost MOTS generates a tube of surfaces which is generally
smooth but may jump from time to time \cite{Mars_et_al_2009}. In the smooth part, this tube has been called
marginally outer trapped tube \cite{Booth2005}. If the foliation is by marginally trapped surfaces
(instead of MOTS), the tube is usually called 
trapping horizon \cite{Hayward1994-2} or dynamical horizon \cite{AshtekarKrishnan2002} (with slight differences
in their definitions). A proper study of the evolution of these tubes, specially their late time
behaviour, is potentially a powerful method for establishing the Penrose inequality \cite{Mars_et_al_2005}.
This is because it is expected that, at late times,
the tube approaches the event horizon of the spacetime. Furthermore, if the MOTS foliating the tube
are in fact marginally trapped surfaces, then their area increases with time \cite{AshtekarKrishnan2002}.
Since this
area is believed to approach that of the event horizon of the final black hole that forms (and this
is greater than the initial ADM mass, as usual) the Penrose inequality would follow from a detailed
understanding of the  late time evolution of the spacetime and of the outermost marginally
outer trapped tube. This approach however, is very different in spirit to the ones discussed above
because understanding the late time behaviour of the tube goes a long way towards establishing
weak cosmic censorship. Thus, in essence, the Penrose inequality would follow because cosmic censorship
would hold. So, instead of looking at the Penrose inequality as an indirect test of cosmic
censorship, as originally envisaged by Penrose, it would become a remarkable corollary of a 
much stronger theorem 
establishing weak cosmic censorship (or something very close to it).
In any case, studying the evolution of the outermost tube is an active area of research, which combines
physical properties, numerical simulations  and rigorous geometric results. The interested
reader is referred to \cite{AshtekarKrishnan2003}, \cite{AshtekarGalloway2005},
\cite{Mars_et_al_2005}, \cite{Andersson2006}, \cite{Schnetter_et_al2006},
\cite{AshtekarKrishnan2007},
\cite{Jaramillo_Kroon}, 
\cite{Gourgoulhon2008}, 
\cite{Williams2008}, 
\cite{AMS08}, 
\cite{Hayward2009}, 
\cite{Mars_et_al_2009} and references therein.

To conclude, as I have tried to show in this review, the Penrose conjecture is a very challenging problem
that requires techniques from geometric analysis, partial differential equations, Riemannian and
Lorentzian geometry, as well as physical intuition. The recent advances in this field have been 
impressive and our understanding of the problem is now better than ever. Nevertheless, many open
problems remain and their study is likely to uncover new and unexpected features in the future.

\section*{Acknowledgments}

I am indebted to Hugh L. Bray, Alberto Carrasco, Jos\'e Luis Jaramillo,
Markus Khuri, Miguel S\'anchez, Jos\'e M.M. Senovilla, Juan
Valiente Kroon and Ra\"ul Vera for 
useful comments on a previous version of this paper.
Financial support under projects
FIS2006-05319 of the Spanish MEC, SA010CO5 of the Junta de Castilla y Le\'on 
and P06-FQM-01951 of the Junta de Andaluc\'{\i}a is gratefully acknowledged.

\end{document}